\documentclass[12pt]{article}
\usepackage{graphicx}
\usepackage{amsmath}
\usepackage{amssymb}
\usepackage{array}
\usepackage{cite}
\usepackage{epsfig}
\usepackage[english]{babel}
\usepackage[latin1]{inputenc}
\usepackage[T1]{fontenc}
\usepackage{amsfonts}
\usepackage{float}
\usepackage{sidecap}
\usepackage{amsbsy}
\usepackage{caption}
\usepackage{multirow}
\usepackage{longtable}
\usepackage{color}
\usepackage{lscape}
\usepackage[dvips,pdftitle={D-Branes at del Pezzo Singularities: Global Embedding and Moduli Stabilisation}, pdfauthor={Michele Cicoli, Sven Krippendorf, Christoph Mayrhofer, Fernando Quevedo, Roberto Valandro}, pdfsubject={string phenomenology}, pdfkeywords={Kaehler moduli stabilisation, string phenomenology, quiver gauge theories}]{hyperref} 
\usepackage{collref}

\newcommand{\bea}{\begin{eqnarray}}
\newcommand{\eea}{\end{eqnarray}}
\newcommand{\vo}{{\cal V}}
\newcommand{\be}{\begin{equation}}
\newcommand{\ee}{\end{equation}}
\newcommand{\mc}{\mathcal}
\newcommand{\mbb}{\mathbb}

\def\ba{\begin{eqnarray}}
\def\ea{\end{eqnarray}}
\def\nn{\nonumber}

\makeatletter
\def\x@arrow{\DOTSB\Relbar}
\def\xlongequalsignfill@{\arrowfill@\x@arrow\Relbar\x@arrow}
\newcommand{\xlongequal}[2]{%
    \ext@arrow 0099\xlongequalsignfill@{#1}{#2}}
\makeatother

\newcommand{\roughly}[1]{\mathrel{\raise.3ex\hbox{$#1$\kern-0.85em
\lower1ex\hbox{$\sim$}}}}

\def\endignore{}
\def\ignore #1\endignore{}
\def\nn{\nonumber}

\def\beq{\begin{equation}}
\def\eeq{\end{equation}}
\def\beqa{\begin{eqnarray}}
\def\eeqa{\end{eqnarray}}

\newcommand{\bmat}{\left(\begin{array}}
\newcommand{\emat}{\end{array}\right)}

\def\endignore{}
\def\ignore #1\endignore{}

\def\-{\hphantom{-}}

\def\s2{\frac{1}{2}}

\def\IF{\relax{\rm I\kern-.18em F}}
\def\II{\relax{\rm I\kern-.18em I}}
\def\IP{\relax{\rm I\kern-.18em P}}
\def\IC{\relax{\rm I\kern-.48em C}}
\def\IR{\relax{\rm I\kern-.18em R}}
\def\IK{\relax{\rm I\kern-.20em K}}
\def\IM{\relax{\rm I\kern-.25em M}}

\def\Dsl{\,\raise.15ex\hbox{/}\mkern-13.5mu D}

\def \one{\relax{\rm 1\kern-.26em I}}

\def\nn{\nonumber}

\def\({\left(}
\def\){\right)}

\newcommand{\G}{\hat{\Gamma}}

\begin{document}
\begin{titlepage}
\begin{flushright} DAMTP-2012-47\\ ZMP-HH/12-10 
\end{flushright}
\vspace{0.02cm}
\begin{center}
{\bf \LARGE D-Branes at del Pezzo Singularities:\\ \vspace{5mm} Global Embedding and Moduli Stabilisation
}\\[0.81cm]
Michele Cicoli${}^{1,2},$ Sven Krippendorf${}^{3},$ Christoph Mayrhofer${}^4,$ \\ Fernando Quevedo${}^{1,5},$ Roberto Valandro${}^6$\\[0.3cm]

{\it \small $^1$ ICTP, Strada Costiera 11, Trieste 34014, Italy.\\
$^2$ INFN, Sezione di Trieste, Italy.\\
$^3$ Bethe Center for Theoretical Physics and Physikalisches Institut der\\ Universit\"at Bonn, Nussallee 12, 53115 Bonn, Germany.\\
$^4$ Institut f\"ur Theoretische Physik, Universit\"at Heidelberg,\\ Philosophenweg 19,  69120 Heidelberg, Germany.\\
$^5$ DAMTP, University of Cambridge, Wilberforce Road, Cambridge, CB3 0WA, UK.\\
$^6$ II. Institut f\"ur Theoretische Physik, Universit\"at Hamburg,\\ Luruper Chaussee 149, 22761 Hamburg, Germany.}
\\[0.5cm]

\date{\today}
\end{center}
\begin{abstract} 
\noindent In the context of type IIB string theory we combine moduli stabilisation
and model building on branes at del Pezzo singularities in a fully consistent global compactification.
By means of toric geometry, we classify all the Calabi-Yau manifolds
with $3<h^{1,1}<6$ which admit two identical del Pezzo singularities mapped into each other under the orientifold involution.
This effective singularity hosts the visible sector containing the Standard Model while the K\"ahler moduli are
stabilised via a combination of D-terms, perturbative and non-perturbative effects supported on hidden sectors.
We present concrete models where the visible sector, containing the Standard Model, gauge and matter content,
is built via fractional D3-branes at del Pezzo singularities and all the
K\"ahler moduli are fixed providing an explicit realisation of both KKLT and LARGE volume scenarios, the latter with D-term uplifting to de Sitter minima.
We perform the consistency checks for global embedding
such as tadpole, K-theory charges and Freed-Witten anomaly cancellation.
We briefly discuss phenomenological and cosmological implications of our models.
\end{abstract}

%\noindent\footnotesize{Email: \texttt{mcicoli@ictp.it}, \texttt{krippendorf@th.physik.uni-bonn.de}, \texttt{c.mayrhofer@thphys.uni-heidelberg.de}, \texttt{f.quevedo@damtp.cam.ac.uk}, \texttt{roberto.valandro@desy.de}.}

\end{titlepage}
\newpage
\tableofcontents
\newpage

%=======================================================================
\section{Introduction}
%=======================================================================

Realistic string model building is a major task~\cite{iu} not that much because there is a very large number of string compactifications,  but mostly because the experimental constraints that have to be satisfied are very restrictive. Conventional field theoretical models neglect the quantisation of gravity, allow for the inclusion of as many particles and symmetries as needed and, usually, focus on one or very few features beyond the Standard Model (SM).
On the contrary, for a string model to be realistic, it has to be consistent with {\it all} experimental observations in particle physics and cosmology. This includes (gauge) symmetries, matter content, couplings and mixings, an explanation of baryogenesis, dark matter, dark energy and the density perturbations of the cosmic microwave background. At the same time it has to avoid unrealistic additional features, such as un-observed long range interactions and fractionally charged particles, or unwanted topological defects like monopoles or domain walls
and avoid cosmological features that could be against the successes of the standard cosmological scenario. Even if only one of these requirements is not satisfied the model must be ruled out.

The bottom-up approach to string model building~\cite{0005067} is an intermediate step between conventional and string model building. Given the magnitude of the problem to find a fully realistic string model, the original motivation was that it is more convenient to separate the construction in several modular steps. This can be done if the SM is hosted at branes localised within the compact manifold.

In these local models, questions regarding the gauge symmetry, matter content and SM couplings  (including the number of chiral families, flavour issues, gauge coupling unification) do not depend on the details of the full compactification and can be treated independently of global questions such as moduli stabilisation, the value of the cosmological constant, supersymmetry breaking and early universe cosmology.
This separation of challenges shares some of the advantages of conventional model building in the sense that only a few features are addressed at a time, while keeping the stringy nature of the models.

Substantial progress has been made in both of these fronts in type IIB string theory:

\begin{itemize}
\item{} Quasi realistic local models have been constructed from D-branes at singularities~\cite{0005067,0105042,0001083,0312051,0508089,0810.5660,1002.1790,1106.6039} and from F-theory constructions (for a review see~\cite{0911.3008,1001.0577,1009.3497}). At the moment there are a few local models with quasi-realistic spectrum, couplings and mixings, which is very encouraging.

\item{} Moduli stabilisation has been achieved by considering flux compactifications combined with perturbative and non-perturbative corrections to the string effective action~\cite{GKP,KKLT,0502058}. These constructions have allowed for the first time to stabilise all moduli, with controllable supersymmetry breaking. The fluxes determine a discrete set of solutions which allow to separate most physical questions from the cosmological constant problem. Early universe scenarios of inflation have also been found (see for recent reviews~\cite{1108.2659,1108.2660}) with features that can be subject to observational scrutiny in the next few years. Moreover one of the main approaches to moduli stabilisation, the LARGE Volume Scenario (LVS)~\cite{0502058,0505076} actually implies that the SM has to be localised. In fact, in such models the SM cannot be hosted by a brane wrapping the exponentially large cycle, giving an independent support to the bottom-up approach.
\end{itemize}
However, at the end of the day these two steps of the local constructions need to be combined in order to have completely realistic string compactifications.  Embedding local models in compact Calabi-Yau (CY) compactifications is highly non-trivial. Identification of manifolds with the right local structure and issues of consistency such as tadpole cancellation for all D-branes need to be addressed. Furthermore, there is a well-known tension between moduli stabilisation of the SM cycle and the existence of chiral matter~\cite{Blumenhagen:2007sm}, which, together with D-term conditions, tend to stabilise some K\"ahler moduli towards a collapsing cycle and therefore having the SM brane at a singularity~\cite{0811.2936,0906.3297,CKM}. This is the challenge that we start to address in this article: constructing explicit models of branes at singularities in a compact setting with the stabilisation of all moduli.

Models of D branes at singularities have several interesting phenomenological features that are worth emphasising:
\begin{enumerate}
\item Despite not having a simple GUT group the gauge couplings are unified to leading order.

\item The well-known problem of Yukawa couplings of brane models with simple GUT groups is not present. There are models with hierarchical masses for all quarks and leptons~\cite{0810.5660}. Furthermore, some of them include realistic flavour features such as the structure of CKM and PMNS matrices~\cite{1002.1790,1106.6039}.

\item The full perturbative superpotentials have been constructed for classes of toric singularities (and all del~Pezzo singularities). Leading to controllable studies of couplings, flat directions, etc. In particular, R-symmetries are identified implying the appearance of a very limited number of couplings (but rich enough to leave to realistic properties).

\item D-terms tend naturally to stabilise the corresponding modulus to the
singular locus. Some of the anomalous and non-anomalous $U(1)$s are broken
without a matter field getting a vacuum expectation value (VEV) and therefore leaving the associated
global symmetry unbroken, a property that can be used for forbidding proton
decay.
\end{enumerate}
Recently, progress has been made in two complementary directions.  In~\cite{1110.3333} explicit chiral models with moduli stabilisation 
were constructed in terms of magnetised D7-branes on a smooth CY. The previously encountered problems, such as tension between chiral matter and moduli stabilisation, tension  between Freed-Witten anomaly cancellation and non-perturbative effects and shrinking induced by D-terms, were overcome.
Regarding branes at singularities, compact models  were constructed in~\cite{DiacEtAl2,0610007,1201.5379}\footnote{One compact F-theory uplift of a realistic brane model was already presented in~\cite{0005067}.}, even though moduli stabilisation was not included. In this article, we extend the analysis of~\cite{1110.3333} to construct quasi realistic models on branes at singularities with a fully consistent global embedding (including tadpole, K-theory charges and Freed-Witten anomaly cancellation) combined with moduli stabilisation.

In order to make progress on this task we use the following systematic approach. Following the general discussion in~\cite{0810.5660} we look for CY manifolds with two identical, non-intersecting singularities that can be mapped to each other under the orientifold involution. This implies products of $U(n)$ gauge groups and bi-fundamental matter in the visible sector which is the most promising area for phenomenology. Further, rigid four-cycles (collapsed or not) are needed to have at least one hidden sector generating a non-perturbative superpotential required for K\"ahler moduli stabilisation. We use the classification of four-dimensional reflexive lattice polytopes made by Kreuzer and Skarke~\cite{0002240}. From this list we take the polytopes corresponding to CY manifolds with $h^{1,1}=4,5$ and identify those manifolds which have two identical non-intersecting del~Pezzo singularities (requiring $h^{1,1}_-\geq 1$ after orientifolding). In order to illustrate the techniques, we study in full detail a model with two dP$_0$ singularities to host the SM and a simple trinification extension. The possible breakdown to the SM gauge
group is outlined and non-vanishing quark and lepton Yukawa couplings are obtained. In Appendix~\ref{dP8appendix}, we also briefly discuss another model with dP$_8$ singularities. In both cases a complete study on moduli stabilisation is done, from a combination of D-terms, perturbative and non-perturbative effects.
A general strategy with interesting examples for obtaining quiver gauge theories without flavour branes useful for phenomenology is presented. In addition, we check which of the CY manifolds allow for a K3 fibration which can be useful to realise cosmological models~\cite{0808.0691,1005.4840,1110.6182,1202.4580,1203.6655} or anisotropic compactifications~\cite{1105.2107}.

The organisation of the article is as follows. In the next section we define the computational challenge of finding the 
CY manifolds with the desired property, spell out how the classification of models with $h^{1,1}\leq 5$ is done 
and review all the consistency conditions that have to be satisfied in order to have a full CY embedding of the model. 
We leave a detailed listing of the manifolds to appendix~\ref{sec:applist}. Section~\ref{sec:GlobalEmbdP0} describes in detail the geometry of a dP$_0$ model (we present a dP$_8$ model in Appendix~\ref{dP8appendix}), to illustrate how all consistency conditions are realised and finally how the SM can arise from D3-branes at dP$_0$ 
singularities without flavour branes. In Section~\ref{sec:HigherdPn:withoutFlBr} we discuss the phenomenology of dP$_n$ models with $n>0$, also without flavour branes. 
In Section~\ref{sec:modstab}, guided by the explicit example, we discuss  how moduli stabilisation is achieved. 
We conclude and list open questions in Section~\ref{sec:concl}.

\section{Global embeddings of D-branes at singularities
}\label{sec:2}

In this paper we want to present a systematic procedure to construct global models at del~Pezzo singularities (dP$_n$ with $n=0,\ldots,8$)\footnote{A dP$_n$ surface is given by
$\mbb{P}^2$ with $n$ generic points blown up to $\mathbb P^1$.} with the stabilisation of all moduli. 

From this class of singularities one can obtain phenomenologically interesting, if not the most promising, models by placing D3-branes on top of point-like del~Pezzo singularities in a CY geometry~\cite{0005067,0508089,0810.5660,1002.1790,1106.6039}. These singularities arise by shrinking a dP$_n$ four-cycle to zero size. One D3-brane on top of such a singularity splits into a collection of fractional branes with chiral modes in bi-fundamental representations of the resulting unitary gauge groups. Their matter and gauge content is conveniently summarised in a quiver diagram, where the nodes represent the gauge group and the arrows between them represent the bi-fundamental fields. The superpotential can be obtained via dimer techniques~\cite{0803.4474,1002.1790}.
In the geometric regime,  when the dP$_n$ has finite size, the fractional branes are roughly described as 7-branes that wrap the shrinking dP$_n$ four-cycle with a vector bundle (more precisely a sheaf) living on its world-volume. This set of 7-branes are stable only at the singular point (when the dP$_n$ has zero size). See~\cite{0711.2451} for a pedagogic review on that topic.

\subsection{Search strategy for compact Calabi-Yau embeddings} \label{sec:mod-stab-in-global-quiver}

Models coming from D-branes at singularities were analysed in detail from a local point of view, studying the theory around the singularity. In order to consider global issues like moduli stabilisation and supersymmetry breaking, one needs to embed such singularities in compact geometries. This is the aim of the present paper: we will consider type IIB orientifold vacua with O3/O7 planes, incorporating both dP$_n$ singularities which lead to a phenomenologically interesting quiver gauge theory and a global set-up realising moduli stabilisation.

To make contact with the real world, often the local models include also fractional flavour D7-branes, passing through the singularities. These branes make the global embedding more complicated, as explained recently in~\cite{1201.5379}. Therefore, we will consider quiver models without flavour branes, showing that also these models lead to interesting phenomenology and leave the study of globally embedding of  flavour branes  for the future.

We also want to restrict to quiver models that are not modded out by the orientifold action, since such models tend to produce more interesting phenomenological set-ups. This implies that the dP$_n$ singularity is not on top of the O7-plane. Hence, we need a CY manifold with two dP$_n$ singularities that are exchanged by the orientifold involution. The orientifold action, then, relates the physics at one singularity with the physics on the other one. Since a dP$_n$ four-cycle is rigid, the two four-cycles considered must be in different homology classes. This already leads to $h^{1,1}(X)>2$, for the CY three-fold $X$, as we also want to have finite volume. 

In addition to requiring a globally embedded quiver gauge theory without flavour D7-branes that is not modded out by the orientifold action, we implement the conditions to have moduli stabilisation. Given that our main interest is to stabilise the moduli realising the LVS, we need, on top of the overall volume mode, at least one rigid divisor wrapped by non-perturbative effects, i.e.\ either 
an E3-instanton or a D7-brane stack where gaugino condensation takes place. This immediately leads to 
the condition $h^{1,1}(X)\geq 4$. As pointed out in~\cite{CKM}, this rigd divisor has to be a `diagonal' dP$_n$ four-cycle (see~\cite{CKM} for a definition of `diagonal' dP$_n$ cycles). Moreover, a limit on the size of $h^{1,1}(X)$ to be less than six simplifies the explicit stabilisation of all K\"ahler moduli. 
Therefore, we look for CY manifolds $X$ satisfying the following properties:
\begin{itemize}
 \item[$\blacktriangleright$] The dimension of the second homology group must be $h^{1,1}(X) = 4$ or $h^{1,1}(X) =5$.
 \item[$\blacktriangleright$] There are two dP$_n$ divisors and a holomorphic involution exchanging them.
 \item[$\blacktriangleright$] The two given dP$_n$ surfaces do not intersect each other and, therefore, they do not touch the orientifold plane either.
 \item[$\blacktriangleright$] There is a further rigid divisor that is invariant under the orientifold involution.
\end{itemize}
For computational reasons, we only consider CY manifolds which are hypersurfaces in four-dimensional toric ambient spaces~\cite{0612307,9806059,0803.1194}. The smooth generic case in this set can be described by a four-dimensional reflexive lattice polytope and its triangulations.
Kreuzer and Skarke  generated a list of all these four-dimensional reflexive lattice polytopes~\cite{ks2} which after triangulation give the toric ambient varieties for the CY hypersurfaces.
We will describe these ambient spaces by giving the weight matrix and the Stanley-Reisner (SR) ideals. The weights give the linear relations between the lattice points of the polytope or rather the rays of the fan. Each ray of the fan corresponds to a homogeneous coordinate of the toric variety and the rows of the weight matrix are the actual weights of the  equivalence relation between them. The information about the triangulation of the polytope,  or rather the fan of the toric variety, is encoded in the SR ideal.
It gives the combinations of the homogeneous coordinates that are not allowed to vanish simultaneously.

By using the data of the toric ambient space, one can compute
for instance, the Hodge numbers and
the triple intersection numbers of the divisors on the CY. Moreover, also the K\"ahler cone of the ambient space can be determined using toric methods. This cone may be only a subspace of the K\"ahler cone of the CY hypersurface. We will discuss this point in detail by means of the dP$_0$ example that we present.

The results of this search are the basis for explicit model building with branes at singularities. The full classification is given in Appendix~\ref{sec:applist}. In numbers, we find 21 (168) models for $h^{1,1}=4$ $(h^{1,1}=5)$ satisfying all requirements with different dP$_n$ singularities. The distribution of singularities after a given step of imposing the above constraints is shown in Table~\ref{tab:searchresults}.
\begin{center}\begin{tabular} {c || c | c c c c c c c c c }
$h^{1,1}=4:$ 1197 polytopes & $\Sigma$ & dP$_0$ & dP$_1$ &dP$_2$ &dP$_3$ &dP$_4$ &dP$_5$ & dP$_6$ & dP$_7$ & dP$_8$ \\ \hline \hline
\footnotesize{There are 2 dP$_n$ + O-involution} & 82 & 9 & 5 & - & - & - &2 &10  &31 &25 \\ 
\footnotesize{The 2 dP$_n$ do not intersect} & 68 & 9 & 2 & - & - & - & 2  &10 &27 &18 \\
\footnotesize{Further rigid divisor} & 21 & 3 & - & - & - & - & - & 4 & 9 & 5
\end{tabular}
\\ \vspace{5mm}
\begin{tabular} {c || c | c c c c c c c c c}
$h^{1,1}=5:$ 4990 polytopes& $\Sigma$ & dP$_0$ & dP$_1$ &dP$_2$ &dP$_3$ &dP$_4$ &dP$_5$ & dP$_6$ & dP$_7$ & dP$_8$ \\ \hline \hline
\footnotesize{There are 2 dP$_n$ \& O-involution} & 386& 27 & 60 & 21 & 7 & 3 & 13 & 40 & 121 & 94 \\	
\footnotesize{The 2 dP$_n$ do not intersect} & 327 & 27 & 55 & 7 & 3 & 1 & 11 & 39 & 112 & 72 \\
\footnotesize{Further rigid divisor} & 168 & 14 & 16 & - & - & - & 5 & 28 & 68 & 37
\end{tabular}
\captionof{table}{\footnotesize{Summary of the search for polytopes allowing for CY-manifolds with our requirements. The numbers indicate the amount of polytopes for a certain dP$_n$ singularity. Different lines show our results after imposing more and more requirements. Note that in some cases the rows do not add up to the numbers presented in Appendix~\ref{sec:applist} because some polytopes might have two kinds of del~Pezzos.}\label{tab:searchresults}}
\end{center}
We shall focus on the minimal situation with $h^{1,1}=4$ and show that all the K\"ahler moduli can be fixed. In this case the second homology group splits into a one-dimensional odd and a three-dimensional even subspaces, i.e.\ $h^{1,1}_- =1$ and $h^{1,1}_+=3$. Consequently we have one complex modulus $G$ coming from reducing the $B_2$ and $C_2$ two-form potentials and three geometric moduli $T_i$ which we can relate to the size of the two shrinking SM dP$_n$ divisors, the size of the `non-perturbative' cycle and the CY volume.
The $G$-modulus and the size of the shrinking dP$_n$ divisors are fixed by the D-terms of the quiver gauge theory, while the other two geometric moduli by $\alpha'$ and non-perturbative effects, leading to the desired situation, i.e.\ a CY with (exponentially) large volume as realisations of LVS or KKLT, and with two dP$_n$ singularities.
For $h^{1,1}=5$, we have both $h^{1,1}_-=1$ and  $h^{1,1}_-=2$. We do not perform a detailed study of moduli stabilisation for these cases but
we expect them to behave similarly to the simplest case with $h^{1,1}=4$
given that the two local moduli are still fixed by the D-terms of the quiver gauge theory, and
the volume and the rigid cycle by an interplay of $\alpha'$ and non-perturbative effects.
In the case $h^{1,1}_-=1$, the remaining geometric modulus, if rigid, can also be fixed non-perturbatively,
while if non-rigid, it will be stabilised by string loop effects \cite{Loops,LoopStab}.
On the other hand, if $h^{1,1}_-=2$, on top of the overall volume mode and the two local moduli, 
we expect to have two rigid cycles mapped into each other under the orientifold involution 
which will both be fixed by non-perturbative effects.

\subsection{Consistency constraints}

Let us see in more detail which kind of brane set-ups the requirements outlined in the previous section allow and what are the consistency constraints one has to check.

\begin{itemize}

\item[$\ast$]{\bf Orientifold involution:} Since we have introduced an orientifold involution, we have to compute the location of the fixed point locus. This gives a number of O7-planes and possibly O3-planes. One has to check that these RR-charges are cancelled. 

\item[$\ast$]{\bf RR-charges:}

\begin{itemize}
\item[$\rhd$]{\emph{D7-tadpole:}} 
We will cancel the charge of the O7-plane by wrapping D7-branes 
on cycles that sum up to eight times the homology class of the O7-plane. The simplest possibility is to consider four D7-branes 
(plus their images) on top of each O7-plane. 

\item[$\rhd$] {\emph{D5-tadpole:}}
A D5-charge might be
induced on the quiver locus in the presence of flavour D7-branes. As we do not have such objects, our quiver model will cancel the D5-charge
locally. Other sources of D5-charge are the fluxes on the D7-branes. Since we will consider D7-branes on orientifold 
invariant divisors, i.e.\ divisors that are mapped to themselves by the orientifold involution, the D5-charge generated by the flux on
one brane is cancelled by the D5-charge generated by the flux on the image brane (${\cal F}' = - {\cal F}$). 

\item[$\rhd$]{\emph{D3-tadpole:}}
We will have contributions from the D3-branes at the dP$_n$ singularity, from O3-planes (if present),  
the geometry of the D7-branes and the O7-plane(s) and the flux on the D7-branes. The sum of these contributions must be negative, since we will need to switch on bulk three-form fluxes to stabilise the complex structure moduli and these contribute positively to the
D3-charge.

\item[$\rhd$]{\emph{K-theory torsion charges:}} Possible K-theoretic torsion charges~\cite{Witten:1998cd,Moore:1999gb} are zero in the constructions we consider here. We check this, by using the probe argument presented in~\cite{Uranga:2000xp}: such charges are cancelled if, in the given set-up, every invariant $SU(2)$ probe brane has an even number of fields in the fundamental representation~\cite{Witten:1982fp}.
\end{itemize}

\item[$\ast$]{\bf Gauge flux and non-perturbative hidden sector:}
The dP$_n$ divisor wrapped by non-perturbative effects is a rigid and invariant cycle. We distinguish the 
following two cases. If the dP$_n$ divisor is one component of the fixed locus of the orientifold involution, we will wrap on it a stack of four 
D7-branes (plus their images), realising an $SO(8)$ gauge group with gaugino condensation. 
On the other hand, if the dP$_n$ divisor is not part of the fixed locus, we will consider the $O(1)$ E3-instanton wrapping such a divisor.\\ 
In order to generate the non-perturbative superpotential, we must have, in both cases, zero gauge invariant flux ${\cal F} = F - B$ 
on the world-volume of the D-brane. Here, $F$ is the gauge flux and $B$ the pull-back of the NSNS two-form potential. On the $SO(8)$ stack, a non-zero flux would generate chiral matter, which could 
destroy the possibility of gaugino condensation. Recall that the rigidity of the cycle forbids the presence of additional matter in the adjoint 
representation of the gauge group. On the E3-instanton, such a flux would destroy its invariance under the orientifold involution, 
projecting it out from the spectrum.
In order to cancel ${\cal F}$, we need to switch on a non-trivial $B$-field. In fact, the dP$_n$ divisors are non-spin, i.e.\ their first Chern-class $c_1$ is odd, and this induces a Freed-Witten anomaly that has to be cancelled leading to the half-integral flux 
quantisation condition $F+ \frac{c_1(D)}{2} \in H^2(D,\mathbb{Z})$~\cite{Freed:1999vc}. This forces $F$ to be non-zero and, hence, 
the $B$-field to be $B=F$, in order to have ${\cal F}=0$.

\item[$\ast$]{\bf{Gauge flux and chiral modes:}} 
Depending on whether the wrapped cycle intersects 
the non-perturbative one or not, there are two possibilities. If the two cycles do not intersect each other, one can choose 
the $B$-field such that it cancels both the gauge flux on the non-perturbative brane and on the other brane stack. 
On the other hand, if the two cycles intersect, generically this double cancellation is not possible and one is forced to have
non-zero ${\cal F}$ on top of the D7-brane stack.   This flux will break the gauge group to a unitary subgroup and will 
generate chiral modes both at the intersection with the non-perturbative
cycle and on the bulk of the D7-brane themselves. These chiral modes will enter the D-term of the unbroken gauge group. Even though for some choice of
the flux one can also cancel the presence of such modes,  this is model-dependent.

\end{itemize}

%%%%%%%%%%%%%%%%%%%%%%%%%%%%%%%%%%%%%%%%%%%%%%%%%%%%%%%
\section{Global embedding of the dP$_0$ quiver gauge theory} \label{sec:GlobalEmbdP0}
%%%%%%%%%%%%%%%%%%%%%%%%%%%%%%%%%%%%%%%%%%%%%%%%%%%%%%%

In this section we describe a global embedding of a quiver model sitting at a dP$_0$ singularity in a type IIB orientifold compactification. In Appendix~\ref{dP8appendix} we will present a different global embedding with dP$_8$ instead of dP$_0$ singularities.
For this example we pick up the CY number two of the list in section \ref{sec:applist-pic4}, that
has two dP$_0$ singularities. This three-fold has a holomorphic involution that exchanges the two singularities or the two dP$_0$'s in the resolved picture. The fixed point locus of the involution does not include the two singular points -- the resolved dP$_0$'s do not intersect the orientifold plane(s).
Placing $N$ D3-branes on top of one of the two singularities and the $N$ image D3-branes on top of the other, we obtain (in the quotient space) a quiver model with gauge group $SU(N)^3$.

\begin{figure}[ht]
\begin{center}
\includegraphics[width=0.8\textwidth]{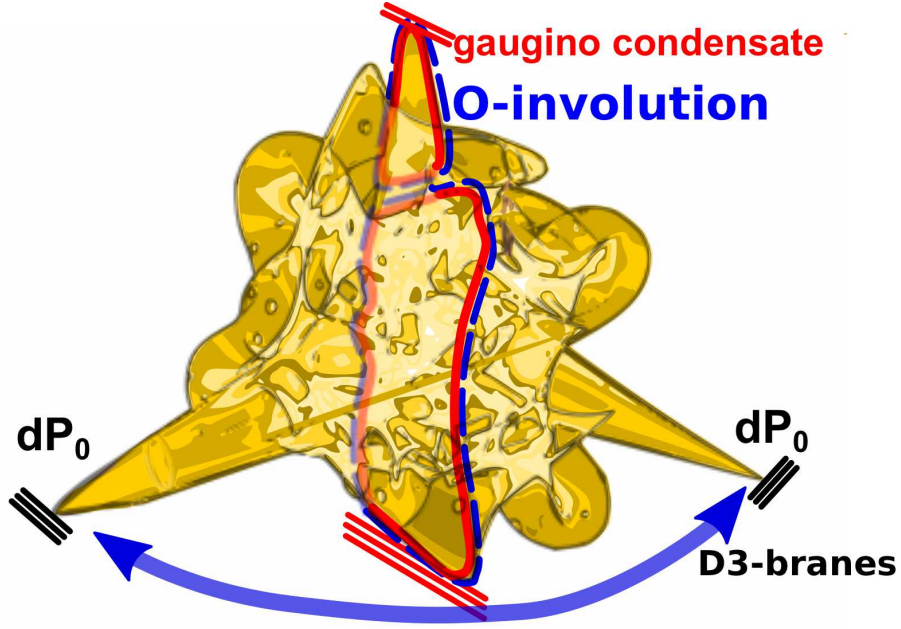}
\captionof{figure}{\footnotesize{A Swiss cheese type CY manifold with four K\"ahler moduli ($h^{1,1}=4$ and $h^{1,1}_-=1$). The CY contains two dP$_0$ 4-cycles that are exchanged under an orientifold involution. On top of the O7-planes on the remaining two four-cycles, D7-branes are wrapped that lead to gaugino condensation needed for moduli stabilisation.}\label{fig:dp0geometry}}
\end{center}
\end{figure}

The D7-charge of the O7-planes is cancelled by putting four D7-branes (plus their orientifold images) on top of each O7-plane. These D7-branes will not intersect the quiver locus and will provide a suitable hidden sector. A cartoon of this geometric set-up can be found in Figure~\ref{fig:dp0geometry}.

%%%%%%%%%%%%%%%%%%%%%%%%%%%%%%%%%%%%%%%%%%%%%%%%%%%%%%%
\subsection{Geometric set-up}
%%%%%%%%%%%%%%%%%%%%%%%%%%%%%%%%%%%%%%%%%%%%%%%%%%%%%%%

In the following, we describe the geometry of the Clabi-Yau number two of section \ref{sec:applist-pic4}. 
All of the following geometric data like SR-ideals, intersection rings, Mori cones etc.\ were computed by means of 
PALP~\cite{Kreuzer:2002uu,Braun:2011ik,Braun:2012vh}. \footnote{For the analysis of appendix~\ref{sec:applist}, 
we use in addition Sage and its toric variety package~\cite{BraunNovoseltsev:toric_variety}.}

The chosen CY three-fold $X$ is built as a hypersurface in a complex four-dimensional
toric ambient variety which is described by its weight matrix 
\be
\begin{array}{|c|c|c|c|c|c|c|c||c|}
\hline z_1 & z_2 & z_3 & z_4 & z_5 & z_6 & z_7 & z_8 & D_{eq_X} \tabularnewline \hline \hline
    1  &  1  &  1  &  0  &  3  &  3  &  0  &  0  & 9\tabularnewline\hline
    0  &  0  &  0  &  1  &  0  &  1  &  0  &  0  & 2\tabularnewline\hline
    0  &  0  &  0  &  0  &  1  &  1  &  0  &  1  & 3\tabularnewline\hline
    0  &  0  &  0  &  0  &  1  &  0  &  1  &  0  & 2\tabularnewline\hline
\end{array}
\label{eq:model3dP0:weightm}\,,
\ee
and its Stanley-Reisner ideal
\be
\label{eq:model3dP0:sr-ideal}
{\rm SR}=\{z_4\, z_6,\,z_4\, z_7, \, z_5\, z_7,\, z_5\, z_8,\, z_6\, z_8,\, z_1\, z_2\, z_3\}\,.
\ee
The CY three-fold $X$ is given by a homogeneous equation $eq_{X}=0$ in the above toric four-fold $Y_4$.
The last column of the table in~\eqref{eq:model3dP0:weightm} indicates the degrees of the homogeneous polynomial $eq_X$.
In the generic case the equation has the following form
\begin{eqnarray}
\label{EqCY3}
 eq_X &\equiv& P^1_{3}(z_4 z_5-z_6 z_7)^2z_8 + P^2_{3}(z_4 z_5+z_6 z_7)^2z_8
      \nonumber \\ &&+ (P_{0}^+ z_5 z_6 + P^+_{6} z_4 z_7 z_8^2)(z_4 z_5+z_6 z_7) + P^+_{9} z_4^2 z_7^2 z_8^3 \\
	&& + (P_{0}^- z_5 z_6 + P^-_{6} z_4 z_7 z_8^2)(z_4 z_5-z_6 z_7)=0\,,\nonumber
\end{eqnarray}
where $P^{\pm,1,2}_{k}$ are polynomials in the coordinates $(z_1,z_2,z_3)$ of degree $k$.
The Hodge numbers of this CY hypersurface are $h^{1,1}=4$ and $h^{1,2}=112$, such that $\chi=-216$.

The toric divisors $D_4$, $D_7$ and $D_8$ are all $\mathbb P^2$ or dP$_0$.
Furthermore, these three del~Pezzo four-cycles do not intersect each other.

As a basis of $H^{1,1}(X)$ we may choose
\be
\Gamma_b = D_4+D_5=D_6+D_7,\qquad \Gamma_{q_1} = D_4,\qquad \Gamma_{q_2} = D_7,\qquad \Gamma_s = D_8\,,
\label{simplebasis}
\ee
where the label `$b$' stays for `big' since this will turn out to be a large four-cycle
controlling the overall size of the CY, `$q_i$' $i=1,2$ for `quiver' since
these will be the two four-cycles exchanged by the orientifold action which
are collapsed to zero size by the D-terms, and `$s$' for `small' since this will
turn out to be a divisor fixed by non-perturbative effects in the geometric regime
but at a size much smaller than the large four-cycle.
Note that this basis of integral cycles is not an `integral basis', i.e.\ it does not generate the full $H^{1,1}(X,\mathbb{Z})$ by integral combination.\footnote{For example, $D_1=\frac{1}{3}(\Gamma_b-\Gamma_{q_1}-\Gamma_{q_2}-\Gamma_s)$. An integral basis is given, for instance, by $D_1,D_4,D_7,D_8$.}
In this basis the intersection form can be diagonalised, implying that the quiver cycle does 
not intersect the small cycle or the big cycle
\be
\label{intersnumb}
I_3 = 27\Gamma_b^3 + 9\Gamma_{q_1}^3 + 9\Gamma_{q_2}^3 + 9\Gamma_s^3 \:.
\ee
Finally, the second Chern class of $X$ is:
\be
c_2(X) = \frac23 (5 \Gamma_b^2 - \Gamma_{q_1}^2 - \Gamma_{q_2}^2 - \Gamma_s^2)
- \frac53 \Gamma_b (\Gamma_{q_1} + \Gamma_{q_2}) + \frac13 \Gamma_s (\Gamma_b - \Gamma_{q_1} - \Gamma_{q_2}),
\ee
where the $1/3$ factor appears because we are not using an integral basis, but $c_2(X)$ is of course an integral four-form.
Expanding the K\"ahler form in the basis \eqref{simplebasis} as
\be
J= t_b \Gamma_b + t_{q_1}\Gamma_{q_1} + t_{q_2}\Gamma_{q_2}+t_s\Gamma_s\,,
\label{ExpJ}
\ee
we obtain simple expressions for the volumes of the three dP$_0$ divisors (${\rm Vol}(D)=\frac 12 \int_D J\wedge J$)
\be
\tau_{q_1} \equiv {\rm Vol}(D_4) = \frac 92 \, t_{q_1}^2 \:, \qquad
\tau_{q_2} \equiv {\rm Vol}(D_7) = \frac 92 \, t_{q_2}^2 \:,\qquad
\tau_s \equiv {\rm Vol}(D_8) = \frac 92 \, t_s^2 \:,
\ee
and the `big' four-cycle is given by
\be
\tau_b \equiv {\rm Vol}(D_4+D_5)={\rm Vol}(D_6+D_7) = \frac{27}{2}\, t_b^2\,.
\ee
Finally the volume of the CY three-fold can be written in the `Swiss-cheese' form:
\be
\vo \equiv {\rm Vol}(X) = \frac 32 \left(3 t_b^3 + t_{q_1}^3 + t_{q_2}^3 + t_s^3\right)
=\frac{1}{9}\sqrt{\frac{2}{3}}\left[\tau_b^{3/2} -  \sqrt{3}\left( \tau_{q_1}^{3/2} + \tau_{q_2}^{3/2} + \tau_s^{3/2}\right)\right]\,,
\label{vol}
\ee
where ${\rm Vol}(X) = \frac 16 \int_X J\wedge J\wedge J.$

%%%%%%%%%%%%%%%%%%%%%%%%%%%%%%%%%%%%%%%%%%%%%%%%%%%%%%%
\subsubsection*{Orientifold involution}
%%%%%%%%%%%%%%%%%%%%%%%%%%%%%%%%%%%%%%%%%%%%%%%%%%%%%%%

As mentioned above, we require an orientifold involution that exchanges two of the three dP$_0$ divisors
present in $X$, and such that these two dP$_0$'s do not intersect the O7-plane(s).
These criteria are met by the following automorphism:
\be
\label{OrInvol}
 z_4\leftrightarrow z_7\qquad\textmd{and}\qquad z_5\leftrightarrow z_6\:.
\ee
To make the CY $X$ symmetric under this holomorphic involution, we have to restrict its complex structure such that the defining equation $eq_X=0$, given in \eqref{EqCY3}, 
is symmetric under the involution. This is realised by setting $P_{0}^-$ and $P^-_{6}$ identically to zero. We call $eq^s_X$ the symmetric form of $eq_X$.
The two exchanged dP$_0$'s are located at $z_4=0$ and $z_7=0$. In this section we study the generic point in the K\"ahler cone,
where these two dP$_0$ four-cycles are of finite size. Eventually we will go to the boundary where both of them shrink to zero size, generating two dP$_0$ singularities.

The (anti-)invariant monomials of the involution \eqref{OrInvol} are
\be
y_4=z_4\,z_7\,,\qquad y_7=z_5\,z_6\,,\qquad y_{5}=z_4\,z_5 + z_6\,z_7\, \qquad \textmd{and}\qquad y_{6}=z_4\,z_5- z_6\,z_7\,.
\ee
To compute the fixed locus of this involution, we rewrite the CY, via a Segre map,
as a complete intersection in the following toric ambient space:
\begin{align}\label{eq:segre-mapped-space}
&\begin{array}{|c|c|c|c|c|c|c|c||c|c|}
\hline z_1 & z_2 & z_3 & y_4 & y_5 & y_6 & y_7 & z_8 & D_{\textmd{H}_1} & D_{\textmd{H}_2}  \tabularnewline \hline \hline
    1  &  1  &  1  &  0  &  3  &  3  &  6  &  0  & 9 & 6\tabularnewline\hline
    0  &  0  &  0  &  1  &  1  &  1  &  1  &  0  & 2 & 2\tabularnewline\hline
    0  &  0  &  0  &  0  &  1  &  1  &  2  &  1  & 3 & 2\tabularnewline\hline
\end{array} \\
&\qquad {\rm SR}=\{y_4\, y_5,\,y_4\, y_6, \, y_7\,z_8,\, z_1\, z_2\, z_3\}\,.\nonumber
\end{align}
The two hypersurface equations defining the CY three-fold are $\tilde{eq}_X^s\equiv y_5^2-y_6^2 - 4\,y_4\,y_7=0$ and the invariant hypersurface equation~$eq_X^s=0$. In this parameterisation, it is straightforward to compute the locus on the CY three-fold that is fixed under the involution.
In the new coordinates, the orientifold involution maps $y_6\mapsto - y_6$. Hence, the divisor $\{y_6=0\}$ is obviously fixed under the orientifold involution. However, the set of fixed points is larger, as we can compensate the minus one factor of $y_6$ by a rescaling. In fact,
consider the locus $y_5=z_8=0$ in the ambient space. Its points are fixed under involution since they are mapped as
\begin{equation}
(z_1,z_2,z_3,y_4,0,y_6,y_7,0) \mapsto (z_1,z_2,z_3,y_4,0,- y_6,y_7,0) \sim (z_1,z_2,z_3,y_4,0,y_6,y_7,0)\:.
\end{equation}
In the last step, we have undone the minus sign using the last row in~\eqref{eq:segre-mapped-space}.
This fixed point set  is the only extension to $y_6=0$. It is given by a co-dimension two locus in the ambient space,
contrary to $\{y_6=0\}$ that has co-dimension one.
When the equation is in the orientifold-symmetric form, the locus $\{y_5=z_8=0\}$
intersects the CY non-transversally in a submanifold of $X$ of co-dimension one.\footnote{In the case of a generic hypersurface equation, it would generate a co-dimension two locus on the CY three-fold. This would be an `O5-plane' instead of an O7-plane.} This happens since $y_5=z_8=0$ implies the vanishing of the invariant CY equation $eq^s_X=0$. Moreover, this locus is equivalent to the locus $z_8=0$ in $X$. In fact, if one sets $z_8=0$ in the symmetric version of \eqref{EqCY3}, one obtains $P_{0}^+\, z_5z_6 \,(z_4z_5+z_6z_7)=0$. Since,  $z_5z_8$ and $z_4z_8$ are in the SR-ideal,
the CY equation just reduces to $y_5\equiv (z_4z_5+z_6z_7)=0$.

Hence, for the chosen involution we have two O7-planes:
\begin{equation}\begin{array}{ccccc}
 \mbox{O7-planes}&& \mbox{Locus in ambient space} && \mbox{Homology class in }X_3 \\
   O7_1: && y_6=z_4z_5-z_6z_7=0 && D_{O7_1}=D_6+D_7=\Gamma_b \\ O7_2: &&   y_5=z_8=0 && D_{O7_2}=D_8=\Gamma_s \\
\end{array}
\end{equation}

%%%%%%%%%%%%%%%%%%%%%%%%%%%%%%%%%%%%%%%%%%%%%%%%%%%%%%%
\subsubsection*{K\"ahler cone and relevant volumes}
%%%%%%%%%%%%%%%%%%%%%%%%%%%%%%%%%%%%%%%%%%%%%%%%%%%%%%%

The K\"ahler form $J$ has to lie within the K\"ahler cone which is equivalent to the statement that
the integral of $J$ over all effective curves has to be positive, i.e.~$\int_{\mc{C}_j}J>0$
for all effective curves $\mc{C}_j$. The effective curves also form a cone,
referred to as the Mori cone, which is dual to the K\"ahler cone. Expanding $J$ in a basis $\{\Gamma_i\}$ of $H^{1,1}(X)$,
$J=\sum_i t_i\Gamma_i$, the K\"ahler cone conditions map to restrictions on the coefficients $t_i$
\be
 0<\int_{\mc{C}_j}J=
 \sum_i t_i\Big(\int_{\mc{C}_j} \Gamma_i\Big) =:\sum_i t_i\,A_{ji}\,.
\ee
The matrix $A_{ji}$ is the intersection of the Mori cone generators
$\mc{C}_j$ with the divisors (Poincar\'e dual of) $\Gamma_i$.
In the following, we will also refer to $A_{ji}$ as the Mori cone.

If we choose, for the moment, the integral basis $\{D_1,D_4,D_7,D_8\}$ for $H^{1,1}(X)$,
then we have the Mori cone of the ambient space $Y_4$ in the following form 
\be
\label{eq:mc-ambient-space}
 A_{ji}=\left(\begin{array}{rrrr}
   0 & -1 &0 & 1\\
   0 & 0 & -1& 1\\
   0 & 1 & 1 & -1\\
   1 & 0 & 0 &-3
\end{array}\right)\,.
\ee
In principle, not all the effective curves of $Y_4$ lie in the Mori cone of $X$.
The Mori cone of $X$ can be smaller than the Mori cone of $Y_4$, implying a larger K\"ahler cone for $X$.
This means that the K\"ahler cone of $Y_4$ is only a subspace of the K\"ahler cone of $X$
and may not include interesting points of the K\"ahler moduli space of $X$.\footnote{In particular,
in our example it does not include the point we are interested in, i.e.\ when two dP$_0$'s collapse to zero size,
while the third is finite.}
In order to obtain a better approximation for the K\"ahler cone of $X$,
we have to discard from the Mori cone of $Y_4$ curves that are not in the Mori cone of $X$.
There are curves for which we can check straightforwardly whether they must be omitted.
They are the curves that have a negative intersection with surfaces in $X$
of which we know all the geometry and hence in particular the Mori cone.\footnote{We thank Sheldon Katz for enlightening discussions on this point.}
The fact that an intersection number in \eqref{eq:mc-ambient-space} between a curve and a divisor is negative,
tells us that such a  curve must lie in that divisor. If we know the Mori cone of that divisor,
we can look whether the curve under consideration belongs to this Mori cone.

In our case, the surfaces $D_4$, $D_7$ and $D_8$ are well known surfaces, i.e.\ $\mathbb P^2$'s;
therefore, we know their (one-dimensional) Mori cones:
\be
\label{eq:mc-p1s}
 (1,-3,0,0),\qquad (1,0,-3,0),\qquad\text{and}\qquad(1,0,0,-3)\,,
\ee
for $D_4$, $D_7$ and $D_8$, respectively.\footnote{We used that the divisor $D_1$ restricted to the three $\mathbb P^2$'s
gives in each case the hyperplane class $H$ on the three $\mathbb P^2$'s. In fact, we have the intersection numbers $D_1^2 D_4=D_1^2 D_7=D_1^2 D_8=1$.}
The only lines in $A_{ji}$ with negative entries on the intersections with $D_4,D_7,D_8$ must be \eqref{eq:mc-p1s}.
The three rays (curves) corresponding to the first three rows in~\eqref{eq:mc-ambient-space} have each
a negative intersection with $D_4$, $D_7$ and $D_8$, respectively, and lie, therefore, in these divisors. However, as we can see from \eqref{eq:mc-p1s},
they are not part of the Mori cones of the three $\mathbb P^2$'s and, therefore, not in the Mori cone of $X$ as well.
So, we have to discard the curves ${\cal C}_1,{\cal C}_2,{\cal C}_3$.
We also omit positive combinations of the ${\cal C}_1,{\cal C}_2,{\cal C}_3$ that give a row in $A_{ji}$ with a negative entry.
Furthermore, as we see by inspecting~\eqref{eq:mc-p1s}, we must have the following rays in the cone: $3{\cal C}_1+{\cal C}_4$ and $3{\cal C}_2+{\cal C}_4$.
For higher combinations --- higher multiples (only) of the first respectively second row --- the curve is either irreducible
and not in the Mori cone of the CY or splits up into a part that is in~\eqref{eq:mc-p1s} and a part that is not in the Mori cone. The final result is
\be
\label{eq:mc-CY}
 \int_{\tilde{\mc C}_j}\Gamma_i=A_{ji}=\left(\begin{array}{rrrr}
   0 & 1 & 0 & 0\\
   0 & 0 & 1 & 0\\
   0 & 0 & 0 & 1\\
   1 & 0 & 0 &-3\\
   1 & 0 & -3 &0\\
   1 & -3 & 0 &0
\end{array}\right)\,.
\ee
where $\tilde{\mc{C}}_j$ are the new generators of the Mori cone.

The result \eqref{eq:mc-CY} is in agreement with the `extended' Mori cone of $\mathbb P_{1,1,1,3,3}$
with just one blow-up (corresponding to keeping only row one and three in table \eqref{eq:model3dP0:weightm}).
The CY $X$ at hand is almost the CY hypersurface in this ambient space.
$\mathbb P_{1,1,1,3,3}$ has a locus of $\mathbb Z_3$-singularity along the curve $z_1=z_2=z_3=0$.
The CY hypersurface always intersects this locus in three $\mathbb{Z}_3$ singularities.
The toric divisor that blows up the singular locus in the ambient space splits into three divisor classes on $X$.
These are the three dP$_0$'s that resolve the three $\mathbb{Z}_3$ singularities
on $X$ and correspond to $D_4$, $D_7$ and $D_8$ in our description. The Mori cone of the resolved ambient space is given in this case by:
\be
 \left(\begin{array}{rr}
   0& 1\\
   1 & -3
\end{array}\right),
\ee
which becomes \eqref{eq:mc-CY} if we take the splitting of the exceptional divisor into account.

For later convenience, we expand $J$ in the basis \eqref{simplebasis} as done in (\ref{ExpJ}).
After implementing the change of basis in the matrix \eqref{eq:mc-CY},
we obtain in the new parameterisation the following K\"ahler cone conditions on the coefficients $t_i$:
\be
t_b+t_{q_1}>0\,, \qquad t_b+t_{q_2}>0\,, \qquad t_b+t_s>0\,, \qquad t_{q_1}<0\,, \qquad t_{q_2}<0\,, \qquad t_s<0\,. \nn
\ee
Under orientifold involution, the K\"ahler form is even and must therefore belong to $H^{1,1}_+(X)$.
This is obtained by taking $t_{q_1}=t_{q_2}$.
Moreover, eventually we want the two dP$_0$ divisors at $z_4=0$ and $z_7=0$ to shrink to zero size
in order to generate the two (exchanged) dP$_0$ singularities.
This is realised on the boundary of the K\"ahler cone given by $t_{q_1}=t_{q_2}=0$.
The remaining K\"ahler cone conditions are then
\be
t_b+t_s >0 \qquad \text{and} \qquad t_s < 0 \:.
\ee

\subsection{Brane set-up}

We consider the case when the two dP$_0$ divisors, $D_4$ and $D_7$, are collapsed to zero size,
generating two $\mbb{C}^3/\mbb{Z}_3$ singularities,
while the other dP$_0$ divisor, $D_8$, is of finite size.
As we shall see in section \ref{sec:dtermstab}, the shrinking of the two K\"ahler moduli $\tau_{q_1}=\tau_{q_2}$
is forced by the D-terms.

We have the following set of D-branes and O-planes:
\begin{itemize}
\item There are two orientifold O7-planes, one at $z_4z_5-z_6z_7=0$,
lying in the class $\Gamma_b$,
and the other at $z_8=0$, in the class $\Gamma_s$.
The two fixed loci are disconnected and do not intersect each other.

\item To cancel the D7-charge of the O7-planes, we put four D7-branes (plus their images)
on top of each O7-plane. This generates a hidden sector gauge group $SO(8)\times SO(8)$.

\item We put $N_{D3}$ D3-branes on the singularity at $z_4=0$ and their
$N_{D3}$ images on the singularity at $z_7=0$  to have an invariant configuration.
The visible sector is given by these D3-branes. The gauge theory is $SU(N_{D3})^3$.\footnote{Two of the three $U(1)$ factors are anomalous, and become massive by eating up
the local axions given by the reduction of the RR forms $C_4$ and $C_2$
on the dP$_0$ divisor and its dual two-cycle (the canonical class of $H^{1,1}({\rm dP}_0)$).
The remaining $U(1)$ factor is an anomaly-free $U(1)$ but it decouples since it becomes a trivial overall $U(1).$ This leads to the
phenomenologically interesting trinification models with gauge group
$SU(3)_c\times SU(3)_L\times SU(3)_R$ as discussed later in Section~\ref{sec:dP0:trinif}.}
\end{itemize}

\subsubsection*{D-brane charges of the chosen set-up}

A global consistent construction must have zero charges along the compact directions.
In our case, we have D7-branes and fractional D3-branes. The RR charges of both of them are formally expressed by the `Mukai' charge vector of D7-branes.
In fact, a D3-brane at a singularity can be roughly seen as a collection of fluxed D7-branes wrapping the shrinking divisor.\footnote{More precisely,
a D3-brane at a singularity splits into a collection of fractional branes. The fractional branes are described by (coherent) sheaves ${\cal E}$.}
The total D-brane charge is given by the sum of the Mukai vectors of each object in the D-brane configuration.

The Mukai charge vector of a D7-brane is given by the following polyform
\begin{equation}
 \Gamma_{{\cal E}} = [D]\wedge \mbox{ch} ({\cal E}) \wedge \sqrt{\frac{\mbox{Td}(TD)}{\mbox{Td}(ND)}} \:,  \qquad
   \mbox{ with } \qquad S_{D7} = \int_{\mathbb{R}^{1,3}\times X} C \wedge e^{-B} \wedge \Gamma_{{\cal E}}.
\label{ChargeVectD7br}
\end{equation}
Here $D$ is the divisor wrapped by the D7-brane,
Td$(V)= 1 + \frac12 c_1(V) + \frac{1}{12}c_1(V)^2+c_2(V)+... $ is the Todd class of the vector bundle $V$,
$TD$ is the tangent bundle of $D$ and $ND$ the normal bundle of $D$ in $X$ while
ch$({\cal E})$ is the Chern character of the vector bundle (more precisely a sheaf) ${\cal E}$ living on the brane.\footnote{The charge vector can also be written in an equivalent way as $\Gamma_{D7} = [D]\wedge \mbox{ch} ({\cal W}) \wedge \sqrt{\frac{\hat{A}(TD)}{\hat{A}(ND)}}$
where $\hat{A}$ is the A-roof genus and ${\cal W}={\cal E}\otimes K_S^{1/2}$ is the sheaf
whose first Chern class is identified with the gauge flux.}

Expanding~\eqref{ChargeVectD7br} and using the fact that $X$ is a CY, we obtain
\be\label{ChVectExp}
 \Gamma_{D7}(D,{\cal F})\equiv  e^{-B}\Gamma_{{\cal E}} = [D] \left( 1 + {\cal F} + \frac12 {\cal F}\wedge {\cal F} + \frac{c_2(D)}{24} \right),
\ee
with ${\cal F}=F-B$ and $F=c_1({\cal E})+\frac{c_1(D)}{2}$. Looking at $S_{D7}$ in \eqref{ChargeVectD7br}, one finds that,
in $\Gamma_{D7}$, the two-form encodes the D7-charge, the four-form the D5-charge and the six-form the D3-charge.
The charge vector of the image-D7-brane $D7'$, wrapping the image divisor $D'$ is given by $\Gamma_{D7}(D',-{\cal F})$.

An orientifold plane wrapping the divisor $D$ has the following charge vector
\be
 \Gamma_{O7} (D)= - 8 \, [D] \wedge \sqrt{\frac{\mbox{L}(\tfrac14 TD)}{\mbox{L}(\tfrac14 ND)}} = [D]\left(-8 + \frac{c_2(D)}{6}\right) \quad
   \mbox{with} \quad S_{O7} = \int_{\mathbb{R}^{1,3}\times X} C \wedge \Gamma_{O7},
\ee
where L$(V)=1+\frac{1}{3}(c_1(V)^2-2c_2(V))+...$ is the Hirzebruch L-genus. 

\

In our set-up, the D7-brane stacks have been chosen in such a way that they trivially cancel (globally and locally) the D5- and the D7-charges:
The D7-charge of the O7-plane is cancelled by D7-branes wrapping divisors of class $[D_{D7}]=8 [D_{O7}]$. This can be realised, as in the chosen set-up, by putting four D7-branes plus their images on top of the orientifold plane.
Given the D7-branes wrap the same divisor as the image-D7-branes (i.e.~$[D']=[D]$), the four-form of the corresponding charge vectors \eqref{ChVectExp} are opposite to each other and hence cancel the D5-charge.\footnote{When $D'$ is in a different class from $D$, one needs to check if it is cancelled.}

It remains to compute the D3-charge. 
The D3-charge is obtained by integrating (minus) the six-form component of the charge vector. In particular, for a set 
of four D7-branes with the same flux background (plus their images) on top of the O7-plane wrapping the divisor $D$, the D3-charge is
\be
\label{D3chO7D7}
Q_{D3} = \int_X ( - 8\, e^{-B}\Gamma_{D7}(D,F) - \Gamma_{O7} (D) )\Big|_\textmd{six-form} =
-\frac{\chi(D)}{2} -8\times\frac{1}{2}\int_D {\cal F}\wedge {\cal F},
\ee
where ${\cal F}=F-B$ is the gauge invariant flux.
In our case, when we have non-zero brane flux ${\cal F}_b$ and ${\cal F}_s$
on the two stacks $D7_b$ and $D7_s$ on top of the two O7-planes wrapped on $\Gamma_b$ and $\Gamma_s$
(and minus the same fluxes on the image-stacks), the total D3-charge is given by
\be\label{D3chdP0D7br}
 Q_{D3}^{(b)} + Q_{D3}^{(s)} 
 =-60-4\left( \int_{\Gamma_b} {\cal F}_b^2 + \int_{\Gamma_s} {\cal F}_s^2\right)\,, 
\ee
where we have used the intersection numbers \eqref{intersnumb} and the fact that
the Euler characteristics of the divisors wrapped by the O7-planes are $\chi(D_6+D_7)=117$ and $\chi(D_8)=3$
(recall that on a CY three-fold $X$ the Euler characteristic of a divisor $D$ is given by $\chi(D)=\int_X (D\cdot c_2(X)+D^3)$).

The quiver locus is given by the two dP$_0$ singularities at $z_4=0$ and $z_7=0$.
Since $b_2({\rm dP}_0)=1$, all the divisors on a dP$_0$ are proportional to the hyperplane class $H$
that generates $H^{1,1}({\rm dP}_0)$. Let us consider the dP$_0$ at $z_7=0$.
From the intersection number $D_7 \cap D_1 \cap D_1 = 1$, we see that $D_1|_{D_7} = H$. A fractional brane $a$ corresponds to a bound state described by a coherent sheaf $F_a$ on the dP$_0$ surface; it is characterised by the charge vector of a 7-brane wrapping the shrinking divisor.
For a dP$_0$ singularity, one has three types of fractional branes.
The geometric part of the vector ($\sqrt{\frac{{\rm Td}(TS)}{{\rm Td}(NS)}}$) is the same for all of them
as they wrap the same divisor. The flux (sheaf) part is different for the three fractional branes and is given by~\cite{Diaconescu:1999dt}
\bea
 \mbox{ch}({F_0}) &=& -1 + \ell\,H -\frac12\, H\wedge H\,,  \\
 \mbox{ch}({F_1}) &=& 2 - \ell\,H -\frac12\, H\wedge H\,, \\
 \mbox{ch}({F_2}) &=& - 1\,.
\label{fluxsheaf}
\eea
We can now compute the total charge vectors \eqref{ChargeVectD7br} for the three fractional branes wrapping the locus at $z_7=0$:
\begin{eqnarray}
 \Gamma_{F_0} &=& [D_7]\wedge\left\{ -1 + (\ell-\tfrac32) [D_1] + (\tfrac32\ell -\tfrac74 ) [D_1] \wedge [D_1]  \right\}\,, \\
 \Gamma_{F_1} &=& [D_7]\wedge\left\{ 2 - (\ell-3) [D_1] - (\tfrac32 \ell-2) [D_1] \wedge [D_1]  \right\}\,, \\
 \Gamma_{F_2} &=&   [D_7]\wedge\left\{ -1 - \tfrac32 [D_1] -\tfrac54  [D_1] \wedge [D_1]  \right\}\,.
\end{eqnarray}
If the three fractional branes have the same multiplicity $N_{D3}$,
their total charge vector is just $N_{D3}$ multiplied by the sum of their charge vectors
\begin{equation}
 N_{D3} \cdot (\Gamma_{F_0} + \Gamma_{F_1} + \Gamma_{F_2}) = - N_{D3} \cdot [D_7] \wedge [D_1]\wedge [D_1]\:.
\end{equation}
This means that their total D7- and D5-charges are zero. Moreover integrating (minus) it over the CY three-fold $X$,
we obtain a D3-charge equal to $N_{D3}$, with $N_{D3}=3$ in the case of trinification models.
The same happens for the fractional branes at $z_4=0$. Since we want an orientifold invariant configuration, we need to take the same multiplicity $N_{D3}$ of fractional branes. We then have for the D3-charge of the quiver locus (in the covering space)
\be
Q_{D3}^{\rm excep}=2\times N_{D3}= 6\:.
\ee

\subsubsection*{Chiral spectrum in the bulk}

On the D7-brane stacks we can have chiral matter only on the bulk of the D7-branes.
In fact, in the chosen configuration there are no intersections among the two stacks.

First we consider possible massless scalars in the adjoint of $SO(8)$.
These are given by reducing the D7-brane gauge field on the $h^{1,0}(D)$ one-cycles of the wrapped divisor $D$
and by the $h^{2,0}(D)$ deformations of the divisor $D$.
In our case we have $h^{1,0}(\Gamma_b)=h^{1,0}(\Gamma_s)=h^{2,0}(\Gamma_s)=0$ and $h^{2,0}(\Gamma_b)=11$.
The 11 scalars living on the $D7_b$-stack on the large cycle $\Gamma_b$ might be lifted by a gauge flux
that is trivial\footnote{
By trivial flux we mean two-forms of $D$ whose Poincar\'e dual two-cycle in $D$ are trivial as two-cycles of the CY.}
within the CY $X$~\cite{0602129,Bianchi:2011qh}.

Chiral massless fields can be generated by switching on a non-zero flux ${\cal F}$.
If ${\cal F}$ is the same for each of the four branes of the stack, then it breaks the gauge group as
\be
SO(8) \rightarrow SU(4) \times U(1)\, .
\ee
This induces the following number of chiral states in the antisymmetric representation of $SU(4)$
\be
I^{A} = \int_X  {\cal F}\wedge [D7]\wedge ([D7]+[O7]) =   2 \int_X  {\cal F}\wedge [O7]\wedge [O7]\:,
\ee
where we used that in our case $[D7]=[O7]$. If we can set ${\cal F}=0$ for both stacks, we will have no chiral states coming from the D7-brane stacks
(and we can have gaugino condensation on both D7-brane stacks if the adjoints on $\Gamma_b$
can be lifted by an appropriate choice of trivial fluxes).

\subsection{Consistency conditions}

As mentioned in Section~\ref{sec:2} the D7 tadpole charges are automatically cancelled by placing the appropriate number of D7-branes on top of the O7-planes. Also the K-theory torsion charges are cancelled by construction as explained before. We now concentrate on the Freed-Witten anomaly and D3-brane charges. 

\subsubsection*{Freed-Witten anomaly and gaugino condensation}

The two D7-stacks wrap non-spin divisors. This implies that we have to switch on non-zero gauge fluxes $F$ on both stacks.
However, this may be dangerous for what we want to do. In fact, as discussed above, it would generate chiral matter that could kill the gaugino condensation
on the given stack. In turn, this would prevent the corresponding terms in the superpotential that are necessary to stabilise the volume of the CY.

The simplest solution of this problem is to set the gauge invariant flux ${\cal F}=F-B$ to zero
by switching on an appropriate $B$-field. Usually, choosing a $B$-field that sets ${\cal F}=0$ on one stack,
makes it impossible to set ${\cal F}=0$ also on the other stack.
However, our case is non-generic as we can choose a $B-$field which cancels the Freed-Witten gauge fluxes on both D7-brane stacks.
This is due to the fact that the two divisors have zero geometrical intersection.

In detail, the two Freed-Witten fluxes are
\be
F_b = \frac{D_6}{2}+\frac{D_7}{2} \qquad\text{and}\qquad F_s = \frac{D_8}{2}\,.
\ee
Vanishing gauge invariant flux ${\cal F}_b={\cal F}_s=0$ can be achieved by the following choice of the $B$-field
\be
B = \frac{D_6}{2} + \frac{D_7}{2} + \frac{D_8}{2}\, ,
\ee
where we used the fact that $D_8$ is trivial when restricted to $D_6+D_7$ and viceversa.
These considerations imply that, if the 11 adjoint scalars on $\Gamma_b$ can be fixed by an appropriate trivial flux,
the hidden sector on both $\Gamma_b$ and $\Gamma_s$ will consist of a pure $\mc{N}=1$ SYM theory which undergoes
gaugino condensation. More precisely, $\Gamma_s$ supports a pure $SO(8)$
theory whereas the fluxes on $\Gamma_b$ break $SO(8)$ down to $SU(4)\times U(1)$ where no chiral matter is generated and the $U(1)$ factor remains massless as the fluxes are trivial. Therefore, the bulk superpotential will take the form
\be
W = W_0 + A_s\,e^{- \,a_s\,T_s} + A_b\,e^{- \,a_b\,T_b}\,,
\label{Wbulk}
\ee
where $W_0$ is the VEV of the tree-level superpotential generated by the
background fluxes $H_3$ and $F_3$ which fix the dilaton and the complex structure moduli.
For a pure $SO(8)$ theory $a_s$ becomes $\pi/3$ and $a_b=\pi/2$ for a pure $SU(4)$ theory. $T_s$ and $T_b$ are the complex
K\"ahler moduli defined as $T_s=\tau_s + {\rm i} c_s$ and $T_b=\tau_b + {\rm i} c_b$
where $c_s = \int_{\Gamma_s} C_4$ and $c_b = \int_{\Gamma_b} C_4$.

Another kind of Freed-Witten anomaly is present if the three-form NS flux $H_3$
has a non-trivial pull-back on the D7-brane world-volume~\cite{Freed:1999vc}.
In our case this cannot happen, whatever $H_3$ we have,
as the four-cycles wrapped by the D7-branes have $b^3(D)=2h^{1,0}=0$
(for both D7-brane stacks).

\subsubsection*{D3-charge}

The D3-brane charge of this configuration is given by $Q_{D3}^{\rm excep} + Q_{D3}^{(b)} + Q_{D3}^{(s)}$,
where $ Q_{D3}^{\rm excep}$ is the contribution from the fractional branes
and $Q_{D3}^{(b,s)}$ come from the two D7-brane stacks on $\Gamma_b$ and $\Gamma_s$.
If we set the fluxes ${\cal F}_b={\cal F}_s=0$, we have the following total D3-charge
\be
Q_{D3}^{\rm excep} + Q_{D3}^{(b)} + Q_{D3}^{(s)} = -60 + 2N_{D3} = - 54\,,
\ee
where $N_{D3}=3$ is the number of D3-branes at the singularity.
We see that there is still room for the D3-charge coming from the three-form fluxes $H_3$ and $F_3$
which are needed to stabilise the dilaton and the complex structure moduli.

\subsubsection*{Possible Whitney-brane}

There is a slightly different set-up that is worthwhile to mention.
We may keep the visible sector on the fractional branes on the dP$_0$ singularity
and four D7-branes (plus their four images) wrapping the O7-plane on $\Gamma_s$ (at $z_8=0$).
The new ingredient could be a `Whitney brane' cancelling the D7-charge of the O7-plane on $\Gamma_b$~\cite{Collinucci:2008pf}.
This means that we would have a fully recombined orientifold-invariant D7-brane in the class
\be
[D7_W] = 8 [O7_1] = 8(D_6+D_7)\:.
\ee
This would generate a large (negative) D3-charge
which may be useful to fix the complex structure moduli. In fact, in this case the total D3-charge would be
(see~\cite{1110.3333} for the procedure to compute this contribution)
\be
Q_{D3}^{\rm excep} + Q_{D3}^{(b)} + Q_{D3}^{(s)} = -438 + 2N_{D3} = -432 \:.
\ee
The given Whitney brane has no intersection with the other branes,
and so cannot generate chiral modes. Moreover there is no gauge group living on it, and so
we would not generate a $T_b$-dependent non-perturbative contribution to the superpotential.
\footnote{The problem may come from a possible FW anomaly of the second kind, arising due to a possible non-zero pull-back of $H_3$ 
on its world-volume (one should the compute $b_3(D7_W)=h^{1,0}(D7_W)$ and check that it is zero).}

%%%%%%%%%%%%%%%%%%%%%%%%%%%%%%%%%%%%%%%%%%%%%
\subsection{dP$_0$ Trinification model}\label{sec:dP0:trinif}

\begin{figure}[ht]
\begin{center}
\includegraphics[width=0.4\textwidth]{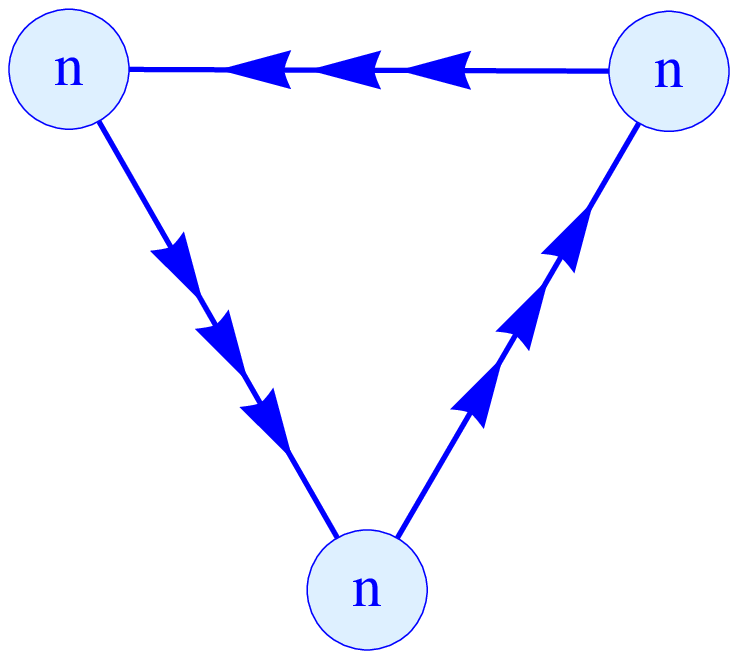}
\includegraphics[width=0.4\textwidth]{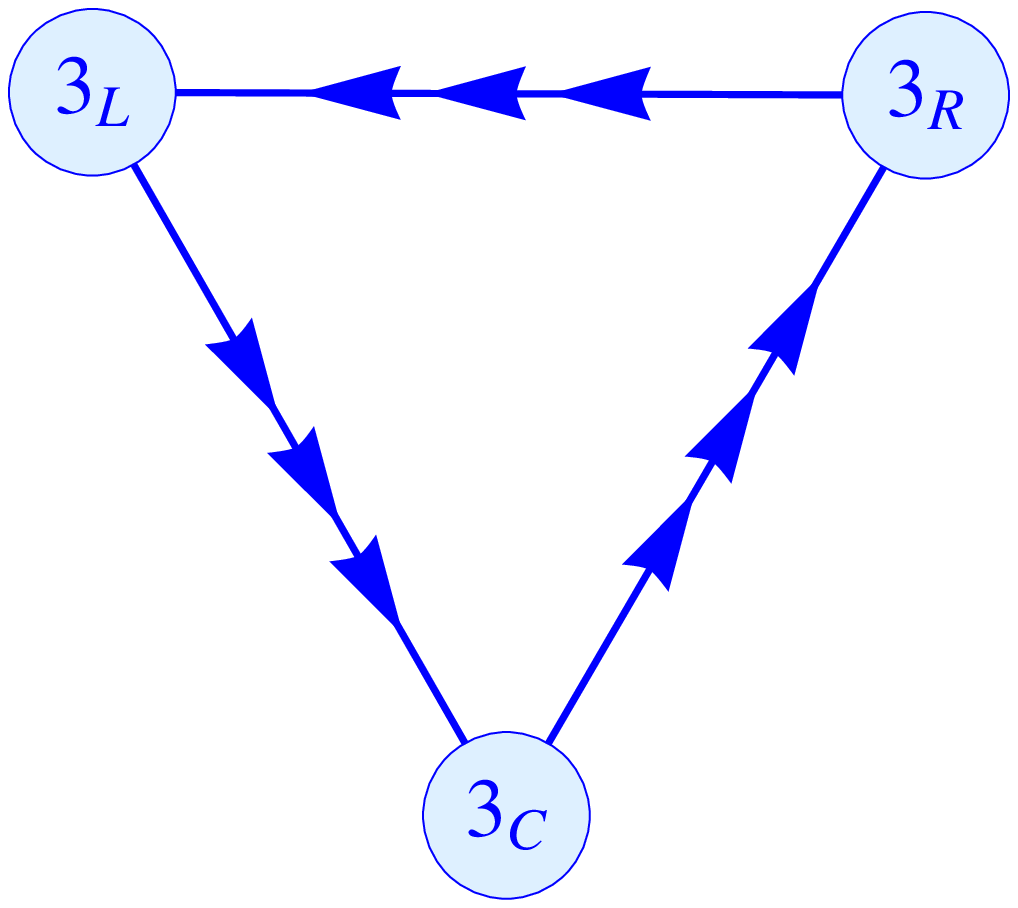}
\captionof{figure}{\footnotesize{{\bf Left:} The dP$_0$ quiver encoding the well-known $SU(n)^3$ gauge theory (in the absence of flavour branes) and three bi-fundamental fields between the different $SU(n)$ factors. {\bf Right:} The specific case of $n=3$ which gives rise to the trinification model based on three copies of $SU(3)$. The labels $SU(3)_C\times SU(3)_L\times SU(3)_R$ indicate which of the $SU(3)$ factors gives rise to which SM gauge groups.}\label{fig:dp0quiver}}
\end{center}
\end{figure}

%%%%%%%%%%%%%%%%%%%%%%%%%%%%%%%%%%%%%%%%%%%%%
The trinification model at the dP$_0$ singularity
naturally leads to three families and
the matter content of the MSSM.
 The chiral matter spectrum under $SU(3)_c\times SU(3)_L\times SU(3)_R$ is (see Figure \ref{fig:dp0quiver}):
\be
3 [ \left(\bf{3}, \bf{\bar{3}}, \bf{1}\right) + \left( \bf{1}, \bf{3},
  \bf{\bar{3}} \right) + \left( \bf{\bar{3}}, \bf{1}, \bf{3} \right) ]\,.
\ee
The superpotential arising from the D3-D3 states has a $SU(3)$ flavour symmetry reflecting the isometry
of the $\mbb{C}^3/\mbb{Z}_3$ geometry, and it is given by
\be
W_{\rm local}=\epsilon_{ijk}\,C^i C^j C^k\,.
\label{Wlocal}
\ee
Generically additional couplings will be induced when the local geometry is embedded in a compact model,
inducing a small breaking of the local $SU(3)$ isometry by compactification effects~\cite{1102.1973,1111.3047}.
From the local perspective the corrections can be understood from non-commutative deformations
of the background geometry~\cite{0512122,0712.1215}.\footnote{Note that non-commutative deformations of the local geometry as described in~\cite{0512122,0712.1215} can lead to $\mc{O}(1)$ deformations of the  superpotential of the gauge theory, i.e.~the induced superpotential couplings due to the non-commutative deformations do not have to be treated as perturbations to the couplings obtained from the geometry without such deformations.}
In a similar fashion it is argued that non-perturbative corrections to the superpotential
can be induced from such bulk effects~\cite{0607050,1001.5028}.
The size and presence of the given bulk induced operators depend on the flux/source background
and generically these operators are suppressed by inverse powers of the bulk volume (cf.~\cite{1111.3047} and references therein).
A detailed study of the presence of these flux/source effects at this stage is beyond the scope of this article
and we take at this stage the phenomenological point of view that such couplings can be present.
We hence are left with the generic gauge invariant superpotential
\be
W=y_{ijk}\,C^i C^j C^k+ \kappa_i \det{C^i}\,.
\label{Wmatter}
\ee

%%%%%%%%%%%%%%%%%%%%%%%%%%%%%%%%%%%%%%%%%%%%%%%%%%%%%%%
\subsubsection*{Towards the SM within the dP$_0$ singularity}
\label{sec:quiverpheno}
%%%%%%%%%%%%%%%%%%%%%%%%%%%%%%%%%%%%%%%%%%%%%%%%%%%%%%%
The restriction to dP$_0$ and no flavour branes limits the possible models leading to a SM gauge theory at low energies.
Here we would like to discuss the phenomenological properties of the trinification model
with the primary focus of obtaining the SM gauge and matter content as well as the Yukawa couplings.
The SM field content is contained in the matter arising at the dP$_0$ singularity as follows:
\begin{eqnarray}
Q_L&=&\left(\begin{array}{c c c}
u_L & d_L & B_L\\
u_L & d_L & B_L\\
u_L & d_L & B_L\\
\end{array}\right),\; Q_R=\left(\begin{array}{c c c}
u_R & u_R & u_R\\
d_R & d_R & d_R\\
B_R & B_R & B_R
\end{array}\right), \; \Phi =\left(\begin{array}{c c c}
H_u & H_d & L\\
H_u & H_d & L\\
e_R & \nu_R & v
\end{array}\right).
\end{eqnarray}
Both fields $B_L$ and $B_R$ are chiral exotics that obtain masses upon breaking to $SU(2)_L\times SU(2)_R\times U(1)_{B-L}$ as discussed later in this section. $v$ is a SM singlet and will be used to break to the intermediate left-right model. Recall that there are three copies of each multiplet. For the phenomenological analysis we assume the following superpotential
\begin{eqnarray}
\nonumber W&=&\left(\begin{array}{c}
Q_L^1\\
Q_L^2\\
Q_L^3
\end{array}\right)^T
\left(
\begin{array}{ccc}
 \lambda_{11} \Phi_1& (1+\lambda_{12}) \Phi_3& (-1+\lambda_{13}) \Phi_2\\
 (-1+\lambda_{21}) \Phi_3 & \lambda_{22} \Phi_2& (1+\lambda_{23}) \Phi_1\\
(1+\lambda_{31}) \Phi_2& (-1+\lambda_{32}) \Phi_1& \lambda_{33} \Phi_3
\end{array}
\right)
\left(\begin{array}{c}
Q_R^1\\
Q_R^2\\
Q_R^3
\end{array}\right)\\
&&
+\kappa_{\Phi,i}\det{\Phi_i}+\kappa_{L,i}\det{Q_L^i}+\kappa_{R,i}\det{Q_R^i}\,.
\end{eqnarray}
Note that the lepton Yukawa couplings and the $\mu$-term are contained in the non-perturbative contribution including $\Phi$:
\be
\kappa\det{\Phi}=\kappa\left(v H_u H_d-\nu_R H_u L+e_R H_d L\right)\, .
\ee
We note that the couplings $\kappa_{L,i}\det{Q_L^i}+\kappa_{R,i}\det{Q_R^i}$ can be dangerous regarding proton decay and a mechanism suppressing these couplings, while keeping the coupling for the $\Phi$-field, needs to be present.

The breaking to the SM gauge groups can be achieved in two stages:
\bea
SU(3)_c\times SU(3)_L\times SU(3)_R&\to& SU(3)_c\times SU(2)_L\times SU(2)_R \times U(1)_{B-L} \nn \\
&\to&  SU(3)_c\times SU(2)_L\times U(1)_Y\,. \nn
\eea
The first breaking is achieved by generating a VEV for one $\Phi^i$-field of the following type:
\be
\langle\Phi\rangle=\left(\begin{array}{ccc}
0 & 0 & 0\\
0 & 0 & 0\\
0 & 0 & v
\end{array}\right).
\label{eq:firstbreaking}
\ee
The second breaking is achieved via a VEV for a right-handed sneutrino $\tilde{\nu}_R.$
Both fields acquire a non-zero VEV when one soft scalar mass becomes negative upon RG-evolution. 
As we discuss later in section~\ref{sec:dtermstab} such VEVs do not resolve the singularity and so the EFT is under control.

Such a radiative breaking can appear with universal boundary soft-masses as discussed in Section~\ref{sec:susybreaking}.
The relevant RG equations for the soft-masses and supersymmetric parameters are well-known as reviewed for example in~\cite{9709356}.
As in the MSSM, it is crucial to identify a large positive contribution to the $\beta$-function 
of the soft-scalar mass for one of the $\Phi$-fields.
We find distinct $\beta$-functions for the soft scalar masses of the following type (see appendix \ref{sec:rgrunning} 
for more explicit expressions):
\be
\beta_{m^2_i}=\frac{1}{24\pi^2}\left[-32g^2|M|^2+(9A^2+27m^2)\left(2+f_i(\lambda_{ab})\right)\right]\,,
\ee
where $g$ denote the gauge coupling, $M$ the associated gaugino mass, $A$ the universal A-parameter, $m$ the universal scalar mass and $f_i(\lambda_{ab})$ is a function of the superpotentials couplings. Given a suitable choice of Yukawa couplings one can easily achieve a significantly larger $\beta$-function for the $\Phi$-field from the above equations (cf. (\ref{eq:rgexample}) for an explicit example). Assuming a negative soft-mass for $\Phi$ at some scale $M_x<M_s$ generated through RG-evolution, 
let us discuss the mass spectrum after this breaking focusing on the scalar potential for the $\Phi_i$ fields:
\begin{equation}
V=\sum_i m^2_i\Phi^\dagger_i\Phi_i+\frac{g_L^2}{2}\left(\sum_i\Phi^\dagger_i T^a \Phi_i\right)^2+\frac{g_R^2}{2}\left(\sum_i\Phi^\dagger_i \tilde{T}^a \Phi_i\right)^2+\sum_i\left|\frac{\partial W}{\partial \Phi_i}\right|^2\,,
\end{equation}
where $g_{L,R}$ denote the gauge couplings for $SU(3)_{L,R}$. The VEV in (\ref{eq:firstbreaking}) is F-flat but leads to a non vanishing D-term potential. Demanding an extremum of the potential $(\partial V/\partial\Phi=0)$ equips us with the following condition
\begin{equation}
m^2_\Phi=-\frac{8}{3}(g_L^2+g_R^2)~v^2\, .
\end{equation}
At this point in the moduli space we find the following three distinct mass eigenvalues for the components of the $\Phi$-field:
\be
\left(0, 2 v^2~(-4( g_L^2+g_R^2)+\kappa\bar{\kappa}),\frac{32}{3}( g_L^2+g_R^2)~v^2  \right),
\ee
where the vanishing eigenvalues (nine of them) correspond to Goldstone bosons and these degrees of freedom are eaten. The second eigenvalue (and hence all eigenvalues) can be positive if $\kappa$ is large enough.\footnote{We note that this can be a severe constraint since $\kappa$ is very suppressed. However, we have not yet explored the r\^ole of other operators, such as higher dimensional couplings in the K\"ahler potential, that could be used for lifting potential negative directions.} Note that without the presence of the non-perturbative effect we would not find a minimum leading to the desired symmetry breaking. The third eigenvalue is unique. Since all masses are positive at this extremum we have found a minimum of the scalar potential that allows the breaking to the SM.

In this breakdown some of the scalars associated to SM fermions obtain large masses, whereas the fermions themselves do not obtain large masses. 
In fact, we have three contributions to fermion masses for the fermionic components of the $\Phi$-fields:
\begin{equation}
{\cal L}_{\rm ferm. mass}=\varphi_{ki}^\star \lambda^a T^a_{ij} \psi_{jk}+\varphi_{ki}^\star \tilde{\lambda}^a \tilde{T}^a_{ij} \psi_{jk}+M_\lambda \lambda^a\lambda^a+M_{\tilde{\lambda}} \tilde{\lambda}^a\tilde{\lambda}^a+\text{Yukawa interactions}\,, \nn
\end{equation}
where $\lambda^a$ (and $\tilde{\lambda}^a$) denote the gauginos for $SU(3)_L$ (and respectively $SU(3)_R).$ 
Given the VEV in eq. (\ref{eq:firstbreaking}) we then find the following mass matrix:
\begin{equation}
{\cal L}_{\rm ferm. mass}=\left(\begin{array}{c}
\vec{\lambda}^+\\
\vec{\tilde{\lambda}}^-\\
\vec{\psi}_L
\end{array}\right)^T
\left(\begin{array}{c c c}
M_\lambda & 0 & v\\
0 & M_{\tilde{\lambda}} & 0\\
0 & v & y_{\rm Lep}
\end{array}\right)
\left(\begin{array}{c}
\vec{\lambda}^-\\
\vec{\tilde{\lambda}}^+\\
\vec{\psi}_R
\end{array}\right),
\end{equation}
where the $\vec{(.)}$ indicates the appropriate flavour family structure and respectively the inclusion of all gauginos. The above mass matrix has eigenvalues $M_\lambda,$ $M_{\tilde{\lambda}},$ and the normal mass eigenvalues arising from the lepton Yukawas. Despite supersymmetry being broken, 
the SM leptons remain light as phenomenologically desired.

%%%%%%%%%%%%%%%%%%%%%%%%%%%%%%%%%%%%%%%%%%%%%%%%%%%%%%
\section{K\"ahler moduli stabilisation}
\label{sec:modstab}

In this section we shall assume that the dilaton and the complex structure moduli can be fixed
by turning on appropriate background fluxes (we showed that in our set-up the D3-tadpole leaves
space to switch on $H_3$ and $F_3$), and focus only on the stabilisation of the K\"ahler moduli.

Given that we are interested in quiver gauge theories from del~Pezzo singularities, we need
to study the effective field theory (EFT) expanded around this singularity
and then show that the moduli can indeed be fixed in this regime.
Even if this regime is different from the standard geometric regime,
where all the cycles have a size larger than the string scale $\ell_s$
and the EFT can be computed in a reliable way via dimensional reduction,
there has still to exist an EFT in the vicinity of the singularity since
we know that string theory is well defined in the singular regime where
most of the worldsheet computations are performed.

\subsection{Effective field theory near the singularity}

For concreteness we shall focus on the dP$_0$ example presented in the previous section.
Furthermore we shall consider the divisors $\Gamma_b$ and $\Gamma_s$ in the geometric regime,
whereas the dP$_0$ divisors $\Gamma_{q_1}$ and $\Gamma_{q_2}$ at the singularity
and expand the EFT around this point following~\cite{0810.5660}.

Let us start by analysing the moduli content after the involution.
Defining $D_+ = D_4 + D_7$ and $D_-=D_4 - D_7$, $D_+$ is invariant under the involution whereas
$D_-$ goes to minus itself. The closed string moduli of the 4D EFT are the axio-dilaton $S = e^{-\phi}+{\rm i} C_0$,
$h^{1,2}_-$ complex structure moduli ($h^{1,2}_+$ vector multiplets), $h^{1,1}_-=1$ K\"ahler modulus given by
\be
G = b + {\rm i} c\,,
\ee
where $b = \int_{\hat{D}_-} B_2$ and $c = \int_{\hat{D}_-} C_2$, and the following $h^{1,1}_+=3$ K\"ahler moduli
\be
T_b = \tau_b + {\rm i} c_b\,, \qquad T_s = \tau_s + {\rm i} c_s\,,
\qquad T_+ = \tau_+ + {\rm i} c_+\,,
\label{T+}
\ee
where $\tau_+ \equiv {\rm Vol}(D_+) = \frac 12 \int_{D_+} J \wedge J$ and $c_+=\int_{D_+} C_4$.
Notice that $T_-$ is projected out.

The tree-level K\"ahler potential with the leading order $\alpha'$ correction takes the form
\be
K = - 2 \ln \left( \hat\vo + \frac{\zeta}{g_s^{3/2}} \right) + \frac{(T_+ + \bar{T}_+ + q_1 V_1)^2}{\hat\vo}
+ \frac{(G + \bar{G} + q_2 V_2)^2}{\hat\vo}+ \frac{C^i\bar{C}^i}{\hat\vo^{2/3}}\,,
\label{KtotSing}
\ee
where from Equation (\ref{vol}) we have that $\hat\vo = \alpha \left(\tau_b^{3/2} -\sqrt{3} \tau_s^{3/2} \right)$ with $\alpha= \frac19\sqrt{\frac23}$.
The coefficient $\zeta = -\chi(X)\zeta(3) /[2 (2\pi)^3]\simeq 0.5$ controls the leading order $\alpha'$ correction~\cite{BBHL},
while $C^i$ are visible sector matter fields. Moreover, $q_1$ and $q_2$ are the charges of the local moduli $T_+$ and $G$
under the two anomalous $U(1)$s with vector multiplets $V_1$ and $V_2$.\footnote{We assumed,
without loss of generality, that only $T_+$ is charged under $U(1)_1$ while only $G$ is charged under $U(1)_2$.
This can always be accommodated at the expenses of introducing a kinetic mixing between the two anomalous $U(1)$s.}

The superpotential is instead given by the sum of the local superpotential (\ref{Wlocal})
(with $\mc{O}(1)$ non-commutative deformations of the background geometry which induce generic
$y_{ijk}$ Yukawa couplings) and the bulk superpotential (\ref{Wbulk})
\be
W = W_{\rm local} + W_{\rm bulk}  =  W_0 + y_{ijk} \,C^i C^j C^k
+  A_s\, e^{-\frac{\pi}{3}\,T_s} +  A_b \,e^{-\frac{\pi}{2}\, T_b}\,,
\label{Wtot}
\ee
where we expect the prefactor $A_s$ to be independent of chiral
matter at the quiver locus due to the geometrical separation between
this four-cycle and the collapsed dP$_0$ divisor.
On the other hand, $A_b$ may in principle depend on visible sector matter fields
but only on gauge invariant combinations of them
since the K\"ahler modulus $T_b$ does not get charged under any
anomalous $U(1)$ at the quiver locus due to the absence of intersections
between the corresponding divisors. Hence the term $\kappa \det{C}$ in (\ref{Wmatter})
might correspond to the term $A_b \,e^{-\frac{\pi}{2}\, T_b}$ in (\ref{Wtot}) with
$\kappa \simeq e^{-\frac{\pi}{2}\, T_b}$ and $\det{C}\simeq A_b$. 
As we have seen in section \ref{sec:quiverpheno},
this term might be important to have a radiative breaking of $SU(3)^3$ to the MSSM.

Notice that we did not write down in $W$ any
non-perturbative term in $T_+$ and $G$ since
these moduli get charged under the two anomalous $U(1)$s of the dP$_0$ quiver. 
Thus gauge invariance of the superpotential will always induce a
prefactor that depends on visible sector fields which 
do not acquire non-zero VEVs at the string scale.
Therefore the four-cycle supporting the visible sector has to be fixed
via either D-terms or perturbative corrections to the K\"ahler potential.
A combination of both effects in the case of intersecting rigid cycles has been
used in~\cite{1110.3333} to fix the visible sector cycle in the geometric regime.
We shall now instead show that, in the case of a diagonal del~Pezzo
divisor which does not intersect other cycles, just the D-terms are sufficient
to force its shrinking to zero size, leading to quiver gauge theories.

\subsection{D-term stabilisation}
\label{sec:dtermstab}

\subsubsection{Shrinking from the singular perspective}

The D-term potential for the two anomalous $U(1)$s can be written as
\be
V_D = \frac{1}{{\rm Re}(f_1)} \left(\sum_i q_{1i} K_i  C_i -\xi_1 \right)^2
+ \frac{1}{{\rm Re}(f_2)} \left( \sum_i q_{2i} K_i  C_i -\xi_2\right)^2,
\label{Dpot}
\ee
where the gauge kinetic function $f$ at the quiver locus is given by the dilaton $S$
plus a shift proportional to the product of the local K\"ahler moduli and their $U(1)$ charges
(i.e.\ $f_{1}= S + q_1 T_+$ and $f_{2}= S + q_2 G$), while the two Fayet-Iliopoulos (FI)
terms $\xi_1$ and $\xi_2$ are given by
\bea
\xi_1 &=& -\frac{\partial K}{\partial V_1} \Bigg\vert_{V_1=0} = -q_1 \frac{\partial K}{\partial T_+}
\Bigg\vert_{V_1=0}=-2 q_1\frac{(T_+ + \bar{T}_+)}{\hat\vo}=-4 q_1\frac{\tau_+}{\hat\vo}\,,
\label{xi1} \\
\xi_2 &=& -\frac{\partial K}{\partial V_2} \Bigg\vert_{V_2=0} = -q_2 \frac{\partial K}{\partial G}
\Bigg\vert_{V_2=0}=-2 q_2\frac{(G + \bar{G})}{\hat\vo}=-4 q_2\frac{b}{\hat\vo}\,.
\label{xi2}
\eea
If all the matter fields acquire vanishing VEVs from F-term contributions involving the
soft scalar masses (i.e.\ terms of the form $m^2_i C^i\bar{C}^i$), the D-term potential (\ref{Dpot})
admits a supersymmetric minimum at $\xi_1=\xi_2=0$, corresponding to the
singular limit $\tau_+ = b = 0$. Notice that the VEV of the visible sector fields
has to be zero at the string scale (the Higgs and other additional matter fields might, as usual, develop a non-zero VEV at
lower scales due to radiative corrections)
not just to avoid the presence of charge or colour breaking minima, but also
to make sure that the D3-branes remain at the singularity. In fact,
if some matter fields could get appropriate non-zero VEVs so to break
the trinification gauge group $SU(3)^3$ to $U(3)$, the fractional D3-branes would recombine
into a stack of three ordinary mobile D3-branes which
can move away from the singularity.

Regarding the axionic partners of $\tau_+$ and $b$, i.e.\ the $C_4$ axion $c_+$
and the $C_2$ axion $c$, they get eaten up by the two anomalous $U(1)$ symmetries which therefore
get masses of the order the string scale and disappear from the EFT.

We stress that we can focus at leading order on the D-terms neglecting the F-terms since
the D-term potential scales as $V_D \sim \mc{O} \left(M_s^4\right)$ whereas the F-term potential
behaves as $V_F \lesssim \mc{O}\left(m_{3/2}^2 M_P^2\right)$ due to the no-scale structure.
Hence their ratio scales as
\be
R\equiv \frac{V_F}{V_D} \lesssim \frac{m_{3/2}^2 M_P^2}{M_s^4} \sim W_0^2\,,
\label{R}
\ee
since $M_s \sim M_P/\sqrt{\vo}$ while $m_{3/2}\sim W_0 M_P/\vo$.
In the KKLT case, $W_0$ is fine tuned to very small values in order to obtain a
trustable vacuum within the regime of validity of the EFT, and so $R\ll 1$. On the other hand, in the case of the LVS, $W_0$ is of order unity,
but the relation (\ref{R}) has really to
be understood just as an upper bound on the ratio between $V_F$ and $V_D$ since, as we will
describe in section \ref{FtermStab}, the F-term potential
scales as $V_F \sim \mc{O}\left(\vo^{-3}\right)$ implying that $R \sim \vo^{-1}\ll 1$
for exponentially large values of $\vo$.

\subsubsection{Shrinking from the geometric perspective}

The shrinking of the dP$_0$ divisor induced by the D-terms can also be understood
from the geometric point of view. Contrary to the singular case where fractional D3-branes can be understood as mutually supersymmetric fluxed D7- 
and anti-D7-branes wrapped on the collapsing divisor, the stable geometric picture involves only
D7-branes with gauge flux $\mc{F}$ that has to satisfy the D-flatness condition
$J \wedge\mc{F} = 0$. This is equivalent to setting the corresponding FI-term
to zero since $\xi = \int_{D_{\rm vs}} J\wedge \mc{F}$ with $D_{\rm vs}$ the divisor wrapped by the
visible sector D7-branes.
However, as pointed out in \cite{CKM}, if $D_{\rm vs}$ is a diagonal del Pezzo as in our case,
the D-terms force its shrinking to zero size, in a regime which is not geometrical anymore.
Let us discuss this shrinking in more detail.

In the four-dimensional EFT in the large volume regime,
the $T_+$ and $G$ moduli have a different definition which involves now a mixing term of the form
\be
G =c + {\rm i} S b \qquad\text{and}\qquad
T_+ = \tau_+ +\frac{S}{2}\,k b^2 + {\rm i} \left(c_+-\frac{k}{2}\, c b\right),
\label{Tplus}
\ee
where $k$ can be determined from (\ref{intersnumb}) to be
\be
k = \int_X \hat{D}_+ \wedge \hat{D}_- \wedge \hat{D}_- = k_{444} + k_{777}\,.
\ee
Notice that only $T_+$ mixes with $G$ since there is no intersection between $D_-$ and the
geometric four-cycles $\Gamma_b$ and $\Gamma_s$. For all our considerations,
we can set the dilaton at its VEV $S=\langle S\rangle = g_s^{-1}$ for $e^{-\langle\phi\rangle}=g_s^{-1}$
and $\langle C_0\rangle=0$. Thus the two expressions (\ref{Tplus}) simplify to
\be
G =c + {\rm i} \frac{b}{g_s} \qquad\text{and}\qquad
T_+ = \left(\tau_+ +\frac{k}{2 g_s}\,b^2\right) + {\rm i} \left(c_+-\frac{k}{2}\, c b\right).
\label{Tpluss}
\ee
The relevant part of the tree-level K\"ahler potential involving the K\"ahler moduli is given by
\be
K= -2\ln\vo\qquad \text{with}\qquad\vo=\frac 16\, k_{ijk} t^i t^j t^k\,.
\ee
In order to write $K$ as a function of $T_s, T_b, T_+$ and $G$, we have first to invert the relation between the $t$'s
and the $\tau$'s so to be able to write $\vo$ as a function of the $\tau$'s. Subsequently we have to
invert the relation (\ref{Tpluss}) to write $\tau_+$ in terms of $T_+$ and $G$.
In our case we obtain (with $\alpha= \sqrt{2/243}$)
\bea
K &=& - 2\ln\left[\alpha\left(\tau_b^{3/2}-\sqrt{3}\,\tau_s^{3/2}- \sqrt{3}\,\tau_+^{3/2} \right)\right]  \nn \\
&=& - 2\ln\left[\alpha\left(\tau_b^{3/2}-\sqrt{3}\,\tau_s^{3/2}\right)- \alpha\sqrt{3}\, \left({\rm Re}(T_+)
- \frac{k}{2 g_s}\, b^2\right)^{3/2} \right]. \nn
\eea
If we wrap a stack of D7-branes around the dP$_0$ divisor $D_4$
and turn on a gauge flux $F = f \hat D_4 = f_+ \hat D_+ + f_- \hat D_-$ with $f_+=f_-=f/2$,
we generate an FI-term of the form
(expanding the K\"ahler form and the $B$ field as $J=t_+ \hat D_+$ and $B=b\hat D_-$)
\be
\xi = \frac{1}{\vo} \int_{D_4} J\wedge \mc{F} =
\frac{t_+}{2\vo}\left(\int_{D_+} \hat D_+ \wedge \mc{F}_++\int_{D_-} \hat D_+ \wedge \mc{F}_-\right)
= \frac{t_+}{2\vo}\, k \left(f_+ + f_- - b\right)\,.
\label{xigeom}
\ee
The previous expression can be rewritten in a way more similar to (\ref{xi1}) and (\ref{xi2}):
\be
\xi = - q_+ \frac{\partial K}{\partial T_+}-{\rm i} q_G \frac{\partial K}{\partial G}\,,
\ee
where $q_+ = k_{+++} f_+ + k_{+--} f_- = k \left(f_+ + f_-\right)$ is the $U(1)$ charge of $T_+$ and
$q_G = 1$ is the $U(1)$ charge of $G$ while the two derivatives read
\be
\frac{\partial K}{\partial T_+} = \frac 12 \frac{\partial K}{\partial \tau_+} = - \frac{t_+}{2\vo}
\qquad\text{and}\qquad \frac{\partial K}{\partial G}=-{\rm i} \frac{e^{\langle\phi\rangle}}{2}\frac{\partial K}{\partial b}
=-{\rm i}\frac{t_+}{2\vo}\,k b\,.
\ee
Notice that $T_+$ gets charged only if a gauge flux is turned on whereas $G$ gets automatically charged
once a stack of D7-branes is wrapped around the dP$_0$ divisor. Thus if $F=0$,
the $C_2$ axion $c$ gets eaten up by the diagonal $U(1)$ via the so-called `geometric' St\"uckelberg
mechanism. Due to supersymmetry, also the partner of $c$, i.e.\ the field $b$, has to get a large mass
of the order the string scale, implying that this is the modulus fixed by imposing $\xi=0$.
Hence the FI-term (\ref{xigeom}) vanishes for $b=0$ without inducing the shrinking of the dP$_0$ divisor.

However the situation changes as soon as we turn on a gauge flux $F\neq 0$.
As can be seen from (\ref{xigeom}), the solution to $\xi=0$ is now $t_+ =0$
since a combination of both $c$ and $c_+$ has to be eaten up, implying that
the corresponding modulus fixed by the D-terms has also to be a combination
of $T_+$ and $G$ like $\tau_+ = {\rm Re}\left(T_+\right) - k\,b\, {\rm Im}\left(G\right)/2$.
Moreover, the expression $\left(f_+ + f_- - b\right)$ is always different from zero
if $f_+$ is an integer (or $f$ is even) given that the component of $B$ along even cycles
can only take half-integer values. We have therefore shown that, whenever a gauge flux
is turned on, the dP$_0$ divisor collapses to the singular regime.

\subsection{F-term stabilisation}
\label{FtermStab}

As we have described in the previous sections, the leading order D-term potential
fixes $b=0$ and the dP$_0$ divisor at the singularity, while the soft masses induce
vanishing VEVs for the chiral fields. It remains therefore to study the F-term
potential for the two bulk moduli $T_b$ and $T_s$. 
Here we can have two different situations depending on the choice of the $B$-field:
\begin{enumerate}
\item If $B$ is chosen such that to cancel the Freed-Witten flux on both divisors $\Gamma_b$ and $\Gamma_s$, 
i.e. $\mc{F}_b=\mc{F}_s=0$, we can have two subcases:
\begin{enumerate}
\item If the 11 adjoints on $\Gamma_b$ are fixed by turning on a trivial flux, 
also non-perturbative effects in $T_b$ get generated since this cycle supports a pure $\mc{N}=1$ 
$SU(4)$ theory that undergoes gaugino condensation. 
This case can give rise to AdS LVS vacua~\cite{0502058} and in principle also to standard AdS KKLT vacua which 
can be uplifted to dS solutions via $\alpha'$ corrections~\cite{0408054,1107.2115}. 

\item If no trivial flux is turned on, the adjoints are unfixed, and so the $T_b$-dependent non-perturbative effects are absent. 
Hence this case can give rise just to AdS LVS minima. In order to uplift these vacua to dS space one would need additional contributions in the model, such as anti-branes at the tip of warped throats~\cite{KKLT} or new non-perturbative effects at hidden sector on different quiver singularities~\cite{1203.1750}. In this last case we would need to consider at least an additional K\"ahler modulus.
\end{enumerate}

\item If $B$ is chosen such that it only cancels the Freed-Witten flux on $\Gamma_s$ (the dP$_0$ divisor in the geometric regime),
i.e. $\mc{F}_s=0$ but $\mc{F}_b\neq 0$, no non-perturbative effect in $T_b$ can be generated, and so no KKLT minimum can exist. 
The only vacua which exist are LVS minima which can now correspond to dS solution since the non-vanishing flux on $\Gamma_b$ will 
induce a non-zero $U(1)$ charge for the large modulus $T_b$, and so a moduli-dependent FI-term. This semi-positive definite
D-term contribution is crucial to realise LVS dS vacua~\cite{1203.1750}.
\end{enumerate}
We shall now focus on the most promising case where $\mc{F}_b\neq 0$ and show explicitly the presence of dS LVS vacua.  

\subsubsection{De Sitter LVS vacua}

If $\Gamma_b$ supports a non-vanishing flux, the large cycle $T_b$ has to be fixed
perturbatively as in the LVS scenario by using the leading order $\alpha'$ correction.
On the other hand, the dP$_0$ divisor $D_8$ is fixed
due to the gaugino condensation.

After fixing the axion of the small modulus $T_s$, the leading order F-term potential for the bulk moduli reads
\be
V_F \simeq \frac 83  (a_s A_s)^2  \sqrt{\tau_s}\, \frac{e^{-2\, a_s \tau_s}}{\hat\vo}
   - 4 \,a_s A_s W_0 \tau_s\frac{e^{-a_s \tau_s}}{\hat\vo^2} + \frac{3}{4} \frac{\zeta W_0^2}{g_s^{3/2} \hat\vo^3}\,.
\label{VO3s}
\ee
where we have approximated $\hat\vo\simeq \alpha\tau_b^{3/2}$.
In the limit $\pi\tau_s/3\gg 1$, this potential admits an AdS global minimum
which breaks supersymmetry spontaneously and is located at exponentially large volume:
\be
\langle\hat\vo\rangle\simeq \frac{3 W_0 \sqrt{\tau_s}}{4 a_s A_s} \,e^{a_s \langle\tau_s\rangle}\qquad \text{and}\qquad
\langle\tau_s\rangle \simeq
\left(\frac{3 \zeta}{2}\right)^{2/3} \frac{1}{g_s}\,.
\label{Vfinal}
\ee
Given that in this case we are not choosing the $B$-field such that to cancel the FW flux on the large four-cycle,
we would generate chiral hidden sector matter and an FI-term which can
be used as an up-lifting term to obtain a dS vacuum.
In fact, being a hidden sector, the matter fields can get non-zero VEVs without breaking any symmetry of the visible sector.
Then the FI-term can be cancelled at leading order (without shrinking the cycle) and
what is left over can give the uplifting~\cite{0701154,0901.0683}.
The tuning to achieve a slightly dS solution is performed on $W_0$ and $g_s$.
Let us see this issue more in detail.

In the presence of a FW flux on the large four-cycle $F_b= \frac 12 \hat{D}_b$, the $D$-term potential takes the form
\be
V_D
=\frac{\pi}{\left(\tau_b-g_s q_{bb}/4\right)} \left( \sum_j q_{bj} \phi_j \frac{\partial K}{\partial \phi_j}
+\frac{q_{bb}}{4\pi}\frac{\partial K}{\partial \tau_b}\right)^2,
\ee
where the diagonal K\"ahler metric for the matter fields living on the large cycle $\phi_j$ scales with the volume as
$K\simeq \sum_j |\phi_j|^2 \tau_b^{-1}$~\cite{0805.2943} while the $U(1)$ charge of the modulus $T_b$ is
\be
q_{bb}= \int_{D_b} \hat{D}_b \wedge \frac{F_b}{2\pi}  = f_b^k k_{bbk} = \frac{27}{2}.
\label{U1charge}
\ee
Therefore, considering canonically normalised matter fields $\phi_{c,j}$ and the corresponding F-term contributions,
the total potential becomes
\be
V_{\rm tot}=V_D+V_F
\simeq \frac{p_1}{\hat\vo^{2/3}} \left( \sum_j q_{bj} |\phi_{c,j}|^2-\frac{p_2}{\hat\vo^{2/3}}\right)^2
+\sum_j \frac{W_0^2}{2 \hat\vo^2}\,|\phi_{c,j}|^2+V_F(T),
\label{Vtotal}
\ee
where $p_1\equiv \pi\,\alpha^{2/3}$ and $p_2\equiv 3 q_{bb} \alpha^{2/3} /\left(4\pi \right)$
and $V_F(T)$ is the potential (\ref{VO3s}) for the K\"ahler moduli.

\begin{figure}[ht]
\begin{center}
\includegraphics[width=0.65\textwidth]{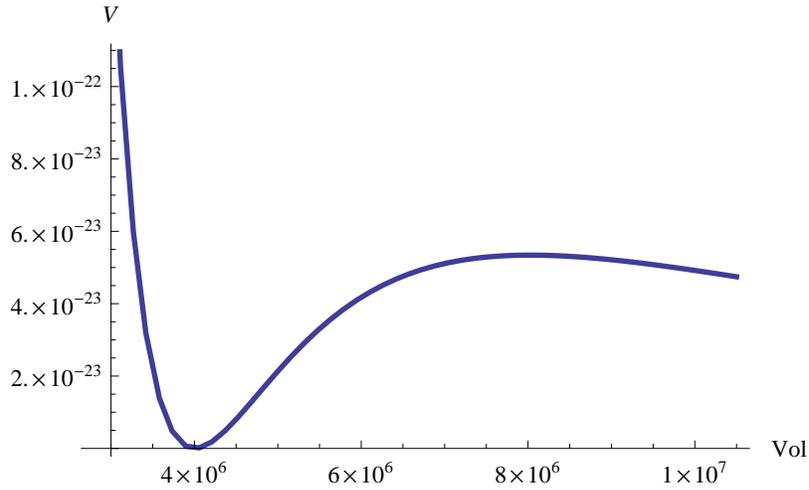} \caption{Minkowski vacuum at $\vo= 4\cdot 10^6$ for natural
values of the underlying parameters $W_0\simeq 0.2$ and $g_s\simeq 0.03$.} \label{FigdS}
\end{center}
\end{figure}

If some matter fields have a positive $U(1)$-charge $q_{bj}$,
the FI-term can be cancelled at leading order by giving a non-zero VEV to these fields, so
that the D-term potential becomes subdominant with respect to the F-term potential
for the matter fields which can provide an interesting source for uplifting to dS vacua.
In fact, focusing just on a single matter field $\phi_c$ with $q_b>0$, the minimisation gives
\be
\langle|\phi_c|^2\rangle= \frac{p_2}{q_b\hat\vo^{2/3}}-\frac{W_0^2}{4 c_b^2 p_1 \hat\vo}\simeq \frac{p_2}{q_b\hat\vo^{2/3}},
\ee
Substituting this VEV in (\ref{Vtotal}), we are left with the potential (\ref{VO3s}) plus an uplifting term:
\be
V \simeq \frac{p \,W_0^2}{\hat\vo^{8/3}}+V_F(T)\,,\qquad
\text{with}\qquad p=\frac{p_2}{2 q_b}=\frac{3 q_{bb} \alpha^{2/3}}{8\pi q_b}\simeq \frac{0.325}{q_b}.
\label{NewV}
\ee
Let us now show that this new term can indeed give rise to Minkowski vacua and in turn to slightly dS minima.
We can start by minimising (\ref{NewV}) with respect to $\tau_s$, obtaining (for $a_s\tau_s\gg 1$)
\be
e^{-a_s \tau_s}= \frac{3\sqrt{\tau_s}}{4 a_s A_s}\frac{W_0}{\hat\vo}
\qquad\Rightarrow\qquad a_s \tau_s =\ln\left(\frac{4 a_s A_s}{3\sqrt{\tau_s}}\right)
+\ln\left(\frac{\hat\vo}{W_0}\right)
\simeq\ln\left(\frac{\hat\vo}{W_0}\right)\,,
\label{tsVEV}
\ee
which substituted back in (\ref{NewV}) yields
\be
V = \frac{W_0^2}{\hat\vo^3} \left\{\frac{3\zeta}{4 g_s^{3/2}}
-\frac 32\left[\frac{\ln\left(\hat\vo/W_0\right)}{a_s}\right]^{3/2}
   + p\,\hat\vo^{1/3} \right\}.
\label{VO3tb}
\ee
Solving the minimisation equation $\partial V/\partial \hat\vo=0$, we find
\be
\frac{3\zeta }{4g_s^{3/2}}=\frac 32 \left[\frac{\ln\left(\hat\vo/W_0\right)}{a_s}\right]^{3/2}
\left(1-\frac{1}{2\ln\left(\hat\vo/W_0\right)}\right)-\frac 89
\,p\,\hat\vo^{1/3}\,,
\label{tbVEV}
\ee
so that the vacuum energy takes the form
\be
\langle V \rangle = \frac{W_0^2}{\langle\hat\vo\rangle^3} \left\{-\frac{3}{4\,a_s^{3/2}} \sqrt{\ln\left(\frac{\langle\hat\vo\rangle}{W_0}\right)}
+ \frac{p}{9}\,\langle\hat\vo\rangle^{1/3} \right\}.
\label{Vmin}
\ee
For a given value of $q_b$ which fixes the value of the parameter $p$,
and a desired value of $\hat\vo$ which gives rise to TeV-scale supersymmetry,
the two equations (\ref{tbVEV}) and $\langle V\rangle =0$ from (\ref{Vmin})
give the values of the underlying parameters $g_s$ and $W_0$ which can be tuned
by varying the background fluxes.
In the next section we will study how supersymmetry is broken by non-vanishing F-terms
of the bulk moduli and we will find that the preferred value of the overall volume is $\hat\vo\simeq 10^{6 - 7}$.
For $q_b=2$, $\zeta \simeq 0.522$, $a_s=\pi/3$ and $\vo= 4\cdot 10^6$,
we obtain a Minkowski vacuum for natural values $W_0\simeq 0.2$ and $g_s\simeq 0.03$ (see Figure \ref{FigdS}).

Larger values of $\vo$ cannot be obtained without tuning. As indicative benchmark points,
we mention $\vo = 4\cdot 10^7$ that requires $W_0\simeq 5\cdot 10^{-27}$ and $g_s\simeq 0.006$,
and $\vo \simeq 10^{15}$ which needs minuscule $W_0\simeq 10^{-10^6}$ and $g_s\simeq 5\cdot 10^{-8}$.
This huge fine-tuning implies that for values of the volume $\vo > 10^8$ such a the D-term uplifting
cannot be achieved in a controlled fashion. A way-out could involve the presence
of a magnetic flux localised in a highly warped region, so that the parameter $p$ would receive
an extra suppression from a warp factor. This would be possible only if $k_{sbb} \neq 0$~\cite{1203.1750}
but this condition is not satisfied in our case. However, as we shall show in the next section,
we only need smaller values of $\vo$ to obtain TeV-scale supersymmetry.

We finally mention that the gauge flux on the large four-cycle generates a coupling between
the axion of $T_b$ and the hyperweak Abelian gauge boson $\gamma'$ living on this divisor.
This coupling gives a St\"uckelberg mass to this $U(1)$ of the order $m_{\gamma'}\sim m_{3/2}\sim 10^9$ GeV.
Moreover this light hidden photon acquires a small kinetic mixing with the ordinary photon
$\chi \simeq 0.5\cdot 10^{-2}/\sqrt{\tau_b}\simeq 10^{-5}$ with very interesting phenomenological
implications~\cite{0909.0515,1103.3705}.

\subsubsection{Supersymmetry breaking and moduli spectroscopy}
\label{sec:susybreaking}
Soft-masses in the LVS have been under continuous investigation and have lead to various phenomenological scenarios~\cite{0505076,0605141,0610129,0906.3297,0912.2950,1002.4633,1003.0388,1011.0999,1012.1858}.
Given that the superpotential does not depend on $T_+$, the corresponding F-term vanishes since $\langle {\rm Re}(T_+)\rangle=0$:
\bea
F^{T_+} & = & e^{K/2} K^{T_+ \bar{i}} D_i W =  e^{K/2} \left( K^{T_+ \bar{s}}\partial_s W + W_0 K^{T_+ \bar{i}} \partial_i K \right)\nn \\
& = &  2 e^{K/2} {\rm Re}(T_+) \left(  \tau_s  \partial_s W - W_0 \right) = 0, \nn
\eea
where $K^{T_+ \bar{s}}= 2 \tau_s {\rm Re}(T_+)$ and $K^{T_+ \bar{i}} \partial_i K = - 2 {\rm Re}(T_+)$.
Similarly $F^G=0$.
Therefore there is no local supersymmetry breaking. This can also be seen from the fact that the F-terms of the local moduli are
proportional to the corresponding FI-terms. Thus supersymmetry is broken far away in the bulk
by $F^{T_b}\sim\mc{O}\left(\vo^{-1/3}\right)$ and $F^S\sim\mc{O}\left(\vo^{-2}\right)$
resulting in suppressed soft terms
with respect to the gravitino mass~\cite{0906.3297} which is given by
\be
m_{3/2} = \sqrt{\frac{g_s}{4\pi}}\,\frac{W_0 M_p}{\hat\vo}\simeq 5\cdot 10^9\,{\rm GeV}.
\ee
The gaugino masses are found to scale as
\be
M_{1/2} \simeq \frac{m_{3/2}}{\hat\vo}\simeq \mc{O}\left(1 - 10\right) \,{\rm TeV}\, .
\ee
Depending on the sub-leading and unknown structure in the K\"ahler matter metric, the scalar masses can be as small as the gaugino masses or as large as
\be
m_0 \simeq \frac{m_{3/2}}{\sqrt{\hat\vo}}\simeq \mc{O}\left(10^{6 - 7}\right) \,{\rm GeV}.
\ee
As far as the moduli mass spectrum is concerned, the two local moduli $\tau_+$ and $b$
fixed by the D-terms, get a mass of the order the string scale whereas the two moduli in the geometric regime,
$\tau_b$ and $\tau_s$, turn out to be lighter:
\be
m_{\tau_b} \simeq m_0 \simeq \frac{m_{3/2}}{\sqrt{\hat\vo}}\simeq \mc{O}\left(10^{6 - 7}\right) \,{\rm GeV},
\qquad m_{\tau_s}\simeq m_{3/2}\ln\hat\vo \simeq\mc{O}\left(10^{10 - 11}\right) \,{\rm GeV}.
\ee
The sequestering of the visible sector is very appealing to get TeV-scale SUSY avoiding any possible cosmological moduli
problem and raising $M_s$ to the GUT scale since $M_s\simeq M_p/\sqrt{\hat\vo}\sim\mc{O}\left(10^{15}\right)$ GeV.
However there are two situations where $F^{T_+}$ can be non-zero, and so soft terms can arise at $\mc{O}(m_{3/2})$:
\begin{enumerate}
\item In the presence of a resolution of the dP$_0$ singularity which takes place if some of the matter fields get a non-zero VEV of the order the string scale cancelling the FI term. However we have seen that this is not possible in the dP$_0$ case since these VEVs will arise at a scale $M\ll M_s$ through RG-evolution as described in Section~\ref{sec:dP0:trinif}.
As in a comparable set-up described in~\cite{1006.3341}, the D-flatness relation hence implies that the blow-up mode $\tau_+$ would still be in the singular regime $\tau_+\ll 1$:
\be
\xi = \frac{\tau_+}{\hat\vo} \simeq \frac{\langle |\Phi|^2\rangle}{M_P^2} = \left(\frac{M}{M_P}\right)^2 \qquad\Rightarrow\qquad
\tau_+ \simeq \left(\frac{M}{M_P}\right)^2 \hat\vo \simeq \left(\frac{M}{M_s}\right)^2 \ll 1\,. \nn
\ee

\item If the local blow-up mode $T_+$ gets redefined due to one-loop corrections to the gauge kinetic function.
However this redefinition does not seem to take place for orbifold singularities with only D3-branes as in our case
(it occurs instead for orientifold singularities and orbifold singularities with D3- and D7-branes)~\cite{0901.4350,0906.1920}.
\end{enumerate}

\subsubsection{The KKLT scenario}

Let us briefly comment on the possibility to get KKLT vacua when the $B$-field 
is chosen such that to cancel the FW flux on the large cycle and the extra adjoint fields 
are fixed via a trivial flux. In this way we would also generate non-perturbative effects in $T_b$ 
which combined with gaugino condensation on $T_s$ are expected to give rise to a typical 
supersymmetric AdS KKLT minimum at smaller values of the volume if $W_0$ is tuned exponentially small~\cite{KKLT}.
The inclusion of $\alpha'$ corrections might then allow an uplifting of this minimum to 
dS space by increasing the value of $W_0$~\cite{0408054,1107.2115}. 

However this situation seems very hard to achieve in our model since, neglecting the $\alpha'$ corrections, 
the solution of the two F-flatness conditions $D_s W=0$ and $D_b W=0$ implies
\be
a_b\tau_b - a_s\tau_s = \ln\left(\frac{a_b A_b}{a_s A_s}\right)+\frac 12 \ln\left(\frac{\tau_s}{\tau_b}\right)\,.
\label{KKLTcondition}
\ee
For natural values of the prefactors $A_b\sim A_s\sim \mc{O}(1)$ the expression on the right hand side of (\ref{KKLTcondition}) 
is of order unity whereas we require both $a_b\tau_b\gg 1$ and $a_s\tau_s\gg 1$ in order to be able to neglect 
higher order non-perturbative effects. Hence (\ref{KKLTcondition}) can very well be approximated as
\be
a_b\tau_b \simeq a_s\tau_s\,.
\label{KKLTcondition2}
\ee
If we want to have a positive volume, $\hat\vo=\alpha\left(\tau_b^{3/2}-\sqrt{3}\tau_s^{3/2}\right)>0$, 
$\tau_b$ has to be larger than $\tau_s$ implying from (\ref{KKLTcondition2}) that we need to impose $a_b<a_s$. 
In general, this condition is satisfied by an appropriate choice of $a_b$ and $a_s$ but in our case tadpole cancellation 
and the requirement of fixing the extra adjoint zero modes, fixes $a_b=\pi/2>a_s=\pi/3$. Hence this crucial condition 
is not satisfied. A way out could be to perform a tuning on the ratio $A_b/A_s$ on the right hand side of (\ref{KKLTcondition}), 
but due to the presence of a logarithm this tuning is huge. As illustrative numerical examples, we mention that 
in order to obtain $\hat\vo\simeq 100$ for $\tau_s\simeq 10$ and $\tau_b\simeq 110$, the tuning is of the order 
$A_b/A_s\simeq 10^{-72}$ and $W_0\sim 10^{-74}$. If we want to push the volume to larger values, $\hat\vo\simeq 200$ 
for $\tau_s\simeq 10$ and $\tau_b\simeq 170$, the tuning increases considerably to
$A_b/A_s\simeq 10^{-114}$ and $W_0\sim 10^{-116}$.
Hence we do not consider these vacua particularly interesting and conclude that in
this model LVS minima seem to arise more naturally than KKLT vacua.

%%%%%%%%%%%%%%%%%%%%%%%%%%%%%%%%%%%%%%%%%%%%%%%%%%%%%%
\section{Higher dP$_{n}$ singularities without flavour branes}\label{sec:HigherdPn:withoutFlBr}
\label{sec:higherdpmodels}
%%%%%%%%%%%%%%%%%%%%%%%%%%%%%%%%%%%%%%%%%%%%%%%%%%%%%%

Even though dP$_0$ provides interesting phenomenological models, such as the trinification model just discussed, the fact that all gauge groups are identical is very restrictive and may be desirable to have more general  models without  flavour D7-branes. Turning to the phenomenology with D-branes at higher dP$_{n}$ singularities with $n>0,$ there is a large class of quiver gauge theories arising that allow for phenomenologically interesting models without flavour branes which we now would like to discuss. Within local model building it is very convenient to introduce flavour branes which allow for very interesting anomaly-free gauge theories serving as candidates for theories beyond the SM (e.g.~\cite{1106.6039}). However there are at least three potential constraints for these types of models:
\begin{enumerate}
\item The  models are not truly local and there is a challenge of decoupling of the dynamics on the flavour brane, i.e.\ generating sufficiently large masses for these states.
\item The field redefinitions found in~\cite{0901.4350,1003.0388,1011.0999} changes substantially the  scales of the soft supersymmetry breaking terms as compared to the case with only D3-branes.
\item The embedding of a large flavour structure into compact models with moduli stabilisation is non-trivial.
\end{enumerate}
With this  in mind we would like to highlight that it is nevertheless possible to construct models based on D3-branes at singularities and no flavour D7s with un-equal ranks, i.e.~gauge theories with ranks not just equal to $SU(n)^k,$ and discuss how to obtain such theories systematically.

%%%%%%%%%%%%%%%%%%%%%%%%%%%%%%%%%%%%%%%%%%%%%%%%%%%%%%
\subsubsection*{Classifying models without flavour branes}
%%%%%%%%%%%%%%%%%%%%%%%%%%%%%%%%%%%%%%%%%%%%%%%%%%%%%%
There is a whole un-explored class of phenomenologically interesting models without flavour branes at higher del~Pezzo singularities (i.e.~$n>0).$ Here we would like to outline a strategy on how to explore this class of models, starting with the toric del~Pezzo surfaces. In particular we are interested in generating anomaly free gauge theories with unequal ranks, i.e.~gauge theories with ranks not just equal to $SU(n)^k.$

\begin{figure}[ht]
\begin{center}
\includegraphics[width=0.4\textwidth]{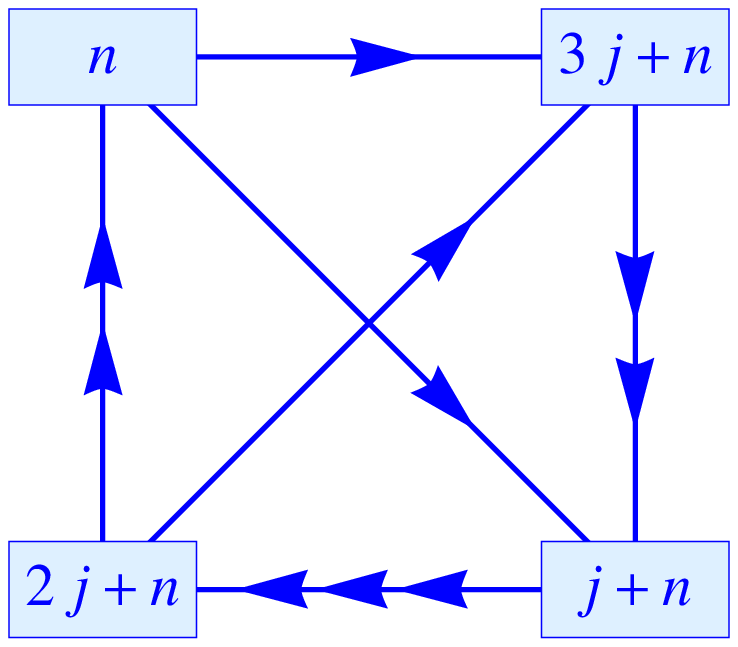}
\includegraphics[width=0.4\textwidth]{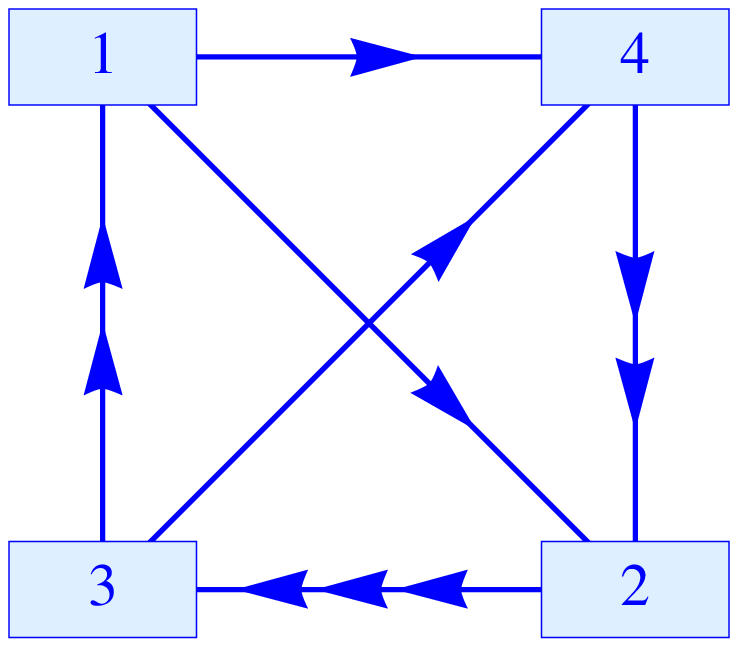}
\captionof{figure}{\footnotesize{{\bf Left:} The general $dP_1$ quiver with unequal ranks, no flavour branes and anomaly cancellation. {\bf Right:} A phenomenologically interesting case of the general quiver on the left with $n=j=1.$}\label{fig:quiversdp1}}
\end{center}
\end{figure}

\begin{figure}[ht]
\begin{center}
\includegraphics[width=0.49\textwidth]{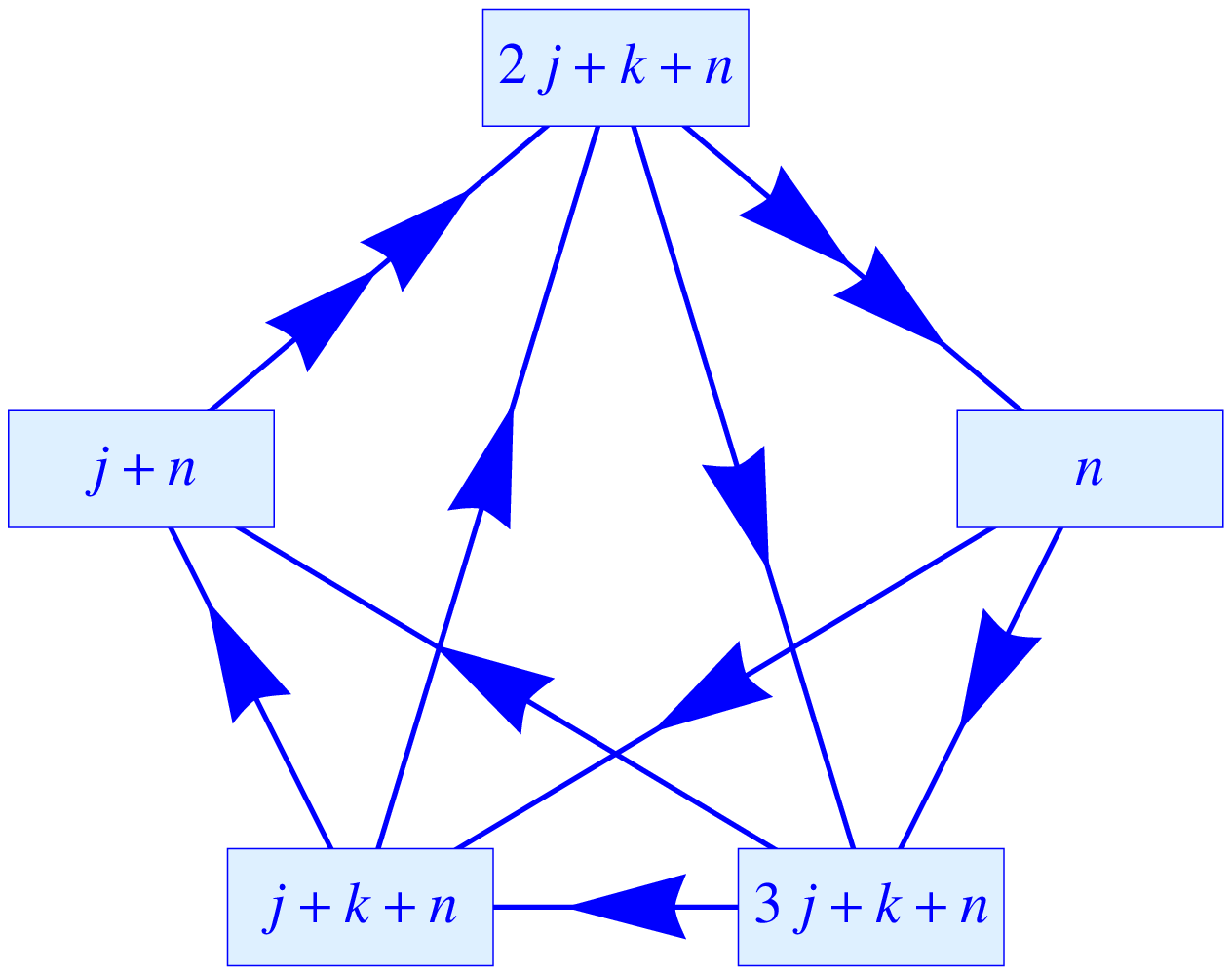}
\includegraphics[width=0.49\textwidth]{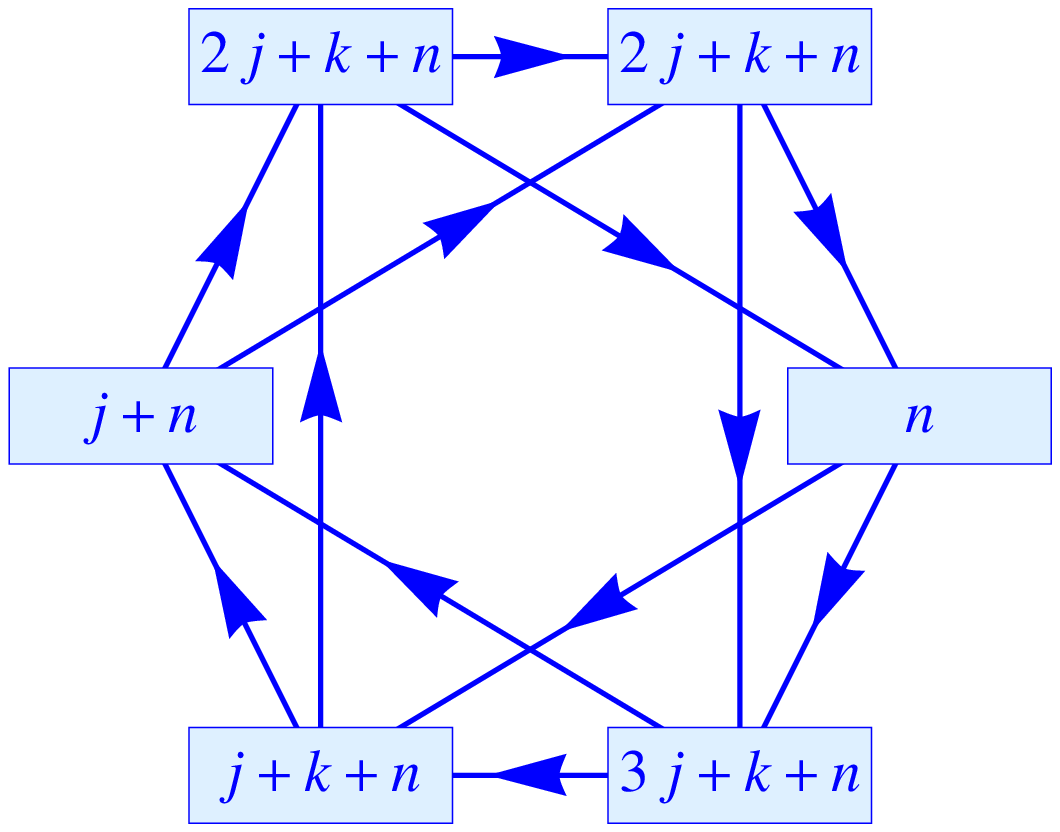}
\captionof{figure}{\footnotesize{{\bf Left:} The general $dP_2$ quiver with unequal ranks, no flavour branes and anomaly cancellation. {\bf Right:} The quiver of $dP_3$ with unequal ranks, no flavour branes and anomaly cancellation.}\label{fig:dp2dp3}}
\end{center}
\end{figure}

For example in the context of dP$_1$ we find an interesting combination based on $SU(3)_c\times SU(2)_L\times SU(4)\times U(1)$ (or more generally $SU(n+k)\times SU(n+2k)\times SU(n+3k)\times SU(n)$) with quivers shown in Figure~\ref{fig:quiversdp1} with three families of left handed quarks. This anomaly-free combination arises from the general anomaly cancellation conditions (without flavour branes)
\begin{eqnarray}
0=2 n_2-n_3-n_4 && 0=-2n_1+3n_3-n_4\\
\nonumber 0=n_1-3n_2+2n_4 && 0=n_1+n_2-2n_3
\end{eqnarray}
as follows. Once one rank is fixed (e.g.~$n_1=n$) and we write $n_2=n+k,$ $n_3=n+j,$ and $n_4=n+l$ one can easily verify that the above conditions are satisfied for 
\begin{equation}
k=2j\text{ and }l=3j\, .
\end{equation}
This is exactly the general solution we mentioned before in the case of dP$_1.$ In a similar fashion we can construct models with unequal ranks on dP$_2$ and dP$_3$ as shown in Figure~\ref{fig:dp2dp3}. Note that all of the above gauge theories with unequal ranks are no longer conformal~\cite{0306092}.
These gauge theories are again obtained by finding a basis of general solutions to the anomaly cancellation conditions. This procedure, applied here for the toric del~Pezzo surfaces, can be applied to all singularities with known quiver gauge theory. In particular it would be interesting to elaborate interesting gauge theories for the toric dual phases of these gauge theories.

In general we expect that such models might lead to appealing models beyond the SM with an interesting flavour physics but a detailed study is required, including the breaking to the SM group.  At this stage we leave a detailed analysis for future work.

\subsubsection*{Wijnholt-Verlinde dP$_8$ model} 

\begin{figure}[ht]
\begin{center}
\includegraphics[width=0.4\textwidth]{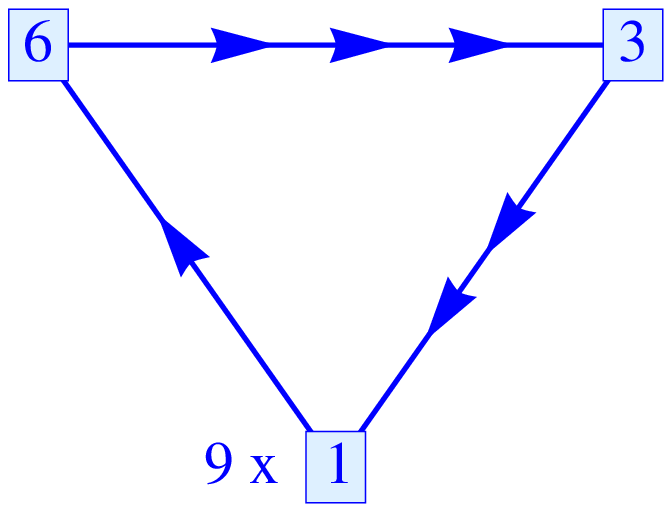}
\includegraphics[width=0.4\textwidth]{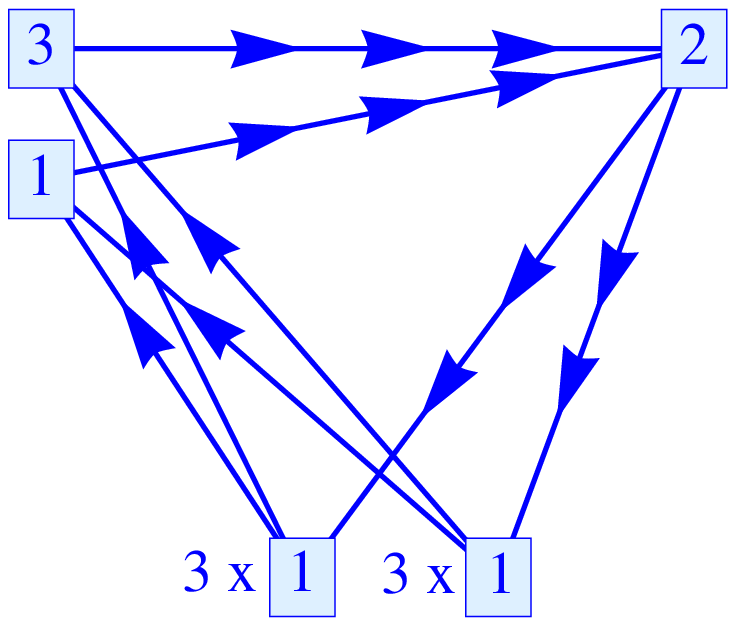}
\captionof{figure}{\footnotesize{{\bf Left:} The dP$_8$ quiver with an $U(6)\times U(3)\times U(1)^9$ gauge group. {\bf Right:} The quiver after breaking to a $SU(3)\times SU(2)\times U(1)^7$ gauge group, containing all SM fields with three generations of Higgses.}\label{fig:dp8quiver}}
\end{center}
\end{figure}

Following the arguments above we can continue with models at higher del~Pezzo surfaces without flavour D7-branes. For completeness we would like to briefly discuss the dP$_8$ model introduced by Wijnholt and Verlinde in~\cite{0508089}.\footnote{We refer the reader for more details on this model to the original paper~\cite{0508089}.} Starting from an $U(6)\times U(3)\times U(1)^9$ gauge group with matter content as shown for completeness in Figure~\ref{fig:dp8quiver} on the left, one can partially resolve the singularity supersymmetrically, corresponding to turning on FI-parameters, that the gauge group is broken down to the $U(3)\times U(2)\times U(1)^7$ as shown in Figure~\ref{fig:dp8quiver} on the right.
Here the Yukawa couplings generally depend on the complex structure moduli and can again as discussed for dP$_0$ in section~\ref{sec:dP0:trinif} be changed by non-commutative deformations.

However in order to be a potentially realistic model to obtain the SM gauge groups all additional $U(1)$ symmetries, apart from hypercharge, need to be massive at low energies. For this to happen upon compactification, the 2-cycle associated to a given $U(1)$ has to be non-trivial in CY three-fold as described in~\cite{0610007}. In this particular model five two-cycles of the dP$_8$ singularity need to be non-trivial within the CY. This leads overall to CY manifolds with $h^{1,1}\geq 9.$ The cycle associated to hypercharge needs to be massless. Since our models are restricted to $h^{1,1}=4,5$ we cannot find a realisation of the Wijnholt-Verlinde model in this class of models, despite featuring various examples with two dP$_8$ singularities such as the model discussed in Appendix~\ref{dP8appendix}. However we do not see an obstacle that for manifolds with large enough $h^{1,1}$ this could be realised explicitly following the formalism developed in the previous sections.

%%%%%%%%%%%%%%%%%%%%%%%%%%%%%%%%%%%%%%%%%%%%%%%%%%%%%%%
\section{Conclusions and open questions}
\label{sec:concl}
%%%%%%%%%%%%%%%%%%%%%%%%%%%%%%%%%%%%%%%%%%%%%%%%%%%%%%%
We have succeeded to make a first approach towards a CY orientifold compactification with a global embedding of D-brane models at singularities in combination with the stabilisation of all moduli and obeying all global consistency conditions.
We consider this a substantial progress in the right direction converting the `local' branes at singularities models into fully-fledged string compactifications. 

We concentrated in this article to the simplest cases, namely D3-branes hosting the visible sector with no flavour D7-branes and a small hidden sector. Besides simplicity, this is an interesting class of models since they are truly local, in the sense that there are no particles charged under the visible sector gauge group outside the singularity and  interactions of the SM states with the other sectors in the CY are highly suppressed. This has important implications for supersymmetry breaking in the observable sector. It was shown in~\cite{0906.3297} that for models at singularities the soft breaking terms can be suppressed as much as $1/{\cal V}$ with respect to the gravitino mass.

Sequestered soft terms allow to address important phenomenological and cosmological issues: the unification scale can be as large as the GUT scale (corresponding to an overall volume of the order ${\cal V}\sim 10^{6-7}$) keeping soft terms in the TeV region, while the cosmological moduli
problem is solved since the moduli are much heavier than the TeV-scale. In particular, the most dangerous modulus, corresponding to the overall volume, has a suppressed mass of order $m_{3/2}/{\cal V}^{1/2}$ compared to the gravitino mass but is still hierarchically heavier than the TeV soft terms. This also ameliorates the Kallosh-Linde problem regarding the relative size of $H$ and $m_{3/2}$ in string models with stabilised moduli~\cite{0411011,0806.0809}. In~\cite{1003.0388,1011.0999} it was shown that a one-loop  field redefinition is needed in order to have a proper supergravity chiral superfield for the `small' K\"ahler moduli. This redefinition brings back the soft breaking terms to be very close to the gravitino mass. However, this field redefinition is absent for precisely the models with only D3-branes at singularities and no flavour D7-branes stretching in the bulk~\cite{0901.4350}. This  differentiates  these models from other local models that include also flavour D7-branes on important physical observables.

From the general class of CY manifolds obtained from toric ambient spaces, we have identified those that allow to embed the SM in fractional D3-branes at del~Pezzo singularities. We have found many examples with $3<h^{1,1}<6$ with a pair of identical del~Pezzo singularities mapped into each other by the orientifold involution and having extra rigid cycles in the geometric regime that give rise to non-perturbative effects either from gaugino condensation on wrapped D7-branes or Euclidean D3-brane instantons. We illustrated this construction in particular models with the observable sector in dP$_0$ and dP$_8$ (this is presented in Appendix \ref{dP8appendix}) singularities for which all K\"ahler moduli were stabilised, giving examples of both KKLT and dS LVS scenarios. Furthermore, for dP$_0$ a quasi-realistic model based on the trinification group $SU(3)^3$ was studied, outlining how breaking to the SM and other realistic features can be achieved.

Despite this success there are several open questions that are worth summarising:
\begin{itemize}

\item{} A more detailed analysis  of the most promising models listed in appendix~\ref{sec:applist} deserves further study. In particular  models that besides having del~Pezzo singularities are also K3 fibrations. This is important in order to have moduli that can play a r\^ole for inflation and for having the potential for  non-gaussianities~\cite{0808.0691,1005.4840,1202.4580}.

\item{} Extension of our general classification to models with $h^{1,1}\geq 6$  will be desirable in order to have a more general class of models that can include several hidden sectors (including hidden sectors at singularities that have been shown recently to give rise to dS vacua~\cite{1203.1750} ) and a larger number of candidates for inflatons or other cosmological applications. The systematic search and explicit construction of the models is technically much more challenging and will be left for future research if needed. 

\item{}  Identifying a fully realistic model that achieves the potential of local models including gauge unification, proton stability, realistic mixings in the quark and lepton sectors as discussed from the local perspective in~\cite{1002.1790,1106.6039}. Such a model also needs to include a dynamical stabilisation of all breaking, respectively flavour, scales within the compact set-up. A first step towards identifying the dynamical breaking scale has been achieved here by outlining how to break to the SM dynamically using the renormalisation group of the soft-masses. In addition, a more concrete scenario of supersymmetry breaking with full control of the structure of soft breaking terms, including flavour issues is still pending~\cite{0912.2950,1012.1858}. Models with only D3-branes at dP$_n$ with $n > 0,$ as those described in section~\ref{sec:higherdpmodels}, can provide a richer spectrum and breaking patterns than the simplest trinification-like models studied here.

\item{} Including flavour D7-branes is needed in order to have a complete description of  this class of models. General consistency constraints may require embedding the models in F-theory as it was done in~\cite{0005067,1201.5379}.  We leave this important challenge for a future project.

\item{} Models with flavour D7-branes would allow many more 
realistic options and could potentially include some promising local models such as the dP$_3$ case studied in~\cite{1106.6039}. The absence of examples, for instance with dP$_3$ singularities, in the class of CY manifolds analysed  is not seen as an obstacle and these models are expected to appear for CY manifolds with $h^{11}\geq 6$ but it might also require additional model building to obtain phenomenologically interesting models by higgsing from higher to lower del~Pezzo singularities.
\end{itemize}
Let us finish with the following  general statements. Even though it is often stated that we may not yet know enough  about string theory in order to approach detailed phenomenological questions, the continuous progress in this area indicates that this statement may be too conservative. Considering the low energy spectrum of type IIB string theory, we have seen that over the years a complete use of each of the fields has been found and each time giving rise to more realistic features. First the background metric $g_{MN}$ provides the basic set-up for a chiral ${\cal N}=1$ four-dimensional theory after CY orientifold compactification. The NS and RR antisymmetric tensors allow for flux compactifications that give rise to moduli stabilisation and the discretuum of vacua. Also, these antisymmetric tensors couple naturally to extended objects, the D-branes. Since they can hold chiral matter, this opened the possibility to have the visible sector inside the D-branes. 

Furthermore, within the implicit weak coupling assumption in these constructions, $\alpha'$ and string perturbative effects have been included and shown to play a r\^ole, also standard and stringy non-perturbative effects have been studied in detail and are computable, giving rise not only to K\"ahler moduli stabilisation but also to corrections of SM couplings. Moreover, local continuous and discrete symmetries are under control, including their breaking and potential r\^ole for providing the gauge symmetries and approximate global symmetries in the SM. Phenomenological issues have been possible to be approached. Not only the spectrum of the  models,  but also hierarchy of Yukawa couplings, mixings in the quark and lepton sectors,  etc.

This continuous progress  give further encouragement for continuing efforts towards models closer and closer to the SM within a full ultraviolet completion. The fact that many ingredients are used towards a realistic model may look too baroque for an outsider. On the other hand, the simplest versions with  zero fluxes, no D-branes, etc. are very non-generic. The apparent baroqueness of the models is actually a positive feature in the sense that, as in any  effective physical model, everything that can be included should be included unless there is a reason of symmetry or otherwise that protects the corresponding quantity to vanish.  Not turning-on fluxes and branes and ignoring perturbative and non-perturbative effects is not an option. Ongoing experimental results from the LHC and cosmological observations regarding density perturbations of the CMB will hopefully provide further guidelines to 
identify fully realistic string models beyond the SM. We hope that  the structure of the models presented here and the formalism used to construct compact local models at singularities is a step in this direction. We  plan to come back to some of the open questions mentioned above in the near future.

\subsection*{Acknowledgements}
We would like to thank Per Berglund, Andreas Braun, Volker Braun, Andres Collinucci, Joe Conlon, I\~naki Garc\'{\i}a-Etxebarria, Mark Goodsell, Sheldon Katz,  Luis Ib\'a\~nez, Anshuman Maharana, Bernd Siebert and Timo Weigand for useful discussions. We would like to thank Sven Krause for help with Sage. The work of C.M. and S.K. was supported by the DFG through TR33 ``The Dark Universe''. S.K. was also supported by the European Union 7th network program Unification in the LHC era (PITN-GA-2009-237920). The work of R.V. was supported by the German Science Foundation (DFG) under the Collaborative Research Center (SFB) 676 and the Research Training Group 1670. MC, SK and FQ thank the Isaac Newton Institute for  providing excellent working conditions  during the last stages of this project.

\appendix
%%%%%%%%%%%%%%%%%%%%%%%%%%%%%%%%%%%%%%%%%%%%%%%%%%%%%%%
\section{Global embedding of the dP$_8$ quiver gauge theory} \label{dP8appendix}
%%%%%%%%%%%%%%%%%%%%%%%%%%%%%%%%%%%%%%%%%%%%%%%%%%%%%%%

In this section we describe a global embedding of a quiver model sitting at a dP$_8$ singularity in a type IIB orientifold compactification.
This is realised by considering a CY three-fold with two dP$_8$ singularities. As for the dP$_0$ model, this three-fold has a holomorphic involution that exchanges the two singularities (or that exchanges the two dP$_8$ divisors in the resolved picture). The fixed locus of the involution does not include the two singular points, as the resolved dP$_8$'s do not intersect the orientifold plane.
Putting D3-branes on top of one of the two singularities, we get (in the quotient space) a quiver gauge theory at the dP$_8$ singularity known from local model building.

The D7-charge of the O7-plane is cancelled by putting four D7-branes (plus their orientifold images) on top of the O7-plane. These D7-branes will not intersect the quiver locus and will provide a suitable hidden sector.

%%%%%%%%%%%%%%%%%%%%%%%%%%%%%%%%%%%%%%%%%%%%%%%%%%%%%%%
\subsection{Geometric set-up}
%%%%%%%%%%%%%%%%%%%%%%%%%%%%%%%%%%%%%%%%%%%%%%%%%%%%%%%
In this section we will study moduli stabilisation for an embedding of a dP$_8$ quiver gauge theory in a compact CY threefold. The CY that we take has already been presented in~\cite{DiacEtAl} in the context of global embedding of quiver gauge theories. We will shortly review the model here and we refer to~\cite{DiacEtAl} for the details. In the table of appendix~\ref{sec:applist-pic4} it is listed under number 21.

This CY $X$ has Hodge numbers $h^{1,1}=4$ and $h^{1,2}=214$.
It is an hypersurface in the ambient space with the following weights\footnote{We use the same conventions as~\cite{DiacEtAl} in order to make easier for the reader to search in the original paper for details. We give the weight matrix in an equivalent form.}
\begin{equation}
\begin{array}{|c|c|c|c|c|c|c|c||c|}
\hline W_1 & W_2 & W_3 & W_4 & W_5 & Z & X & Y & D_\textmd{H} \tabularnewline \hline \hline
    0  &  0  &  0  &  0  &  0  &  1  &  2  &  3  & 6\tabularnewline\hline
    1  &  1  &  1  &  0  &  0  &  0  &  6  &  9  & 18\tabularnewline\hline
    0  &  1  &  0  &  1  &  0  &  0  &  4  &  6  & 12\tabularnewline\hline
    0  &  0  &  1  &  0  &  1  &  0  &  4  &  6  & 12\tabularnewline\hline
\end{array}\label{eq:model3dP8:weightm}\,,
\end{equation}
and Stanley-Reisner ideal
\begin{equation}
\label{eq:model3dP8:sr-ideal}
{\rm SR}=\{W_1\, W_2\,W_3,\, W_2\, W_4,\, W_3 \, W_5,\, W_4\, W_5, \, W_1\,W_2\,X\,Y, \, W_1\,W_3\,X\,Y, \,  W_4\,Z, \,W_5\,Z, \, X\, Y\, Z\}\,.
\end{equation}
A basis of $H^{1,1}(X)$ is given by\footnote{This is not an integral basis: for example $D_{W_1}=\frac16(\G_1-3\G_2-3\G_3-\G_4)$.}
\begin{equation}\label{dP8model-basis}
 \G_1 = 3D_{W_3} + 3D_{W_4} + D_{Z} \qquad \G_2 = D_{W_4} \qquad \G_3 = D_{W_5} \qquad \G_4 =  D_{Z} \:.
\end{equation}
The intersection form in this basis is diagonal:
\begin{equation}
  I_3 = 9\G_1^3 + \G_2^3 + \G_3^3 + 9\G_4^3 \:.
\end{equation}
This CY threefold has one $dP_0$ at $Z=0$ and two $dP_8$'s at $W_4=0$ and $W_5=0$. The two $dP_8$'s are exchanged by the following involution~\cite{DiacEtAl}:
\begin{equation}
 W_2 \leftrightarrow W_3 \qquad \mbox{and} \qquad W_4 \leftrightarrow W_5 \:.
\end{equation}
The corresponding fixed locus is given by a (complex-)codimension one object at $W_3W_4-W_2W_5=0$ and by four isolated fixed points (one at the intersection $W_3W_4+W_2W_5=W_1=Z=0$ and three at the intersection $W_3W_4+W_2W_5=W_1=Y=0$~\cite{DiacEtAl}). So by implementing this orientifold involution, one obtains one O7-plane in the class $[D_{O7}]=[D_{W_3}]+[D_{W_4}]$ and four O3-planes.

Expanding the K\"ahler form on the basis \eqref{dP8model-basis}, $J=\sum_i t_i\G_i$, one has the following volumes of the three del~Pezzo divisors
\be
{\rm Vol}(D_Z) = \tfrac92 t_4^2 \:, \qquad
{\rm Vol}(D_{W_4}) = \tfrac12 t_2^2 \:,\qquad
{\rm Vol}(D_{W_5}) = \tfrac12 t_3^2 \:,
\ee
and the volume of the CY three-fold is:
\be
{\rm Vol}(X) = \frac16 (9 t_1^3 + t_2^3 + t_3^3 + 9 t_4^3) \:.
\ee
The K\"ahler cone of the ambient space is:
\begin{equation}
 t_2<0 \qquad t_3<0 \qquad t_1+t_2+t_4>0 \qquad t_1+t_3+t_4>0 \qquad t_4<0 \:.
\end{equation}
This space is a priori only a subspace of the K\"ahler cone of the CY. On the other hand,
in this case, the point we want to consider, i.e.\ a CY with two dP$_8$ singularities, is included in this subspace.
The orientifold invariant K\"ahler form is given by taking $t_2=t_3$. Moreover, from the volumes of the divisors we see that the limit we want to consider is $t_2=t_3\rightarrow 0$. In this case, the K\"ahler cone reduces to
\begin{equation}
 t_1+t_4>0 \qquad \mbox{and} \qquad -t_4>0 \:.
\end{equation}

%%%%%%%%%%%%%%%%%%%%%%%%%%%%%%%%%%%%%%%%%%%%%%%%%%%%%%%
\subsection{Brane set-up and consistency conditions}
%%%%%%%%%%%%%%%%%%%%%%%%%%%%%%%%%%%%%%%%%%%%%%%%%%%%%%%
The set of branes that we consider is:
\begin{itemize}
 \item On the two singularities we put $N_{D3}$ D3-branes. This can lead to the quiver gauge theory of Wijnholt and Verlinde~\cite{0508089} or other quiver gauge theories obtained as outlined in Section~\ref{sec:higherdpmodels}.  
 \item To cancel the D7-brane of the O7-plane, we consider four D7-branes (plus their images) on top of the O7-plane. This D7-brane stack does not intersect the two $dP_8$ singularities.
 \item We will have an E3-instanton wrapping the rigid and invariant $dP_0$ cycle at $Z=0$.
\end{itemize}

Let us consider the E3-instanton first. It wraps a non-spin cycle, so a non-zero gauge flux is needed to cancel the Freed-Witten anomaly. This flux would render the instanton configuration non-invariant and must then be cancelled by a proper choice of the $B-$field, i.e.\ $B= \sum_i b_i \G_i + \frac12 D_Z$ where $b_i\in\mathbb{Q}$ are chosen such that $\sum_i b_i \G_i$ is an integral two-form and such that ${\cal F}_{E3}=0$.\footnote{This $B-$field makes the option to have a Whitney brane wrapping a cycle in the class $8[O7]$ be not available. In fact, with such a $B-$field the flux ${\cal F}_W$ on the Whitney brane would make it split into one brane and its image (see~\cite{Collinucci:2008pf}).}

With this choice of $B-$field, we are not able to cancel the Freed-Witten gauge flux on the $SO(8)$ D7-brane stack wrapping the divisor $[D_{O7}]=\tfrac13(\G_1-\G_4)$ (note that $D_{O7}$ is an integral cycle; the factor $1/3$ comes from that fact that we have not chosen an integral basis for $H^2(X)$). The gauge flux on this stack is given by
\begin{eqnarray}
 {\cal F}_{D7} &=& (F_{D7}^{\rm integral} + \tfrac12 D_{O7} )- B \nonumber \\
  &=& (f_1-b_1+\tfrac16)\G_1 + (f_2-b_2)\G_2 + (f_3-b_3)\G_3 + (f_4-b_4-\tfrac23)\G_4 \:,
\end{eqnarray}
where $F_{D7}^{\rm integral}$ is an integral two-form. No choice of $F_{D7}^{\rm integral}$ can make ${\cal F}_{D7}=0$. We have chosen it such that it is equal for all the four D7-branes in the stack. This makes the $SO(8)$ gauge group to be broken to $U(1)\times SU(4)$, where the $U(1)$ gauge boson will get a mass by a St\"uckelberg mechanism.

The flux generates chiral modes at the intersection with the E3-instanton wrapping $D_{E3}=D_Z$. Their number is given by
\begin{equation}
 I_{D7-E3}=\int_{D_{O7}\cap D_{E3}} {\cal F}_{D7} = 2 -3f_4+3b_4 \:.
\end{equation}
We will choose $b_4 = f_4 -\frac23$, such that we have $I_{D7-E3}=0$. \footnote{One can check that with this choice of $b_4$,  is integral once we impose that $F_{D7}^{\rm integral}$ is an integral two-form.}
Then there are no further instanton zero modes and the instanton contribution to the superpotential is generated.

The flux will induce also chiral modes on the bulk of the D7-brane stack. In particular there will be the following number of chiral states in the antisymmetric representation of the unbroken $U(4)$ gauge group:
\begin{equation}
 I_{D7}^{\rm antisym} = \int_X D_{O7} \wedge D_{O7} \wedge 2{\cal F}_{D7} = 2f_1-2b_1+\tfrac13   \:,
\end{equation}
where we used $b_4 = f_4 -\frac23$.
These chiral fields are charged under the diagonal $U(1)$ factor of the $U(4)$ gauge group and will enter in its D-term together with the K\"ahler moduli dependent and flux generated FI-term
\begin{equation}
\xi_{D7}=\tfrac{1}{{\cal V}} \int_{D_{O7}}{\cal F}_{D7}\wedge J = \tfrac{1}{2{\cal V}}  (1 - 6 b_1 + 6 f_1) t_1\:.
\end{equation}
Note that when the CY has finite volume, this FI-term is always different from zero (there is no value of $f_1$ that makes both $(1 - 6 b_1 + 6 f_1)=0$ and $F_{D7}^{\rm integral}$ be integral.

Finally, let us compute the D3-charge of the chosen configuration.
The D3-charge of the quiver locus is $2\times N_{D3}$. Each of the four O3-plane contributes as $-1/4$.
Together with the D3-charge of the O7-plane and the fluxed D7-brane stack we have
\be
Q_{D3}^{\rm excep} + Q_{D3}^{D7} + Q_{D3}^{O3} = -2N_{D3}  -19 -\tfrac13 (1 - 6 b_1 + 6 f_1)^2 \:.
\ee
Note that in this case we have a negative contribution coming from the flux. This is due to the fact that its FI-term is not cancelled and then ${\cal F}_{D7}$ is not forced to be anti-selfdual.

\subsection{K\"ahler moduli stabilisation}

The model we have just described has just one O7-plane, on top of which we wrap four D7-branes (plus their images).
Furthermore we have a rigid dP$_0$ cycle that is transversally invariant under the orientifold involution.
Hence we shall wrap an E3 instanton on this rigid divisor. Both the O7 and the E3 do not intersect the two dP$_8$'s,
and so they are completely decoupled from them. However, the O7-plane intersects the E3. We can cancel the FW flux on the E3,
but the flux on the D7 on top of the O7 cannot be cancelled. This implies that there is an FI-term on this stack.
For a generic flux on the D7, we have also chiral intersections with the E3. We can however cancel these chiral modes
with an appropriate flux choice. Once we perform this choice, the resulting FI-term cannot be set to zero without shrinking the whole
Calabi-Yau. Nonetheless we have also some modes coming from the bulk of the D7 which are charged under the corresponding $U(1)$
(these modes are in the anti-symmetric representation of $SU(4)$  ---  the flux breaks $SO(8)$ to $SU(4)\times U(1)$).
These modes enter into the D-terms and they can get a non-zero VEV partially cancelling the FI-term.
What is left over can again give rise to an uplifting term similarly to the dP$_0$ case studied in the main text.
Thus in this model the E3-instanton and the $\alpha'$ corrections fix the volume
and the size of the dP$_0$ supporting the non-perturbative effects, while the two dP$_8$'s are fixed
in the singular regime by using the corresponding D-terms.

\section{List of varieties with global embeddings of quiver gauge theories}\label{sec:applist}

In this appendix we give a list of the models that fulfil the criteria of section~\ref{sec:mod-stab-in-global-quiver}. As we mentioned already in the introduction, our search was performed in a subclass of all CY threefolds, namely those related to reflexive lattice polytopes, which is motivated as follows. Firstly, we know the Hodge data of the CY hypersurfaces coming from such polytopes~\cite{Batyrev:1994hm} and that these CYs are smooth. In addition, this class is already quite large~\cite{0002240} and one assumes that the properties of these CY manifolds represent those of all CY manifolds. The technically most important reason is that we can access a lot of CY data via toric methods which one can easily put on a computer.

The toric varieties in which the CY hypersurfaces are embedded are described by fans. These fans are in one-to-one relation to the coherent star triangulations of the four dimensional reflexive lattice polytopes. Since, there is usually more than just one triangulation for such a polytope, we may obtain several toric varieties per polytope. Note however that not all triangulations lead to different CYs, since there may exist flop-transitions of the ambient spaces which do not effect the hypersurface. Since, we did not care about this issue in our scan, only in the examples we presented in the main text, we present here the list of polytopes for which one or several triangulations meets our constraints. 

The calculations were done by means of  PALP~\cite{Kreuzer:2002uu} with the \verb|mori.x| extension~\cite{Braun:2011ik,Braun:2012vh} and the toric variety package of Sage~\cite{BraunNovoseltsev:toric_variety}.
In our analysis we checked whether the CY has for a given triangulation two toric del Pezzo divisors of the same kind which are related by a $\mathbb Z_2$-symmetry of the lattice polytope. As explained in~\cite{Braun:2012vh}, \verb|mori.x| only tests necessary conditions for the divisor to be del Pezzo. Further, we do not check whether the involution of the polytope leads to an involution on the CY itself, i.e.\ whether the intersection ring on the CY is invariant under it. However, for all the cases that we worked out in detail, the involution was also respected by the CY. An additional technical constraint is that we did not treat divisors which factorise on the hypersurface differently. Therefore, in this scan we miss examples like \verb|9 1 1 1 3 3 0  3 0 0 0 1 1 1| and \verb|16 1 1 2 4 8|.

Our findings can be summarised as follows: starting from 1197 and 4990 polytopes for $h^{1,1}=4$ and $h^{1,1}=5$, respectively, we find 80 and 346, respectively, with two equal del Pezzos and an involution which interchanges them. If we demand, in addition, that the two divisors do not intersect, the numbers go down to 68 and 305. Imposing the final constraint, of having an additional non-intersecting rigid divisor, we obtain 21 and 168 polytopes, respectively.

\subsection{Results for the four moduli case}\label{sec:applist-pic4}
The following table lists all the polytopes that give rise to a CY with $h^{1,1}=4$ and with at least one triangulation such that the hypersurface fulfils all the conditions of section~\ref{sec:mod-stab-in-global-quiver}. The third column indicates the kind of del Pezzo surfaces we find. The fourth column gives the dimension of the anti-invariant classes which we obtain for the involutions that map the two del Pezzos into each other. If  there is a check mark in the fifth column, then the CY admits a K3-fibration  on the level of the polytope and the (would-be) fibre is a K3 surface. Note that like the case of the del Pezzo property test these are just necessary conditions. We did not analyse the SR-ideal of the CY to see whether the projection is well-defined.
\begin{longtable}{c||c|c|c|c|c}
  \# & weights & dP$_n$ & $h^{1,1}_-$ & $h^{2,1}$ & K3\\
  \hline\hline
$1$ & {\scriptsize\verb|9 1 3 3 1 1 /Z3: 0 2 1 0 0|} & $0$ & $1$ &112&  \\
$2$ & {\scriptsize\verb|9 3 3 1 1 1 0 0  2 1 0 0 0 0 0 1  2 0 1 0 0 0 1 0|} & $0$ & $1$ &112&  \\
$3$ & {\scriptsize\verb|9 3 3 1 1 1 0 0  2 0 0 0 0 0 1 1  12 6 0 1 1 1 0 3|} & $0$ & $1$ &112&  \\
$4$ & {\scriptsize\verb|6 2 1 1 1 1 0 0 0  3 1 0 0 1 0 1 0 0  4 1 1 1 0 0 0 1 0  6 2 1 1 0 0 1 0 1|} & $6$ & $1$ &64&  \\
$5$ & {\scriptsize\verb|6 2 1 0 1 1 1 0 0  3 1 0 1 0 0 1 0 0  3 1 0 0 1 0 0 1 0  4 1 1 0 0 1 0 0 1|} & $6$ & $1$ &64&  \\
$6$ & {\scriptsize\verb|6 2 1 0 1 0 1 1 0  3 1 1 0 0 0 0 0 1  3 1 0 1 0 0 0 1 0  3 1 0 0 0 1 1 0 0|} & $6$ & $1$ &70&  \\
$7$ & {\scriptsize\verb|7 2 2 0 1 1 1 0 0  3 1 1 0 0 0 0 1 0  4 1 1 1 0 0 1 0 0  4 1 1 0 0 1 0 0 1|} & $6$ & $1$ &73&  \\
$8$ & {\scriptsize\verb|4 1 1 1 0 0 1  2 0 0 0 1 1 0 /Z2: 1 1 1 1 0 0|} & $7$ & $1$ &52& $\checkmark$ \\
$9$ & {\scriptsize\verb|5 1 1 1 0 1 0 1 0  4 1 1 0 0 0 0 1 1  5 1 1 1 0 0 1 0 1  5 1 1 0 1 0 1 1 0|} & $7$ & $1$ &52&  \\
$10$ & {\scriptsize\verb|8 1 1 3 0 0 1 2 0  2 0 0 1 0 0 0 0 1  4 1 0 1 0 1 0 1 0  4 0 1 1 1 0 0 1 0|} & $7$ & $1$ &72&  \\
$11$ & {\scriptsize\verb|10 5 1 1 0 1 0 2 0  4 2 0 0 1 1 0 0 0  6 3 1 0 0 0 1 1 0  6 3 0 1 0 0 0 1 1|} & $7$ & $1$ &82&  \\
$12$ & {\scriptsize\verb|8 4 0 0 1 0 1 1 1  6 3 1 0 0 0 0 1 1  6 3 0 1 0 0 1 1 0  6 3 0 0 1 1 0 1 0|} & $7$ & $1$ &94&  \\
$13$ & {\scriptsize\verb|8 4 1 1 0 1 0 0 1  4 2 0 1 0 0 0 1 0  4 2 0 0 0 0 1 0 1  6 3 1 0 1 1 0 0 0|} & $7$ & $1$ &94& $\checkmark$ \\
$14$ & {\scriptsize\verb|8 1 4 0 0 1 1 1 0  4 1 2 0 0 0 0 0 1  4 0 2 1 0 0 0 1 0  4 0 2 0 1 0 1 0 0|} & $7$ & $1$ &98&  \\
$15$ & {\scriptsize\verb|10 5 2 1 0 0 1 0 1  4 2 1 0 0 0 0 1 0  6 3 1 0 0 1 0 0 1  10 5 2 0 1 1 1 0 0|} & $7$ & $1$ &98&  \\
$16$ & {\scriptsize\verb|10 5 2 0 1 1 1 0 0  4 2 1 0 0 0 0 1 0  6 3 1 1 0 0 1 0 0  6 3 1 0 0 1 0 0 1|} & $7$ & $1$ &110&  \\
$17$ & {\scriptsize\verb|12 2 0 2 4 1 0 3  4 0 2 0 0 1 0 1  12 2 0 2 0 3 4 1|} & $8$ & $1$ &44&  \\
$18$ & {\scriptsize\verb|7 3 1 1 0 1 1 0 0  2 1 0 0 0 0 0 0 1  2 0 0 0 1 0 0 1 0  8 3 1 1 1 0 2 0 0|} & $8$ & $1$ &64&  \\
$19$ & {\scriptsize\verb|6 1 3 0 1 1 0  4 0 2 1 0 0 1 /Z2: 0 0 1 1 0 0|} & $8$ & $1$ &70&  \\
$20$ & {\scriptsize\verb|8 4 1 1 1 0 1 0 0  4 2 1 0 0 0 0 1 0  6 3 0 0 1 0 0 1 1  8 4 1 1 0 1 0 0 1|} & $8$ & $1$ &74&  \\
$21$ & {\scriptsize\verb|18 6 9 0 1 1 1 0  12 4 6 1 0 0 1 0  12 4 6 0 0 1 0 1|} & $8$ & $1$ &214&
\end{longtable}

\subsection{Results for the five moduli case}\label{sec:applist-pic5}
The following table lists all the polytopes that give rise to a CY with $h^{1,1}=5$ and with at least one triangulation such that the hypersurface fulfils all the conditions of section~\ref{sec:mod-stab-in-global-quiver}.
\begin{center}
\begin{longtable}{c||c|c|c|c|c}
  \# & weights & dP$_n$ & $h^{1,1}_-$ & $h^{2,1}$  & K3 \\
  \hline\hline
%dP0
$1$ & {\tiny\verb|6 1 1 1 1 2 0 0 0 0  2 1 0 0 0 0 0 1 0 0  2 0 1 0 0 0 1 0 0 0  8 1 3 1 1 0 0 0 2 0  8 3 1 1 1 0 0 0 0 2|} & $0$ & $2$ &51&  \\
$2$ & {\tiny\verb|6 1 2 1 0 1 0 1 0 0  2 1 0 0 0 0 0 0 0 1  2 0 1 0 0 0 0 0 1 0  6 2 1 1 0 0 1 1 0 0  8 1 3 1 1 2 0 0 0 0|} & $0$ & $2$ &55&  \\
$3$ & {\tiny\verb|8 1 3 0 1 1 2 0 0 0  2 1 0 0 0 0 0 0 0 1  2 0 1 0 0 0 0 0 1 0  3 1 0 1 0 0 0 1 0 0  3 0 1 1 0 0 1 0 0 0|} & $0$ & $2$ &59&  \\
$4$ & {\tiny\verb|5 1 1 1 1 1 0 0 0 0  5 1 2 0 1 0 1 0 0 0  6 1 2 1 1 0 0 0 0 1  6 1 2 1 0 1 0 0 1 0  6 1 2 0 1 1 0 1 0 0|} & $0$ & $1$ &75& $\checkmark$ \\
$5$ & {\tiny\verb|6 1 3 1 0 0 1  2 0 0 0 1 1 0 /Z2: 0 1 0 1 0 0|} & $0$ & $1$ &77&  \\
$6$ & {\tiny\verb|4 1 1 0 1 0 0 1 0  3 0 1 0 1 1 0 0 0  4 1 1 0 0 1 0 0 1  6 3 1 1 0 0 1 0 0|} & $0$ & $1$ &77&  \\
$7$ & {\tiny\verb|9 1 0 1 3 1 3 0 0  2 0 0 0 1 0 0 1 0  2 0 0 0 0 0 1 0 1  3 0 1 0 0 0 0 1 1|} & $0$ & $1$ &77& $\checkmark$ \\
$8$ & {\tiny\verb|5 1 2 0 1 1 0 0 0 0  2 0 0 1 0 0 1 0 0 0  3 0 1 1 0 0 0 0 0 1  3 0 1 0 0 0 1 0 1 0  6 1 3 0 0 1 0 1 0 0|} & $0$ & $1$ &79& $\checkmark$ \\
$9$ & {\tiny\verb|9 3 2 2 1 1 0 0 0  3 1 1 0 0 0 0 0 1  3 1 0 1 0 0 0 1 0  3 0 1 1 0 0 1 0 0|} & $0$ & $1$ &80&  \\
$10$ & {\tiny\verb|4 1 1 0 1 0 1 0 0 0  5 1 2 1 0 1 0 0 0 0  5 1 2 0 1 0 0 0 0 1  5 1 2 0 0 0 1 0 1 0  6 1 3 0 0 1 0 1 0 0|} & $0$ & $1$ &83&  \\
$11$ & {\tiny\verb|6 1 2 0 1 1 1 0 0 0  5 1 2 1 0 1 0 0 0 0  6 1 3 0 0 1 0 1 0 0  7 1 3 0 1 1 0 0 1 0  7 1 3 0 0 1 1 0 0 1|} & $0$ & $1$ &87& $\checkmark$ \\
$12$ & {\tiny\verb|7 1 2 2 1 1 0 0 0  2 0 1 0 0 0 0 0 1  2 0 0 1 0 0 0 1 0  9 1 3 3 0 1 1 0 0|} & $0$ & $1$ &89&  \\
$13$ & {\tiny\verb|10 2 2 4 1 1 0 0  11 2 2 5 1 0 0 1  11 2 2 5 0 1 1 0|} & $0$ & $1$ &95& $\checkmark$ \\
$14$ & {\tiny\verb|6 1 1 0 0 0 1 0 1 2  2 0 0 0 0 1 0 0 0 1  2 0 0 0 0 0 0 1 1 0  6 1 0 1 0 0 1 0 2 1  8 1 2 0 1 0 0 0 1 3|} & $0$ & $2$ &55&  \\
%dP1
$15$ & {\tiny\verb|5 1 1 0 1 1 1 0  2 0 0 1 0 0 0 1  8 2 0 0 1 2 0 3|} & $1$ & $2$ &33&  \\
$16$ & {\tiny\verb|7 2 2 1 0 0 1 1 0 0  3 1 1 0 0 0 0 0 1 0  4 1 1 0 0 0 0 1 0 1  7 2 2 0 1 1 0 0 0 1  9 2 3 1 0 1 0 0 0 2|} & $1$ & $2$ &44&  \\
$17$ & {\tiny\verb|6 0 1 2 1 1 0 1 0 0  2 0 1 0 0 0 0 0 0 1  2 0 0 1 0 0 0 0 1 0  4 1 1 0 1 0 1 0 0 0  4 1 0 1 1 1 0 0 0 0|} & $1$ & $2$ &55&  \\
$18$ & {\tiny\verb|5 1 1 1 0 1 1 0 0 0  2 0 1 0 0 0 0 0 0 1  2 0 0 0 0 1 0 1 0 0  6 1 1 0 1 2 1 0 0 0  6 1 2 0 0 1 1 0 1 0|} & $1$ & $2$ &65&  \\
$19$ & {\tiny\verb|6 1 2 0 1 0 1 1 0 0  2 1 0 0 0 0 0 0 0 1  2 0 1 0 0 0 0 0 1 0  4 1 1 1 0 0 0 1 0 0  6 2 1 0 0 1 1 1 0 0|} & $1$ & $2$ &67&  \\
$20$ & {\tiny\verb|5 1 2 1 0 0 1 0 0 0  2 1 0 0 0 0 0 0 1 0  2 0 1 0 0 0 0 1 0 0  3 1 1 0 0 0 0 0 0 1  5 2 1 0 1 1 0 0 0 0|} & $1$ & $2$ &84&  \\
$21$ & {\tiny\verb|9 3 3 0 1 1 1  6 2 2 1 0 0 1 /Z3: 1 2 0 0 0 0|} & $1$ & $1$ &101& $\checkmark$ \\
$22$ & {\tiny\verb|6 2 2 1 0 0 1  6 2 2 0 1 1 0 /Z3: 1 2 0 0 0 0|} & $1$ & $1$ &101& $\checkmark$ \\
$23$ & {\tiny\verb|9 3 3 0 1 1 1 0 0  2 1 0 0 0 0 0 0 1  2 0 1 0 0 0 0 1 0  6 2 2 1 0 0 1 0 0|} & $1$ & $1$ &101& $\checkmark$ \\
$24$ & {\tiny\verb|6 2 2 1 0 0 1 0 0  2 1 0 0 0 0 0 0 1  2 0 1 0 0 0 0 1 0  6 2 2 0 1 1 0 0 0|} & $1$ & $1$ &101& $\checkmark$ \\
$25$ & {\tiny\verb|9 3 3 0 1 1 1 0 0  2 0 0 0 0 0 0 1 1  6 2 2 1 0 0 1 0 0  8 4 0 1 0 0 1 0 2|} & $1$ & $1$ &101& $\checkmark$ \\
$26$ & {\tiny\verb|6 2 2 1 0 0 1 0 0  2 0 0 0 0 0 0 1 1  6 2 2 0 1 1 0 0 0  8 4 0 1 0 0 1 0 2|} & $1$ & $1$ &101& $\checkmark$ \\
$27$ & {\tiny\verb|14 3 2 0 1 1 0 7  6 1 1 0 0 0 1 3  14 3 0 1 1 0 2 7|} & $1$ & $1$ &73& $\checkmark$ \\
$28$ & {\tiny\verb|10 2 0 1 1 0 0 1 5  8 2 1 0 0 0 1 0 4  8 2 0 1 0 1 0 0 4  14 3 1 1 0 0 0 2 7|} & $1$ & $1$ &73& $\checkmark$ \\
$29$ & {\tiny\verb|14 3 2 0 0 1 0 1 7  6 1 1 0 0 0 1 0 3  8 2 0 1 0 0 0 1 4  8 2 0 0 1 1 0 0 4|} & $1$ & $1$ &73& $\checkmark$ \\
$30$ & {\tiny\verb|10 1 1 2 0 0 1 0 0 5  4 0 1 0 0 1 0 0 0 2  4 0 0 1 1 0 0 0 0 2  6 0 1 1 0 0 0 0 1 3  10 1 2 1 0 0 0 1 0 5|} & $1$ & $2$ &97&  \\
%dP5
$31$ & {\tiny\verb|5 1 1 1 0 0 0 1 1 0  3 1 0 1 0 0 1 0 0 0  3 0 1 1 0 1 0 0 0 0  4 1 1 0 1 0 0 1 0 0  4 1 1 0 0 0 0 0 1 1|} & $5$ & $1$ &53&  \\
$32$ & {\tiny\verb|4 1 0 0 0 1 0 1 0 1  2 0 0 1 0 0 0 0 0 1  2 0 0 0 0 0 1 1 0 0  3 1 0 0 0 1 0 0 1 0  3 0 1 0 1 0 0 0 1 0|} & $5$ & $1$ &59& $\checkmark$ \\
$33$ & {\tiny\verb|4 1 1 0 0 0 1 0 1 0  2 1 0 0 0 0 0 0 0 1  2 0 0 1 0 0 0 1 0 0  2 0 0 0 0 1 1 0 0 0  4 1 0 0 1 0 1 1 0 0|} & $5$ & $1$ &63& $\checkmark$ \\
$34$ & {\tiny\verb|6 2 1 0 1 0 1 1 0 0  2 0 1 0 0 0 0 0 1 0  4 1 1 1 0 0 0 1 0 0  4 1 1 0 0 0 1 0 0 1  6 1 2 0 0 1 1 1 0 0|} & $5$ & $1$ &63&  \\
$35$ & {\tiny\verb|5 1 1 0 1 0 1 1 0 0  2 0 1 0 0 0 0 0 1 0  3 0 1 1 0 0 0 1 0 0  3 0 1 0 0 0 1 0 0 1  5 0 2 0 0 1 1 1 0 0|} & $5$ & $1$ &69& $\checkmark$ \\
%
%dP6
$36$ & {\tiny\verb|3 1 0 0 0 1 0 0 0 1  3 1 0 0 0 0 1 0 1 0  3 0 1 0 1 0 0 0 0 1  3 0 1 0 0 0 1 1 0 0  3 0 0 1 1 0 0 0 1 0|} & $6$ & $1$ &50& $\checkmark$ \\
$37$ & {\tiny\verb|5 1 1 0 0 0 1 1 0 1  3 1 0 1 0 0 1 0 0 0  3 0 0 0 0 0 0 1 1 1  4 0 0 1 1 0 1 1 0 0  5 1 0 0 0 1 1 1 1 0|} & $6$ & $1$ &52&  \\
$38$ & {\tiny\verb|6 0 2 1 0 1 1 1 0 0  3 1 0 0 1 0 0 0 1 0  3 1 0 0 0 1 1 0 0 0  3 0 1 1 0 0 0 0 1 0  4 0 1 0 1 0 0 0 1 1|} & $6$ & $1$ &53&  \\
$39$ & {\tiny\verb|5 1 1 0 0 0 1 0 1 1  3 1 1 0 0 0 0 1 0 0  3 0 0 1 0 0 1 0 1 0  3 0 0 0 1 0 1 0 0 1  4 1 0 0 0 1 1 1 0 0|} & $6$ & $1$ &55&  \\
$40$ & {\tiny\verb|5 1 1 1 0 0 0 0 1 1  3 1 0 0 1 0 0 1 0 0  3 1 0 0 0 1 1 0 0 0  4 1 1 0 0 1 0 0 1 0  4 1 1 0 0 0 0 1 0 1|} & $6$ & $1$ &56& $\checkmark$ \\
$41$ & {\tiny\verb|7 2 1 0 0 2 1 1 0 0  3 1 0 1 0 0 0 1 0 0  3 1 0 0 0 1 0 0 1 0  4 1 1 0 0 1 0 0 0 1  4 1 0 1 1 0 0 0 0 1|} & $6$ & $1$ &56&  \\
$42$ & {\tiny\verb|7 0 2 0 1 1 1 2 0 0  2 1 0 0 0 0 0 0 1 0  3 1 0 1 0 0 1 0 0 0  3 1 0 0 0 1 0 0 0 1  3 0 1 0 0 0 0 1 1 0|} & $6$ & $1$ &57&  \\
$43$ & {\tiny\verb|5 1 1 0 0 1 1 0 1 0  2 1 0 0 0 0 0 0 0 1  2 0 0 1 0 0 0 1 0 0  2 0 0 0 1 0 1 0 0 0  3 0 1 1 0 0 0 0 1 0|} & $6$ & $1$ &57& $\checkmark$ \\
$44$ & {\tiny\verb|5 0 1 1 0 0 0 1 1 1  2 0 0 1 0 0 1 0 0 0  2 0 0 0 0 1 0 0 0 1  3 1 0 0 1 0 0 1 0 0  3 0 0 0 1 1 1 0 0 0|} & $6$ & $1$ &58& $\checkmark$ \\
$45$ & {\tiny\verb|5 1 1 0 1 0 1 0 1  2 0 0 0 0 0 0 1 1  3 0 1 0 1 1 0 0 0  5 0 1 1 1 0 1 1 0|} & $6$ & $1$ &59& $\checkmark$ \\
$46$ & {\tiny\verb|6 2 1 0 0 1 1 1 0 0  3 1 0 1 0 0 0 1 0 0  3 1 0 0 1 0 1 0 0 0  3 1 0 0 0 1 0 0 1 0  4 1 1 0 0 0 0 1 0 1|} & $6$ & $1$ &59&  \\
$47$ & {\tiny\verb|5 1 0 1 0 1 0 1 0 1  2 0 0 0 0 0 0 0 1 1  3 1 1 0 0 0 0 1 0 0  4 1 0 1 0 0 1 1 0 0  4 1 0 0 1 1 0 1 0 0|} & $6$ & $1$ &59& $\checkmark$ \\
$48$ & {\tiny\verb|5 1 0 0 1 0 1 0 1 1  2 0 0 1 0 0 0 0 0 1  3 1 0 0 0 0 0 1 1 0  3 0 1 0 0 0 1 0 1 0  3 0 0 1 0 1 0 0 1 0|} & $6$ & $1$ &59&  \\
$49$ & {\tiny\verb|5 0 1 1 1 1 0 0 1 0  2 0 1 0 0 0 0 1 0 0  2 0 0 0 1 0 0 0 0 1  3 1 0 1 0 1 0 0 0 0  4 0 1 1 1 0 1 0 0 0|} & $6$ & $1$ &59&  \\
$50$ & {\tiny\verb|5 1 0 1 0 0 1 1 1 0  2 1 0 0 0 0 0 0 0 1  3 0 1 0 0 0 1 1 0 0  3 0 0 0 1 0 1 0 1 0  6 1 0 0 0 1 2 1 1 0|} & $6$ & $1$ &59&  \\
$51$ & {\tiny\verb|5 1 0 1 1 0 0 0 1 1  3 1 0 0 1 1 0 0 0 0  4 1 1 0 1 0 0 0 1 0  4 1 0 1 1 0 0 1 0 0  4 1 0 0 1 0 1 0 0 1|} & $6$ & $1$ &62& $\checkmark$ \\
$52$ & {\tiny\verb|6 2 1 0 0 1 1 0 0 1  3 1 0 1 0 0 0 0 0 1  3 1 0 0 1 0 1 0 0 0  4 1 0 0 0 1 1 0 1 0  6 2 0 1 0 1 1 1 0 0|} & $6$ & $1$ &62&  \\
$53$ & {\tiny\verb|5 1 1 1 0 0 0 0 1 1  3 1 0 0 0 0 1 0 1 0  3 0 1 0 0 1 0 0 1 0  3 0 0 1 1 0 0 0 1 0  6 1 1 1 0 0 0 1 2 0|} & $6$ & $1$ &63&  \\
$54$ & {\tiny\verb|7 2 2 1 1 0 1 0  6 2 2 0 1 1 0 0  7 2 2 1 0 1 0 1|} & $6$ & $1$ &65& $\checkmark$ \\
$55$ & {\tiny\verb|7 2 2 1 0 1 0 0 1  4 1 1 1 0 0 1 0 0  4 1 1 0 0 0 0 1 1  6 2 2 0 1 1 0 0 0|} & $6$ & $1$ &65& $\checkmark$ \\
$56$ & {\tiny\verb|7 2 1 1 0 0 2 1 0  4 1 0 0 1 0 1 1 0  6 2 1 0 0 1 2 0 0  7 2 1 0 1 0 2 0 1|} & $6$ & $1$ &65&  \\
$57$ & {\tiny\verb|5 1 0 0 0 1 1 1 1 0  2 1 0 0 0 0 0 0 0 1  2 0 1 0 0 0 0 0 1 0  2 0 0 0 1 0 0 1 0 0  4 1 0 1 0 0 0 1 1 0|} & $6$ & $1$ &65& $\checkmark$ \\
$58$ & {\tiny\verb|6 2 0 1 0 1 1 1 0 0  3 1 0 0 0 1 0 0 0 1  4 1 1 0 0 1 1 0 0 0  4 1 0 0 0 1 0 1 1 0  7 2 0 0 1 2 1 1 0 0|} & $6$ & $1$ &67&  \\
$59$ & {\tiny\verb|6 2 0 1 0 1 1 1 0 0  2 1 0 0 0 0 0 0 0 1  3 1 1 0 0 0 1 0 0 0  3 1 0 0 0 0 0 1 1 0  5 2 0 0 1 0 1 1 0 0|} & $6$ & $1$ &77& $\checkmark$ \\
$60$ & {\tiny\verb|7 3 0 0 1 1 1 1 0 0  2 1 0 0 0 0 0 0 0 1  5 2 1 0 0 1 0 1 0 0  5 2 0 0 0 1 1 0 1 0  8 3 0 1 0 2 1 1 0 0|} & $6$ & $1$ &79&  \\
$61$ & {\tiny\verb|7 3 0 0 1 1 1 1 0  4 2 1 0 0 0 0 1 0  4 2 0 0 0 0 1 0 1  6 3 0 1 0 0 1 1 0|} & $6$ & $1$ &97& $\checkmark$ \\
$62$ & {\tiny\verb|6 1 0 0 1 0 1 1 0 2  3 1 0 0 0 0 0 0 1 1  3 0 1 0 0 0 0 1 0 1  3 0 0 1 0 0 1 0 0 1  3 0 0 0 1 1 0 0 0 1|} & $6$ & $1$ &59& $\checkmark$ \\
$63$ & {\tiny\verb|6 1 1 0 1 0 1 0 2  3 1 0 0 0 0 0 1 1  3 0 0 0 1 1 0 0 1  6 0 1 1 0 1 0 1 2|} & $6$ & $1$ &59&  \\
%dP7
$64$ & {\tiny\verb|5 1 1 1 0 0 0 1 1 0  3 1 0 0 0 0 1 0 0 1  4 1 1 0 0 0 0 1 0 1  4 1 0 0 1 1 1 0 0 0  5 1 1 0 0 1 0 0 1 1|} & $7$ & $1$ &44&  \\
$65$ & {\tiny\verb|4 1 0 1 1 0 0 1  4 1 0 1 0 1 1 0  4 0 2 0 0 1 0 1|} & $7$ & $1$ &47& $\checkmark$ \\
$66$ & {\tiny\verb|8 2 0 2 1 0 0 1 2  4 1 0 1 0 1 0 1 0  4 1 0 0 1 0 1 0 1  4 0 2 0 1 0 0 1 0|} & $7$ & $1$ &47& $\checkmark$ \\
$67$ & {\tiny\verb|6 1 1 1 1 2 0 0 0 0  2 1 0 0 0 0 0 0 0 1  2 0 1 0 0 0 0 1 0 0  4 0 1 1 0 1 1 0 0 0  4 0 1 0 1 1 0 0 1 0|} & $7$ & $1$ &51&  \\
$68$ & {\tiny\verb|5 1 1 0 0 1 0 1 1 0  2 0 1 0 0 0 0 0 0 1  3 0 0 0 1 0 0 1 0 1  4 1 1 1 0 0 0 0 1 0  5 0 1 1 1 0 1 0 1 0|} & $7$ & $1$ &53&  \\
$69$ & {\tiny\verb|4 1 0 1 0 0 0 0 1 1  2 0 0 0 0 1 1 0 0 0  3 1 0 0 1 0 0 0 1 0  3 1 0 0 0 0 1 0 0 1  3 0 1 0 1 0 0 1 0 0|} & $7$ & $1$ &53& $\checkmark$ \\
$70$ & {\tiny\verb|5 0 1 0 0 0 1 1 1 1  2 0 0 0 0 1 0 0 1 0  3 1 0 0 0 0 1 0 0 1  3 0 1 0 1 0 0 0 0 1  3 0 0 1 0 0 0 1 0 1|} & $7$ & $1$ &53&  \\
$71$ & {\tiny\verb|4 1 1 0 0 0 1 0 1  4 1 0 1 0 1 0 0 1  4 0 1 1 1 0 0 0 1  4 0 1 1 0 1 0 1 0|} & $7$ & $1$ &55& $\checkmark$ \\
$72$ & {\tiny\verb|8 3 0 1 1 2 0 0 1  4 1 1 0 0 0 1 1 0  4 2 1 0 0 0 0 0 1  4 1 0 1 0 1 0 1 0|} & $7$ & $1$ &55&  \\
$73$ & {\tiny\verb|8 1 0 0 3 2 1 1 0 0  2 0 0 0 1 0 0 0 0 1  4 1 0 0 1 1 0 0 1 0  4 0 1 1 0 1 0 0 1 0  4 0 1 0 1 1 0 1 0 0|} & $7$ & $1$ &55& $\checkmark$ \\
$74$ & {\tiny\verb|4 1 0 1 1 0 0 1 0 0  2 0 1 0 0 0 0 0 0 1  2 0 0 1 0 0 0 0 1 0  2 0 0 0 0 1 1 0 0 0  4 1 0 0 1 0 1 0 0 1|} & $7$ & $1$ &55& $\checkmark$ \\
$75$ & {\tiny\verb|6 1 1 1 0 0 1 2 0 0  2 0 0 1 0 0 0 0 0 1  3 1 0 0 0 1 0 1 0 0  3 0 1 0 1 0 0 1 0 0  4 1 0 1 0 1 0 0 1 0|} & $7$ & $1$ &55& $\checkmark$ \\
$76$ & {\tiny\verb|6 1 2 0 0 0 1 1 0 1  2 0 0 1 0 0 0 1 0 0  3 1 1 0 0 0 0 0 1 0  3 0 1 0 0 1 0 0 0 1  4 0 1 0 1 1 0 0 1 0|} & $7$ & $1$ &55&  \\
$77$ & {\tiny\verb|5 1 1 0 1 0 1 1 0 0  2 0 1 0 0 0 0 0 1 0  3 1 0 0 0 0 0 1 0 1  4 1 1 1 0 0 1 0 0 0  4 1 1 0 1 1 0 0 0 0|} & $7$ & $1$ &57&  \\
$78$ & {\tiny\verb|4 1 1 0 0 0 0 1 0 1  2 0 1 0 0 0 0 0 1 0  2 0 0 0 1 1 0 0 0 0  2 0 0 0 0 0 1 0 0 1  3 1 0 1 0 0 0 1 0 0|} & $7$ & $1$ &57& $\checkmark$ \\
$79$ & {\tiny\verb|4 1 0 0 0 1 1 1 0 0  2 1 0 0 0 0 0 0 0 1  2 0 1 0 0 0 0 0 1 0  2 0 0 1 0 0 0 1 0 0  2 0 0 0 1 0 1 0 0 0|} & $7$ & $1$ &59&  \\
$80$ & {\tiny\verb|6 2 1 1 0 0 0 1 1 0  3 1 1 0 0 0 1 0 0 0  3 1 0 1 0 1 0 0 0 0  3 1 0 0 1 0 0 1 0 0  7 2 1 1 0 0 0 2 0 1|} & $7$ & $1$ &59&  \\
$81$ & {\tiny\verb|6 1 0 0 2 0 1 1 1 0  2 0 0 0 1 1 0 0 0 0  4 1 0 0 1 0 1 0 0 1  4 0 1 0 1 0 1 0 1 0  4 0 0 1 1 0 1 1 0 0|} & $7$ & $1$ &59&  \\
$82$ & {\tiny\verb|6 2 1 0 0 0 1 1 0 1  2 1 0 0 0 0 0 0 1 0  2 0 0 1 0 0 1 0 0 0  4 1 0 0 0 1 1 0 0 1  6 2 0 0 1 1 1 1 0 0|} & $7$ & $1$ &59&  \\
$83$ & {\tiny\verb|5 1 0 0 1 1 1 0 1 0  2 1 0 0 0 0 0 0 0 1  4 1 1 0 0 0 1 0 1 0  4 1 0 1 0 1 0 0 1 0  5 2 0 0 1 0 0 1 1 0|} & $7$ & $1$ &61&  \\
$84$ & {\tiny\verb|6 1 0 1 2 1 0 1 0 0  2 0 0 0 1 0 1 0 0 0  3 1 0 0 1 0 0 0 1 0  3 0 0 0 1 0 0 1 0 1  4 0 1 1 1 1 0 0 0 0|} & $7$ & $1$ &63& $\checkmark$ \\
$85$ & {\tiny\verb|6 0 1 1 0 1 0 1 2 0  2 0 0 1 0 0 0 0 0 1  3 1 0 0 0 0 0 1 1 0  3 0 0 1 0 0 1 0 1 0  3 0 0 0 1 1 0 0 1 0|} & $7$ & $1$ &63&  \\
$86$ & {\tiny\verb|6 2 1 1 0 1 0 0 1  2 1 0 0 0 0 0 1 0  2 0 0 0 0 0 1 0 1  6 2 1 0 1 1 1 0 0|} & $7$ & $1$ &65& $\checkmark$ \\
$87$ & {\tiny\verb|5 2 0 0 1 1 1 0 0 0  2 1 0 0 0 0 0 0 1 0  2 0 1 0 0 0 0 1 0 0  3 1 0 1 0 0 1 0 0 0  3 1 0 0 0 1 0 0 0 1|} & $7$ & $1$ &65&  \\
$88$ & {\tiny\verb|6 2 1 0 1 0 1 0 0 1  2 1 0 0 0 0 0 0 1 0  2 0 0 0 0 0 0 1 0 1  5 2 1 0 0 1 1 0 0 0  5 2 0 1 1 0 1 0 0 0|} & $7$ & $1$ &65& $\checkmark$ \\
$89$ & {\tiny\verb|6 2 0 0 1 0 1 1 1 0  2 1 0 0 0 0 0 0 0 1  4 1 1 0 0 0 1 0 1 0  5 2 0 1 0 0 1 1 0 0  5 2 0 0 1 1 1 0 0 0|} & $7$ & $1$ &67&  \\
$90$ & {\tiny\verb|8 1 3 1 1 0 2 0 0 0  4 0 2 1 0 0 0 0 1 0  5 1 2 0 0 0 1 1 0 0  5 0 2 0 1 1 1 0 0 0  6 1 3 0 0 0 0 1 0 1|} & $7$ & $1$ &67&  \\
$91$ & {\tiny\verb|6 2 1 0 1 1 1 0 0 0  2 1 0 0 0 0 0 0 0 1  2 0 1 0 0 0 0 0 1 0  4 1 1 1 0 0 1 0 0 0  4 1 1 0 0 1 0 1 0 0|} & $7$ & $1$ &67&  \\
$92$ & {\tiny\verb|6 1 2 0 0 1 1 1 0 0  2 0 1 0 0 0 0 0 1 0  3 1 1 0 0 0 0 0 0 1  3 0 1 1 0 0 0 1 0 0  3 0 1 0 1 0 1 0 0 0|} & $7$ & $1$ &69&  \\
$93$ & {\tiny\verb|6 2 1 1 0 0 1 0 1 0  2 1 0 0 0 0 0 0 0 1  3 1 0 1 0 0 0 1 0 0  3 1 0 0 0 1 0 0 1 0  5 2 1 0 1 0 1 0 0 0|} & $7$ & $1$ &69& $\checkmark$ \\
$94$ & {\tiny\verb|8 1 0 3 0 1 0 1 2  4 0 1 1 0 0 0 1 1  4 0 0 1 1 1 0 0 1  4 0 0 2 0 0 1 0 1|} & $7$ & $1$ &71&  \\
$95$ & {\tiny\verb|8 1 0 3 0 0 1 1 2  4 0 0 2 1 0 0 0 1  4 0 0 1 0 1 0 1 1  8 0 1 3 0 1 1 0 2|} & $7$ & $1$ &71&  \\
$96$ & {\tiny\verb|7 2 1 0 0 1 2 1 0 0  2 1 0 0 0 0 0 0 0 1  3 1 0 0 1 0 1 0 0 0  4 1 1 0 0 0 1 0 1 0  4 1 0 1 0 0 1 1 0 0|} & $7$ & $1$ &71&  \\
$97$ & {\tiny\verb|6 2 1 1 0 0 0 1 1  4 2 0 0 0 1 1 0 0  5 2 1 0 1 0 0 1 0  5 2 1 0 0 0 1 0 1|} & $7$ & $1$ &73&  \\
$98$ & {\tiny\verb|7 3 0 0 1 0 1 1 1 0  2 1 0 0 0 0 0 0 0 1  5 2 1 0 0 0 0 1 1 0  5 2 0 1 0 0 1 1 0 0  5 2 0 0 1 1 0 1 0 0|} & $7$ & $1$ &73&  \\
$99$ & {\tiny\verb|7 0 3 1 1 0 0 1 1 0  5 1 2 0 1 0 0 0 1 0  5 0 2 1 1 0 1 0 0 0  5 0 2 0 1 1 0 1 0 0  6 1 3 0 0 0 0 0 1 1|} & $7$ & $1$ &73&  \\
$100$ & {\tiny\verb|7 3 1 1 0 1 0 0 1  3 1 0 0 1 1 0 0 0  5 2 1 0 0 0 0 1 1  6 3 1 1 0 0 1 0 0|} & $7$ & $1$ &77&  \\
$101$ & {\tiny\verb|6 1 2 1 1 1 0 0 0 0  2 0 1 0 0 0 0 0 1 0  3 1 1 0 0 0 0 0 0 1  3 0 1 1 0 0 1 0 0 0  7 1 3 1 0 1 0 1 0 0|} & $7$ & $1$ &77&  \\
$102$ & {\tiny\verb|8 1 4 1 0 0 1 0 1  4 0 2 1 0 0 0 1 0  4 0 2 0 0 1 0 0 1  10 2 5 0 1 0 2 0 0|} & $7$ & $1$ &81& $\checkmark$ \\
$103$ & {\tiny\verb|7 1 3 1 0 1 0 0 0 1  4 0 2 1 0 0 1 0 0 0  4 0 2 0 0 0 0 1 0 1  5 1 2 0 1 1 0 0 0 0  6 0 3 0 1 1 0 0 1 0|} & $7$ & $1$ &81& $\checkmark$ \\
$104$ & {\tiny\verb|8 1 4 1 0 0 1 0 1 0  4 0 2 1 0 0 0 1 0 0  4 0 2 0 0 1 0 0 1 0  5 1 2 0 1 0 1 0 0 0  6 1 3 0 0 0 1 0 0 1|} & $7$ & $1$ &81& $\checkmark$ \\
$105$ & {\tiny\verb|8 1 3 0 0 1 1 2 0 0  3 0 1 1 0 0 0 1 0 0  4 0 2 1 0 0 0 0 0 1  5 1 2 0 0 0 0 1 1 0  5 0 2 0 1 0 1 1 0 0|} & $7$ & $1$ &83&  \\
$106$ & {\tiny\verb|8 3 1 0 0 1 1 2 0 0  2 1 0 0 0 0 0 0 0 1  3 1 0 1 0 0 0 1 0 0  5 2 1 0 0 0 0 1 1 0  5 2 0 0 1 0 1 1 0 0|} & $7$ & $1$ &83&  \\
$107$ & {\tiny\verb|7 1 3 1 1 0 0 1 0 0  3 0 1 0 1 1 0 0 0 0  4 1 2 0 0 0 0 0 1 0  4 0 2 1 0 0 1 0 0 0  4 0 2 0 0 1 0 0 0 1|} & $7$ & $1$ &83&  \\
$108$ & {\tiny\verb|9 2 4 1 0 0 1 0 1 0  4 1 2 0 0 0 0 1 0 0  6 1 3 0 0 1 0 0 1 0  9 2 4 0 1 1 1 0 0 0  10 2 5 1 0 0 0 0 1 1|} & $7$ & $1$ &83&  \\
$109$ & {\tiny\verb|10 1 5 0 1 1 2 0 0  6 0 3 0 1 0 1 1 0  6 0 3 0 0 1 1 0 1  10 2 5 1 0 0 2 0 0|} & $7$ & $1$ &85&  \\
$110$ & {\tiny\verb|7 1 3 0 1 0 1 1 0 0  5 1 2 1 0 0 1 0 0 0  6 1 3 1 0 0 0 0 0 1  6 1 3 0 1 1 0 0 0 0  6 1 3 0 0 0 0 1 1 0|} & $7$ & $1$ &85&  \\
$111$ & {\tiny\verb|9 4 0 1 1 1 0 2 0  4 2 1 0 0 0 0 1 0  5 2 0 1 0 0 1 1 0  5 2 0 0 1 0 0 1 1|} & $7$ & $1$ &89&  \\
$112$ & {\tiny\verb|8 4 1 1 0 1 1 0 0  4 2 0 0 1 0 1 0 0  5 2 1 0 0 1 0 1 0  8 4 1 0 1 1 0 0 1|} & $7$ & $1$ &89& $\checkmark$ \\
$113$ & {\tiny\verb|8 1 4 0 1 1 1 0 0  4 0 2 0 0 1 0 0 1  8 1 4 0 0 0 1 1 1  10 2 5 1 0 0 2 0 0|} & $7$ & $1$ &89& $\checkmark$ \\
$114$ & {\tiny\verb|7 1 3 0 1 1 1 0 0 0  4 1 2 0 0 0 0 0 1 0  4 0 2 1 0 1 0 0 0 0  4 0 2 0 0 0 1 1 0 0  8 1 4 0 0 1 1 0 0 1|} & $7$ & $1$ &89&  \\
$115$ & {\tiny\verb|8 1 4 0 0 1 1 0 1  10 2 5 1 0 0 2 0 0  12 2 6 0 1 1 2 0 0  12 2 6 0 0 0 2 1 1|} & $7$ & $1$ &97& $\checkmark$ \\
$116$ & {\tiny\verb|9 2 4 0 1 1 1 0 0 0  4 1 2 0 0 0 0 1 0 0  6 1 3 1 0 0 1 0 0 0  6 1 3 0 0 1 0 0 1 0  10 2 5 0 0 1 1 0 0 1|} & $7$ & $1$ &101&  \\
$117$ & {\tiny\verb|10 2 1 0 0 1 1 5  4 0 0 1 1 0 0 2  12 2 1 1 0 0 2 6|} & $7$ & $2$ &77&  \\
$118$ & {\tiny\verb|8 1 0 0 1 0 1 1 0 4  4 1 0 0 0 0 0 0 1 2  4 0 1 0 0 0 0 1 0 2  4 0 0 1 0 0 1 0 0 2  4 0 0 0 1 1 0 0 0 2|} & $7$ & $1$ &81&  \\
$119$ & {\tiny\verb|8 1 1 1 0 0 0 1 4  4 0 1 0 0 0 1 0 2  4 0 0 1 0 1 0 0 2  8 1 0 0 1 1 1 0 4|} & $7$ & $1$ &81&  \\
$120$ & {\tiny\verb|8 1 0 1 1 0 0 0 1 4  4 0 1 0 0 0 1 0 0 2  6 1 1 0 0 0 0 0 1 3  6 1 0 1 0 0 0 1 0 3  6 1 0 0 1 1 0 0 0 3|} & $7$ & $1$ &81&  \\
$121$ & {\tiny\verb|8 0 0 1 1 1 0 0 1 4  4 1 0 0 0 0 0 1 0 2  4 0 1 0 0 0 1 0 0 2  6 1 0 0 1 0 0 0 1 3  6 0 1 1 0 0 0 0 1 3|} & $7$ & $1$ &81& $\checkmark$ \\
$122$ & {\tiny\verb|10 0 1 1 0 1 0 0 2 5  4 1 0 0 0 0 0 0 1 2  4 0 1 0 0 0 0 1 0 2  6 0 0 1 0 0 1 0 1 3  6 0 0 0 1 1 0 0 1 3|} & $7$ & $1$ &81&  \\
$123$ & {\tiny\verb|10 1 0 2 1 1 0 0 5  6 1 0 1 0 0 0 1 3  6 0 0 1 1 0 1 0 3  10 0 1 2 0 2 0 0 5|} & $7$ & $1$ &85& $\checkmark$ \\
$124$ & {\tiny\verb|8 1 0 0 1 0 1 0 1 4  4 1 0 0 0 0 0 1 0 2  4 0 1 0 0 0 1 0 0 2  4 0 0 1 0 0 0 0 1 2  6 0 0 1 0 1 0 1 0 3|} & $7$ & $1$ &85&  \\
$125$ & {\tiny\verb|8 1 1 0 0 1 0 1 4  4 0 1 0 0 0 1 0 2  4 0 0 1 0 0 0 1 2  8 0 1 1 1 1 0 0 4|} & $7$ & $1$ &89&  \\
$126$ & {\tiny\verb|8 0 1 1 0 0 0 1 1 4  4 1 0 0 0 0 0 1 0 2  4 0 0 0 1 0 0 0 1 2  6 0 1 0 0 0 1 1 0 3  6 0 0 1 0 1 0 1 0 3|} & $7$ & $1$ &89& $\checkmark$ \\
$127$ & {\tiny\verb|8 0 1 1 1 0 0 1 0 4  4 0 1 0 0 0 1 0 0 2  6 1 1 0 0 0 0 1 0 3  6 0 1 1 0 0 0 0 1 3  6 0 1 0 1 1 0 0 0 3|} & $7$ & $1$ &93& $\checkmark$ \\
$128$ & {\tiny\verb|10 2 1 1 0 0 1 5  8 2 0 1 1 0 0 4  10 2 1 0 1 1 0 5|} & $7$ & $1$ &97& $\checkmark$ \\
$129$ & {\tiny\verb|10 2 1 0 0 1 0 1 5  6 1 1 0 0 0 1 0 3  6 1 0 0 1 1 0 0 3  8 2 0 1 0 0 0 1 4|} & $7$ & $1$ &97& $\checkmark$ \\
$130$ & {\tiny\verb|10 2 1 1 0 0 1 0 5  6 1 0 1 1 0 0 0 3  8 2 1 0 0 0 0 1 4  10 2 1 0 1 1 0 0 5|} & $7$ & $1$ &97&  \\
$131$ & {\tiny\verb|10 2 1 1 0 0 0 0 1 5  4 1 0 0 0 0 1 0 0 2  6 1 1 0 0 0 0 1 0 3  6 1 0 1 0 1 0 0 0 3  10 2 0 1 1 0 0 1 0 5|} & $7$ & $1$ &97&  \\
%dP8
$132$ & {\tiny\verb|8 2 1 2 1 0 2  4 0 1 0 1 2 0 /Z2: 0 1 1 0 0 0|} & $8$ & $1$ &39& $\checkmark$ \\
$133$ & {\tiny\verb|6 1 0 0 1 1 0 1 0 2  2 0 0 1 0 0 0 0 1 0  2 0 0 0 1 0 1 0 0 0  4 1 0 0 0 1 0 0 1 1  4 0 1 0 0 1 0 1 1 0|} & $8$ & $1$ &43&  \\
$134$ & {\tiny\verb|4 1 0 0 1 1 0 1  2 0 1 0 0 0 0 1  2 0 0 1 0 0 1 0 /Z2: 0 0 1 1 0 0 0|} & $8$ & $1$ &45& $\checkmark$ \\
$135$ & {\tiny\verb|6 1 1 0 1 2 0 1 0  4 1 0 0 1 1 0 0 1  4 2 0 0 0 0 1 0 1  6 1 0 1 2 1 0 1 0|} & $8$ & $1$ &47&  \\
$136$ & {\tiny\verb|5 1 1 0 1 1 1 0  2 0 0 1 0 0 0 1  6 2 0 0 1 2 0 1|} & $8$ & $2$ &49&  \\
$137$ & {\tiny\verb|5 0 1 1 1 0 0 1 0 1  3 1 0 0 1 0 0 1 0 0  4 1 0 0 1 0 1 0 0 1  4 0 1 0 1 0 0 1 1 0  5 0 1 0 1 1 1 1 0 0|} & $8$ & $1$ &50&  \\
$138$ & {\tiny\verb|4 1 1 1 0 0 1  4 0 2 0 1 1 0 /Z2: 0 0 1 1 0 0|} & $8$ & $1$ &51& $\checkmark$ \\
$139$ & {\tiny\verb|6 2 1 1 1 0 1  6 3 1 0 0 1 1 /Z2: 0 1 1 0 0 0|} & $8$ & $1$ &53& $\checkmark$ \\
$140$ & {\tiny\verb|6 3 1 1 0 0 1  2 0 0 0 1 1 0 /Z2: 0 1 0 1 0 0|} & $8$ & $1$ &53& $\checkmark$ \\
$141$ & {\tiny\verb|4 1 0 1 0 0 0 0 1 1  2 0 0 0 1 0 0 1 0 0  3 1 0 1 0 0 1 0 0 0  3 0 1 0 1 0 0 0 1 0  3 0 0 0 1 1 0 0 0 1|} & $8$ & $1$ &53&  \\
$142$ & {\tiny\verb|5 1 1 0 1 1 0 0 1 0  2 0 0 0 1 0 1 0 0 0  3 1 1 0 0 0 0 1 0 0  3 1 0 1 0 0 0 0 1 0  3 0 1 0 0 1 0 0 0 1|} & $8$ & $1$ &53&  \\
$143$ & {\tiny\verb|5 1 1 0 0 1 0 1 0 1  2 0 1 0 0 0 0 0 1 0  2 0 0 0 1 0 0 0 0 1  2 0 0 0 0 1 1 0 0 0  3 1 0 1 0 0 0 1 0 0|} & $8$ & $1$ &53&  \\
$144$ & {\tiny\verb|4 1 0 1 0 0 0 1 0 1  2 0 0 1 0 0 0 0 1 0  2 0 0 0 0 0 1 0 0 1  3 1 0 0 0 1 0 1 0 0  3 0 1 0 1 0 0 1 0 0|} & $8$ & $1$ &56& $\checkmark$ \\
$145$ & {\tiny\verb|5 1 1 0 1 0 0 1 1 0  3 1 0 1 0 0 0 0 1 0  3 1 0 0 1 1 0 0 0 0  3 0 0 0 1 0 1 1 0 0  5 1 0 0 1 0 1 0 1 1|} & $8$ & $1$ &56&  \\
$146$ & {\tiny\verb|4 1 0 0 0 1 0 1 0 1  2 0 1 0 0 0 0 0 0 1  2 0 0 0 1 0 0 1 0 0  2 0 0 0 0 1 1 0 0 0  3 0 0 1 0 1 0 0 1 0|} & $8$ & $1$ &61& $\checkmark$ \\
$147$ & {\tiny\verb|5 1 1 0 1 0 1 0 1 0  2 1 0 0 0 0 0 0 0 1  3 1 0 1 0 0 0 1 0 0  4 1 1 1 0 0 0 0 1 0  5 2 0 0 0 1 1 0 1 0|} & $8$ & $1$ &61&  \\
$148$ & {\tiny\verb|8 3 2 0 0 1 1 1 0 0  2 1 0 0 0 0 0 0 0 1  3 1 1 0 0 0 0 0 1 0  4 1 1 1 0 0 0 1 0 0  4 1 1 0 1 0 1 0 0 0|} & $8$ & $1$ &71&  \\
$149$ & {\tiny\verb|5 2 0 0 1 1 1 0 0 0  2 1 0 0 0 0 0 0 1 0  3 1 1 0 0 0 0 1 0 0  3 1 0 1 0 0 1 0 0 0  3 1 0 0 0 1 0 0 0 1|} & $8$ & $1$ &73& $\checkmark$ \\
$150$ & {\tiny\verb|7 3 0 0 1 1 1 1 0  3 1 1 0 0 0 0 0 1  5 2 1 0 1 1 0 0 0  6 3 0 1 0 1 1 0 0|} & $8$ & $1$ &77&  \\
$151$ & {\tiny\verb|7 3 0 0 0 1 1 1 1 0  2 1 0 0 0 0 0 0 0 1  3 1 1 0 0 0 0 0 1 0  5 2 0 1 0 0 0 1 1 0  5 2 0 0 1 0 1 0 1 0|} & $8$ & $1$ &83&  \\
$152$ & {\tiny\verb|6 3 0 1 0 0 1 1 0  3 1 0 0 1 1 0 0 0  4 2 1 0 0 0 0 1 0  4 2 0 0 0 0 1 0 1|} & $8$ & $1$ &89& $\checkmark$ \\
$153$ & {\tiny\verb|8 4 0 1 0 1 1 1 0 0  3 1 0 1 0 0 0 0 1 0  4 2 1 0 0 0 0 1 0 0  4 2 0 1 0 0 0 0 0 1  4 2 0 0 1 0 1 0 0 0|} & $8$ & $1$ &89&  \\
$154$ & {\tiny\verb|12 6 3 0 0 1 1 1 0  6 3 1 1 0 0 0 1 0  6 3 1 0 1 0 1 0 0  6 3 2 0 0 0 0 0 1|} & $8$ & $1$ &107& $\checkmark$ \\
$155$ & {\tiny\verb|12 6 3 1 0 0 1 1 0  3 1 1 0 0 0 0 0 1  6 3 1 0 1 0 0 1 0  6 3 1 0 0 1 1 0 0|} & $8$ & $1$ &107&  \\
$156$ & {\tiny\verb|10 2 5 0 1 1 1 0 0 0  3 1 1 0 0 0 0 0 0 1  4 1 2 0 0 0 0 0 1 0  6 1 3 1 0 0 1 0 0 0  6 1 3 0 0 1 0 1 0 0|} & $8$ & $1$ &107&  \\
$157$ & {\tiny\verb|12 6 3 0 1 1 1 0 0  3 1 1 0 0 0 0 1 0  8 4 2 1 0 0 1 0 0  8 4 2 0 0 1 0 0 1|} & $8$ & $1$ &131&  \\
$158$ & {\tiny\verb|12 6 3 0 1 1 1 0 0  2 0 1 0 0 0 0 1 0  8 4 2 1 0 0 1 0 0  8 4 2 0 0 1 0 0 1|} & $8$ & $1$ &131&  \\
$159$ & {\tiny\verb|12 6 3 0 1 1 1 0 0  8 4 2 1 0 0 1 0 0  8 4 2 0 0 1 0 0 1  12 0 6 1 0 0 1 4 0|} & $8$ & $1$ &131&  \\
$160$ & {\tiny\verb|14 7 4 0 1 1 1 0 0  6 3 2 0 0 0 0 1 0  8 4 2 1 0 0 1 0 0  8 4 2 0 0 1 0 0 1|} & $8$ & $1$ &137&  \\
$161$ & {\tiny\verb|16 5 8 0 1 1 1 0 0  6 2 3 0 0 0 0 1 0  10 3 5 1 0 0 1 0 0  10 3 5 0 0 1 0 0 1|} & $8$ & $1$ &173& $\checkmark$ \\
$162$ & {\tiny\verb|6 1 0 1 0 1 0 3  6 0 1 1 1 0 0 3  8 1 1 0 0 0 2 4|} & $8$ & $1$ &65&  \\
$163$ & {\tiny\verb|12 1 1 2 0 0 2 0 6  6 1 0 0 1 0 1 0 3  6 0 1 1 0 1 0 0 3  8 1 1 0 0 0 0 2 4|} & $8$ & $1$ &65&  \\
$164$ & {\tiny\verb|8 1 0 1 0 1 0 1 4  6 0 0 1 1 1 0 0 3  8 1 0 0 0 2 1 0 4  8 0 1 1 1 0 0 1 4|} & $8$ & $1$ &73& $\checkmark$ \\
$165$ & {\tiny\verb|8 1 1 1 0 0 0 1 4  8 1 0 1 1 0 1 0 4  8 1 0 2 0 1 0 0 4  10 1 1 0 1 0 2 0 5|} & $8$ & $1$ &73&  \\
$166$ & {\tiny\verb|8 1 1 1 0 0 0 0 1 4  4 0 1 0 0 0 1 0 0 2  6 1 1 0 0 0 0 1 0 3  6 0 0 1 1 0 1 0 0 3  8 1 1 0 1 1 0 0 0 4|} & $8$ & $1$ &73&  \\
$167$ & {\tiny\verb|18 9 1 1 0 1 0 6  12 6 1 0 0 0 1 4  18 9 0 1 1 0 1 6|} & $8$ & $1$ &185& $\checkmark$ \\
$168$ & {\tiny\verb|18 1 0 9 0 1 1 0 6  12 1 0 6 0 0 0 1 4  12 0 1 6 0 1 0 0 4  12 0 0 6 1 0 1 0 4|} & $8$ & $1$ &185& $\checkmark$
\end{longtable}
\end{center}

%%%%%%%%%%%%%%%%%%%%%%%%%%%%%%%%%%%%%%%%%%%%%%%%%%%%%%%
\section{RG details}
\label{sec:rgrunning}
%%%%%%%%%%%%%%%%%%%%%%%%%%%%%%%%%%%%%%%%%%%%%%%%%%%%%%%
Below the detailed $\beta-$functions for the soft-scalar masses, following the conventions of~\cite{9709356}
\begin{eqnarray}
\beta_{m^2_{QL1}}&=&\frac{1}{24\pi^2}\left[-32g^2|M|^2+(9A^2+27m^2)\left(2+\lambda_{11}^2+2 \lambda_{23}+\lambda_{23}^2-2 \lambda_{32}+\lambda_{32}^2\right)\right]\\
\nonumber \beta_{m^2_{QL2}}&=&\frac{1}{24\pi^2}\left[-32g^2|M|^2+(9A^2+27m^2)\left(2-2 \lambda_{13}+\lambda_{13}^2+\lambda_{22}^2+2 \lambda_{31}+\lambda_{31}^2\right)\right]\\
\nonumber \beta_{m^2_{QL3}}&=&\frac{1}{24\pi^2}\left[-32g^2|M|^2+(9A^2+27m^2)\left(2+2 \lambda_{12}+\lambda_{12}^2-2 \lambda_{21}+\lambda_{21}^2+\lambda_{33}^2\right)\right]\\
\nonumber\beta_{m^2_{\Phi 1}}&=&\frac{1}{24\pi^2}\left[-32g^2|M|^2+(9A^2+27m^2)\left(2+\lambda_{11}^2+2 \lambda_{12}+\lambda_{12}^2-2 \lambda_{13}+\lambda_{13}^2\right)\right]\\
\nonumber\beta_{m^2_{\Phi 2}}&=&\frac{1}{24\pi^2}\left[-32g^2|M|^2+(9A^2+27m^2)\left(2-2 \lambda_{21}+\lambda_{21}^2+\lambda_{22}^2+2 \lambda_{23}+\lambda_{23}^2\right)\right]\\
\nonumber\beta_{m^2_{\Phi 3}}&=&\frac{1}{24\pi^2}\left[-32g^2|M|^2+(9A^2+27m^2)\left(2+2 \lambda_{31}+\lambda_{31}^2-2 \lambda_{32}+\lambda_{32}^2+\lambda_{33}^2\right)\right]\\
\nonumber\beta_{m^2_{QR1}}&=&\frac{1}{24\pi^2}\left[-32g^2|M|^2+(9A^2+27m^2)\left(2+\lambda_{11}^2-2 \lambda_{21}+\lambda_{21}^2+2 \lambda_{31}+\lambda_{31}^2\right)\right]\\
\nonumber\beta_{m^2_{QR2}}&=&\frac{1}{24\pi^2}\left[-32g^2|M|^2+(9A^2+27m^2)\left(2+2 \lambda_{12}+\lambda_{12}^2+\lambda_{22}^2-2 \lambda_{32}+\lambda_{32}^2\right)\right]\\
\nonumber\beta_{m^2_{QR3}}&=&\frac{1}{24\pi^2}\left[-32g^2|M|^2+(9A^2+27m^2)\left(2-2 \lambda_{13}+\lambda_{13}^2+2 \lambda_{23}+\lambda_{23}^2+\lambda_{33}^2\right)\right]
\end{eqnarray}
For example taking the following non-vanishing values for $\lambda_{11}=\lambda_{12}=\lambda_{21}=1$ and $\lambda_{13}=\lambda_{31}=-1$ leads to an enhanced Yukawa contribution to $\beta_{m^2_{\Phi 1}}$ as required for a breakdown to the left-right model:
\begin{eqnarray}
\label{eq:rgexample}
\beta_{m^2_{QL1}}&=&\frac{1}{24\pi^2}\left[-32g^2|M|^2+3\,(9A^2+27m^2)\right]\\
\nonumber \beta_{m^2_{QL2}}&=&\frac{1}{24\pi^2}\left[-32g^2|M|^2+4\,(9A^2+27m^2)\right]\\
\nonumber \beta_{m^2_{QL3}}&=&\frac{1}{24\pi^2}\left[-32g^2|M|^2+4\,(9A^2+27m^2)\right]\\
\nonumber \beta_{m^2_{\Phi 1}}&=&\frac{1}{24\pi^2}\left[-32g^2|M|^2+9\,(9A^2+27m^2)\right]\\
\nonumber \beta_{m^2_{\Phi 2}}&=&\frac{1}{24\pi^2}\left[-32g^2|M|^2+(9A^2+27m^2)\right]\\
\nonumber \beta_{m^2_{\Phi 3}}&=&\frac{1}{24\pi^2}\left[-32g^2|M|^2+(9A^2+27m^2)\right]\\
\nonumber \beta_{m^2_{QR1}}&=&\frac{1}{24\pi^2}\left[-32g^2|M|^2+(9A^2+27m^2)\right]\\
\nonumber \beta_{m^2_{QR2}}&=&\frac{1}{24\pi^2}\left[-32g^2|M|^2+4\,(9A^2+27m^2)\right]\\
\nonumber \beta_{m^2_{QR3}}&=&\frac{1}{24\pi^2}\left[-32g^2|M|^2+4\,(9A^2+27m^2)\right]
\end{eqnarray}

\begin{footnotesize}
\bibliographystyle{utphys}
\bibliography{CKMQVv12}
\end{footnotesize}
\end{document}